\documentclass{article}
\usepackage{graphicx} 
\usepackage[square,numbers]{natbib}
\usepackage{hyperref}
\usepackage{amsmath}
\usepackage{amssymb}
\usepackage{graphicx}
\usepackage[margin=1in]{geometry}
\usepackage{booktabs}

\usepackage{pifont}

\newcommand{\cmark}{\ding{51}} 
\newcommand{\xmark}{\ding{55}} 
\newcommand{\healpix}{HEALPix\ }

\DeclareMathOperator*{\argmin}{arg\,min}
\title{Climate in a Bottle: Towards a Generative Foundation Model for the Kilometer-Scale Global Atmosphere}
\author{
  Noah D. Brenowitz\footnote{Corresponding author. Email nbrenowitz@nvidia.com} \and
  Tao Ge \and
  Akshay Subramaniam \and
  Peter Manshausen \and
  Aayush Gupta \and
  David M. Hall \and
  Morteza Mardani \and
  Arash Vahdat \and
  Karthik Kashinath \and
  Michael S. Pritchard
}
\date{NVIDIA, Santa Clara, CA, USA}

\begin{document}
\maketitle


\begin{abstract}
    Climate modeling is reaching unprecedented resolution, producing petabytes of climate data \citep{DestinE}. AI climate model emulators offer a path to computationally cheap analysis of this data, enabling new scientific insight and scenario planning \citep{EVE}. Recent advances have shown promise in faithfully emulating climate data \citep{ACE2,CAMulator, Cresswell2024,ClimBench}. However, prevailing auto-regressive paradigms are challenging to train on climate time horizons due to drifts, instabilities and component-coupling challenges, they are hard to scale to high resolution, and pose the challenge of having to sift through troves of output to identify rare climate extremes of outsize interest. 
    Here, we present Climate in a Bottle (\textit{cBottle}), a generative diffusion-based framework \citep{Ho2020,Song2020,Karras2022-edm} emulating global 5 km climate simulations and reanalysis on the \healpix grid. cBottle directly samples from the full distribution of atmospheric states without requiring a previous time step and therefore auto-regressive rollout, as well as the first to reach this global resolution (12.5M-pixel). 
    \textit{cBottle} consists of two model stages \citep{Ho2021}: a global coarse-resolution generator conditioned on sea surface temperatures and solar position, followed by a patch-based 16x super-resolution stage. 
    \textit{cBottle} passes a battery of climate model tests, including diurnal-to-seasonal scale variability, large-scale modes of variability, tropical cyclone statistics, and trends of climate change and weather extremes.
    Moreover, \textit{cBottle} is a step towards a foundation model: It bridges multiple data modalities (reanalysis and simulation), and enables zero-shot bias correction, climate downscaling, and data in-filling. 
     It also enables a new form of human interactivity by leveraging guided diffusion \citep{Song2020}. For example, users can request samples of extreme weather located over regions of interest -- we demonstrate this by training a tropical cyclone (TC) classifier alongside the generator, guiding towards TC states, and obtaining physically credible samples. We anticipate this to kickstart developments of guidance methods for a wide array of user queries and new ways of interacting with climate data. 

\end{abstract}

\section{Introduction}

Policymakers, infrastructure planners, the insurance industry and the public rely on accurate and high-resolution information about the present and possible future climate. This need is driving tremendous advances in climate simulation and Earth observation technology. With exascale computing, energy-efficient computational systems now generate petabytes of kilometer-scale climate data. Digital-twin efforts like the Destination Earth initiative \citep{DestinE} and the Earth Virtualization Engines (EVE) project \citep{EVE} envision such high-fidelity physical climate projections as underpinning next-generation climate informatics infrastructure, which should eventually enable a degree of access and interactivity to allow widespread climate-informed decision making. 

Easy access and low-latency interactivity, however, are nearly impossible with the current paradigm of petabyte-scale datasets generated and hosted by a handful of climate centers. Even if the access problem is unthrottled by moving data, interacting with petabytes of data is computationally expensive. 

Recent advances in AI to replace the \textit{simulation} task do not solve this challenge. While AI has proved efficient and faithful at emulating models and climate data evaluated across a range of metrics on a wide variety of tasks \citep{ACE2,CAMulator,Cresswell2024,ClimBench}, most of these developments have resorted to the auto-regressive modeling paradigm, much like numerical weather and climate models that step forward in time evolving the state of the system modeled by governing systems of equations. Such systems have not been scaled to the km-scale global resolution envisioned by EVE. Even if this were possible it would not solve the informatics challenge of nimbly guiding users to samples or statistics of interest, such as rare extreme weather events impacting a particular region. That is, autoregressive AI emulators share a common challenge with physical climate models -- the need to sift through troves of simulation output to uncover information of interest.

We propose that generative artificial intelligence (AI) for \textit{sampling} offers a solution to both the data volume and interactivity problems by fully leveraging state-of-the-art algorithms optimized for performance on accelerated computing hardware. For our purposes, we define a generative foundation model as a self-supervised climate simulator capable of producing realistic-looking samples that are climate-faithful measured by a set of diagnostics, and can be used to examine \textit{what-if} scenarios not explicitly built into the model. Specifically, diffusion modeling \citep{Ho2020,Song2020,Karras2022-edm} offers an alternative to the auto-regressive time-stepping approach by learning---and sampling---from a distribution of states. This is especially attractive for rapid analyses that do not require the full temporal evolution of the system. By the same token, they are not susceptible to instabilities or drifts inherent to auto-regressive models, though admittedly many auto-regressive emulators perform surprisingly well for large numbers of steps \citep{ACE2,Cresswell2024}. Further, diffusion models can be controlled by the user in a variety of ways, using \textit{guidance} \citep{Song2020} to generate conditional distributions.

Generative diffusion modeling has been used in weather and climate for ensemble generation and uncertainty quantification \citep{Li2024-fs}, probabilistic forecasting \citep{Gencast}, super-resolution (or downscaling) \citep{Mardani2023-cd}, and data-assimilation \citep{Rozet2023,Manshausen2024}. 

A less explored yet promising direction is its application towards learning a compressed representation of massive climate datasets that generates, on demand, climate states steered by user-specified conditions and scenarios. Early examples included training deterministic emulators to predict the mean-state response of climate models to various forcings \citep{ClimBench}. More recent efforts have extended this approach to stochastic emulators in the diffusion framework \citep{bassetti2024diffesm,quilcaille2022showcasing} as well as traditional dimensionality reduction approaches \citep{wang2025stochastic}. But these approaches have been limited to a small number of channels (typically surface temperature). And, as with current AI simulators, none have been scaled to the ambitious km-scale resolution of next-generation climate predictions such as proposed by EVE. 

In this context, we present ``Climate in a bottle" (\textit{cBottle}), a first-of-its-kind generative diffusion-based framework, faithfully emulating multimodal high-resolution climate data. \textit{cBottle} offers pathways towards: (i) solving the data and informatics challenges of km-scale climate data; and (ii) building a generative foundation model for an \textit{interactive} Earth digital twin.

\section{cBottle}
\begin{figure}
    \centering
    \includegraphics[width=0.6\linewidth]{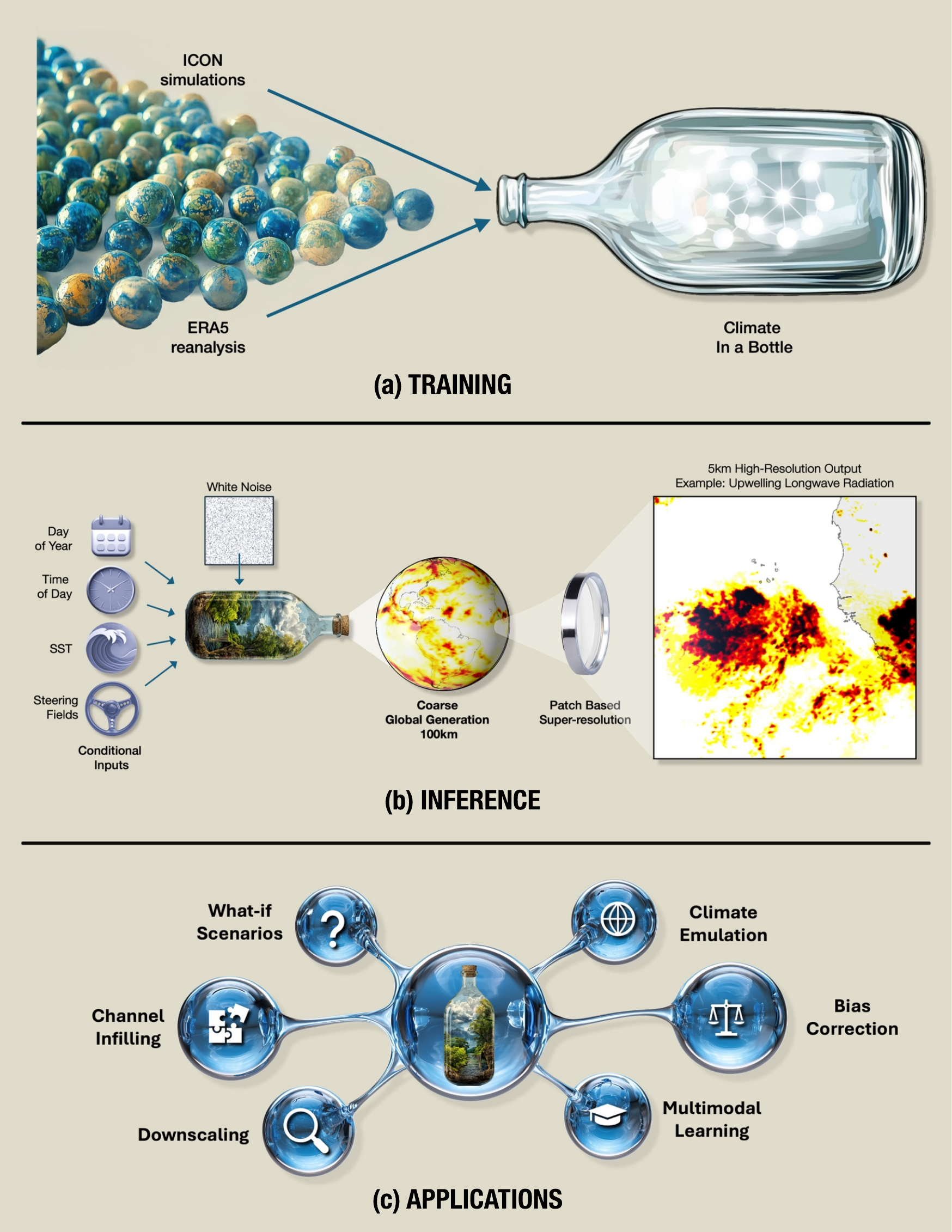}
    \caption{ Climate in a Bottle: A generative foundation model for kilometer-scale climate simulation. (a) Training phase: Petabytes of high-resolution climate data from ICON global cloud resolving simulations and ERA5 reanalysis are compressed into a compact neural network representation ($\sim$ few GB). (b) Inference phase: The trained model generates realistic 5km-resolution climate states from minimal conditional inputs (time, location, sea surface temperature) through a two-stage process: Coarse global generation (100km) followed by patch-based super-resolution. (c) Applications: The foundation model enables diverse climate science applications including interactive what-if scenarios, bias correction between datasets, channel in-filling for missing variables, statistical downscaling, multimodal learning across data sources, and high-fidelity climate emulation—all achievable in minutes rather than hours and accessible without supercomputing resources.}
    \label{fig:schematic}
\end{figure}

\begin{figure}
    \centering
    \newlength{\wo}
\setlength{\wo}{3in}
\begin{tabular}{ll}
Surface air temperature (tas) Ground truth & Generated \\
\includegraphics[width=\wo]{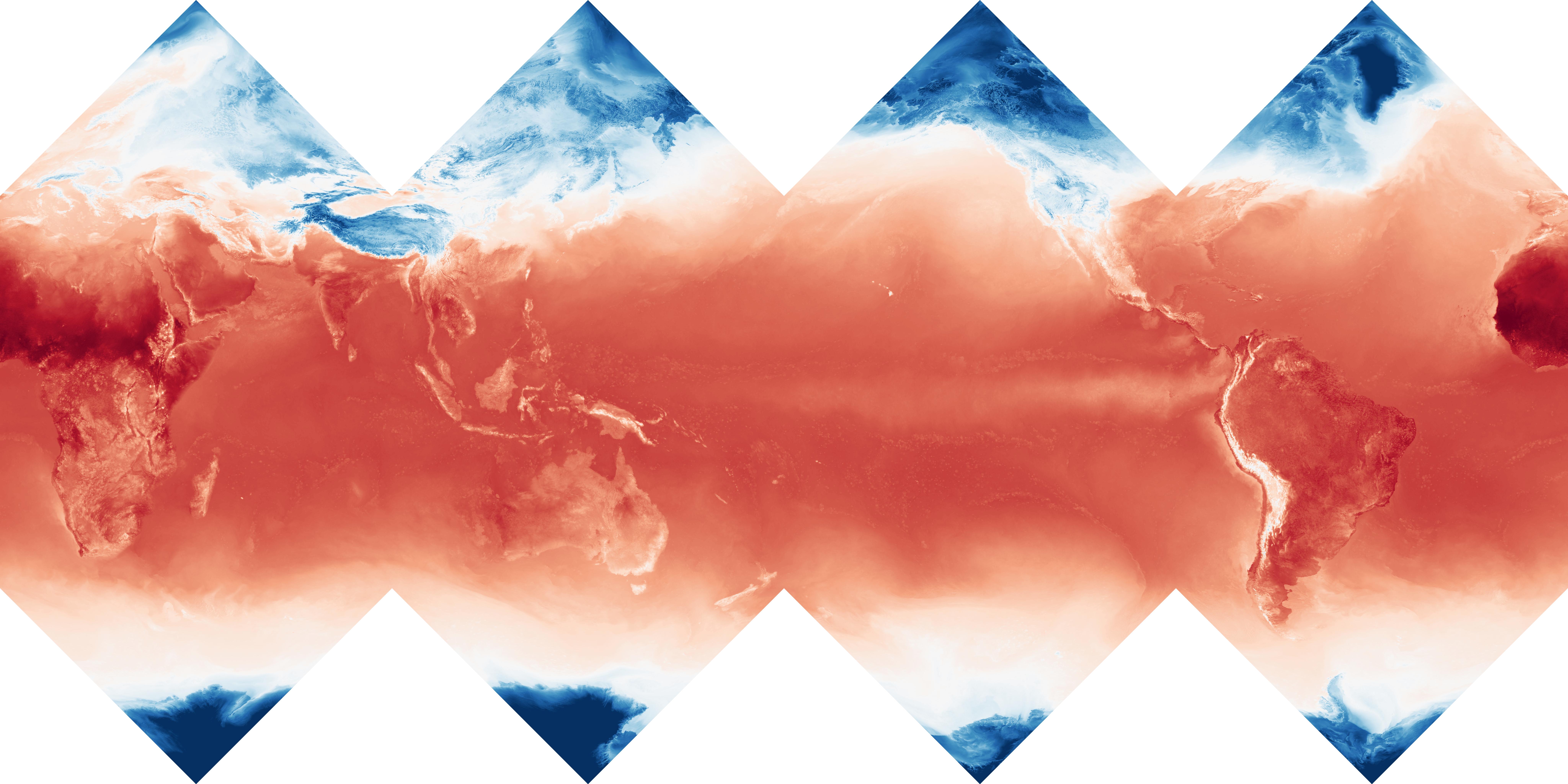} & \includegraphics[width=\wo]{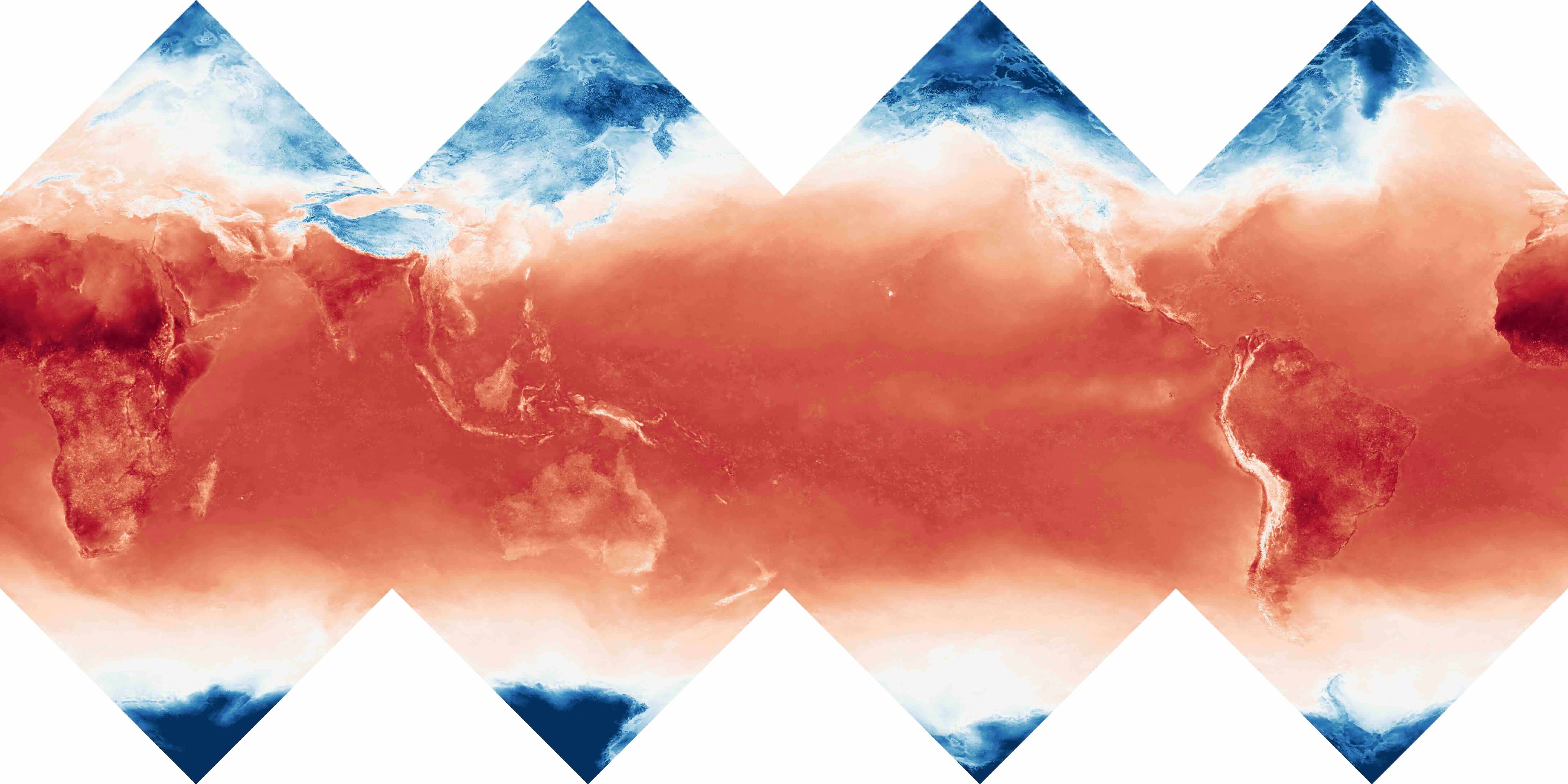} \\
Infrared (rlut) Ground truth &  Generated \\
\includegraphics[width=\wo]{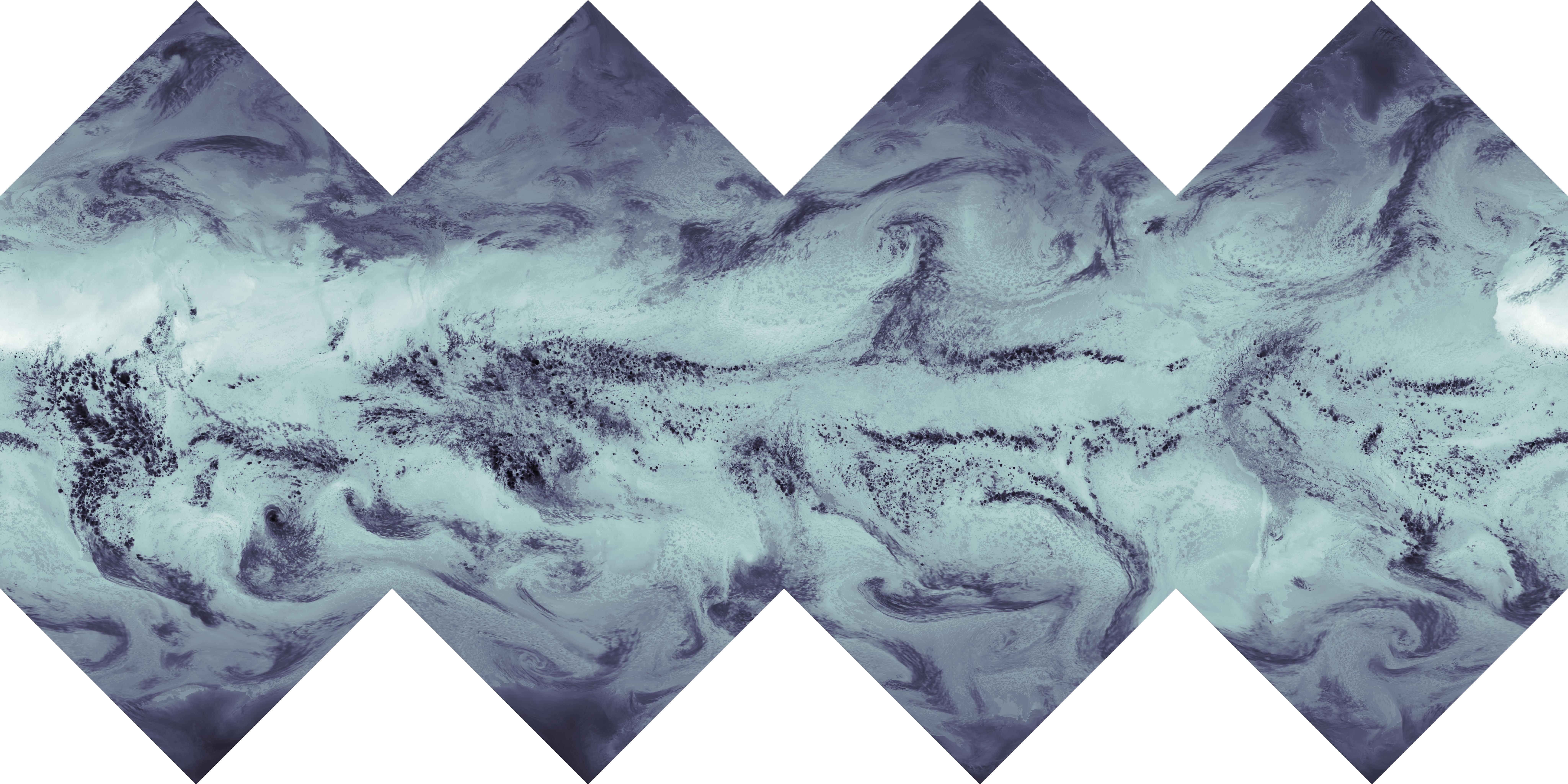} & \includegraphics[width=\wo]{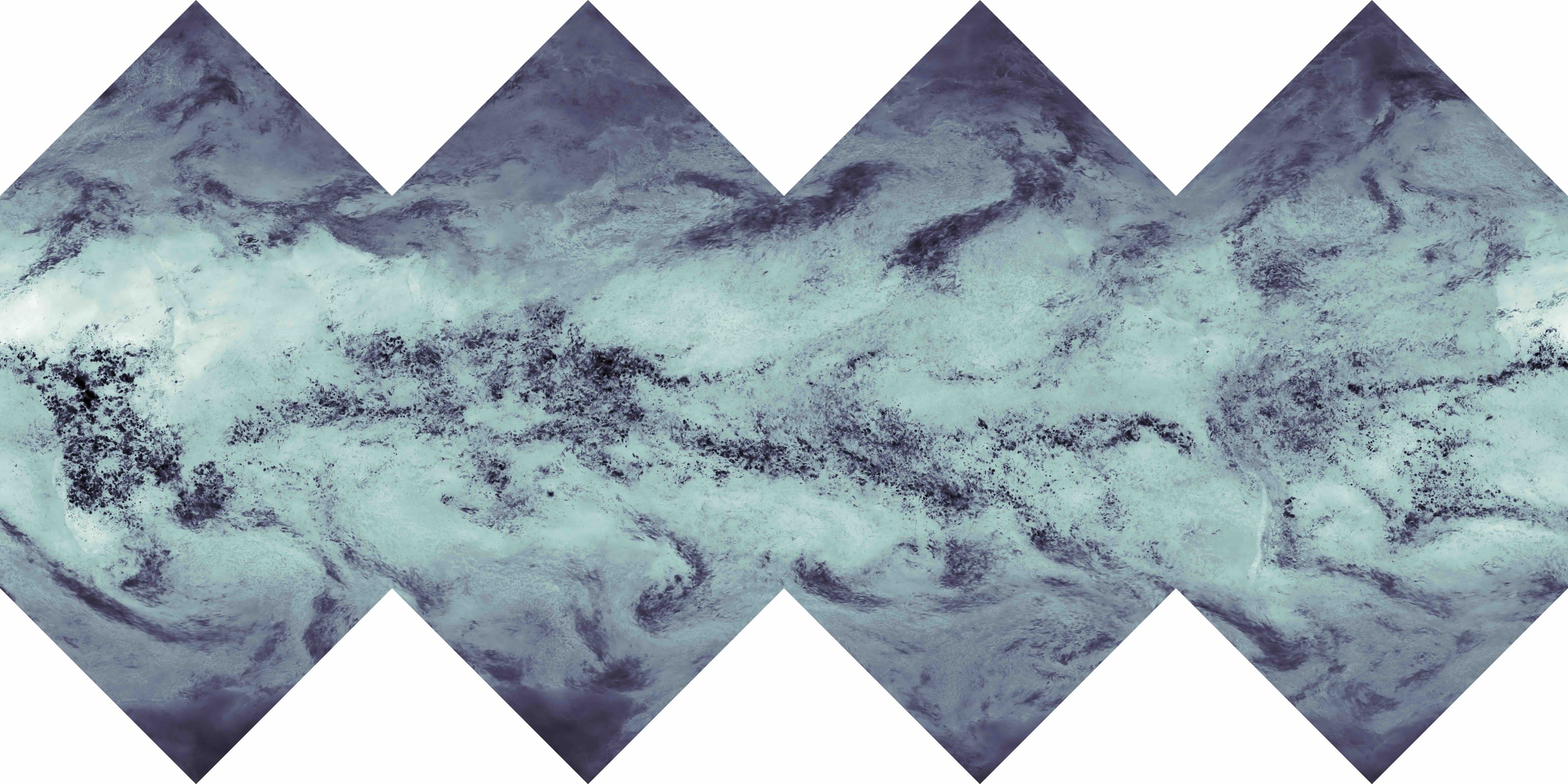}
\end{tabular}
\caption{High resolution (HPX1024) samples from the ICON data (left) and synthesized by cBottle given sea surface temperature and time conditionings (right). Note these are not identically paired but use the same input SST condition for 2025-03-16T15:00:00. For each field the same colormap is used, but to highlight the visual similarity we do not include color-bars. More quantitative comparisons will follow. See Figure \ref{fig:si:1} for more fields.\label{fig:1}}
\end{figure}

\textit{cBottle} generates km-scale samples from the climate distribution using a cascaded approach \citep{Ho2021} as shown in Fig.~\ref{fig:schematic}b): A coarse-resolution generator conditioned on user-specified inputs (time of day, time of year, and monthly-averaged sea surface temperature (SST)) and a 16x super-resolution module capable of generating over ten million pixels globally (see Fig.~\ref{fig:1} for a visualization of the high-resolution output) for dozens of channels (see Table \ref{tab:fields}). Both steps use generative diffusion models, and either can be used independently, enabling generation of coarse-resolution samples and/or super-resolution to km-scale. Special care was required to train the model to produce the largest modes of variability without over-fitting to them. See Methods for details of the model formulation. Briefly, 
\textit{cBottle}'s key properties are: 
\newenvironment{tight_enumerate}{
\begin{enumerate}
  \setlength{\itemsep}{0pt}
  \setlength{\parskip}{0pt}
}{\end{enumerate}}
\begin{tight_enumerate}
    \item Multi-modality and domain transfer: Trained on reanalysis data and 5-km global atmospheric simulations with a convection-resolving climate model, enables generation of km-scale climate states that exploits the information content contained in both modalities. Specifically, we use the ECMWF Reanalysis v5 (ERA5) \citep{ERA5} ---a 40-year observationally-constrained dataset available at coarse resolution, and a 5-year simulation with the ICON global 5-km model \citep{Icon2023}.
    \item Versatility across several useful tasks: removing bias in climate simulations, diagnosing some fields (e.g. precipitation) given others (e.g. winds), or down-scaling coarse-resolution climate data to km-scale.
    \item Climate faithfulness: Generates realistic climatology, climate variability across timescales, large-scale modes, and extreme weather statistics,
    \item Data efficiency: The super-resolution model can be trained on as little as 4 weeks of simulation output, reducing the need for very long simulations with expensive convection-resolving models.
    \item Extreme distillation: Encapsulates km-scale climate model and reanalysis outputs into a few GB of neural network weights, and offers 256x compression ratio per channel.
    \item Speed and scale: Low-latency generation of km-scale samples and scalability to higher resolutions and more modalities.
    \item Interactivity: Can be guided flexibly to yield extreme events like Tropical Cyclones in a specific location on demand.
\end{tight_enumerate}

\section{Results}
Below, the acronyms HPX64 (HPX1024) refer to the low-resolution 100-km (high-resolution 5-km) components of \textit{cBottle} (see Section \ref{sec:healpix}). 

\subsection{ICON in a bottle}


Generation of ICON states at HPX1024 resolution -- nearly 13M pixels, a formidable task -- is achieved through a two-stage process in which the macroscale generator of the ICON modality at HPX64 is fed through a patch-based super-resolution module (see Methods), which we validate below. Figure~\ref{fig:1} shows a comparison of (unpaired) samples of ICON and \textit{cBottle} end-to-end-generated surface temperature and infrared radiation---these are visually similar, and the generated samples show what fine-grained detail \textit{cBottle} can achieve. While we have chosen to highlight the end-to-end synthesis task -- where high-dimensional ICON weather states are generated from low-dimensional climate-controlling inputs -- it is not the only use-case. The compression application, where coarsened ICON output are input to the super-resolution model also is relevant as a means of compressing large global storm-resolving simulations. Validation of super-resolution in this context is provided in Section \ref{sec:si:icon-sr} where we show cBottle successfully super-resolves the different length scales of ICON's convective organization from shallow cloud systems influencing its shortwave radiation channel to deep tropical convection clusters in its precipitation channel (Fig. \ref{fig:superres}), and provide quantitative validation of PDFs and spectra (Fig. \ref{fig:superres_spectra_pdf}).

\subsection{ERA5 in a Bottle}
\paragraph{Realistic conditional samples}

To be useful as a climate sampler, cBottle must demonstrate realistic variability across independent samples from phenomena operating on diurnal, synoptic, seasonal, and interannual time scales that, when averaged, produce a realistic climatology. To validate this we perform an Atmosphere Model Intercomparison Project (AMIP) control experiment with prescribed monthly-averaged SSTs \citep{cmip6,amip}. Samples from AMIP experiments are typically produced by integrating physical climate models (or autoregressive AI simulators) forward in time subject to SST boundary conditions. In cBottle, analogous samples can be obtained by simply generating independent samples every hour from 1940 to 2021 (inclusive)—a large-scale inference, requiring $\sim 1024$ GPU-hours on the H100 GPU, but embarrassingly parallel since each atmospheric state (timestep) is completely independent and only shares the same conditioning.

The following common diagnostics assess the quality of the generated samples and statistics. Additional analyses are presented in Section \ref{sec:si-climate-diagnostics}.

\begin{figure}
    \centering
    \includegraphics[width=\linewidth]{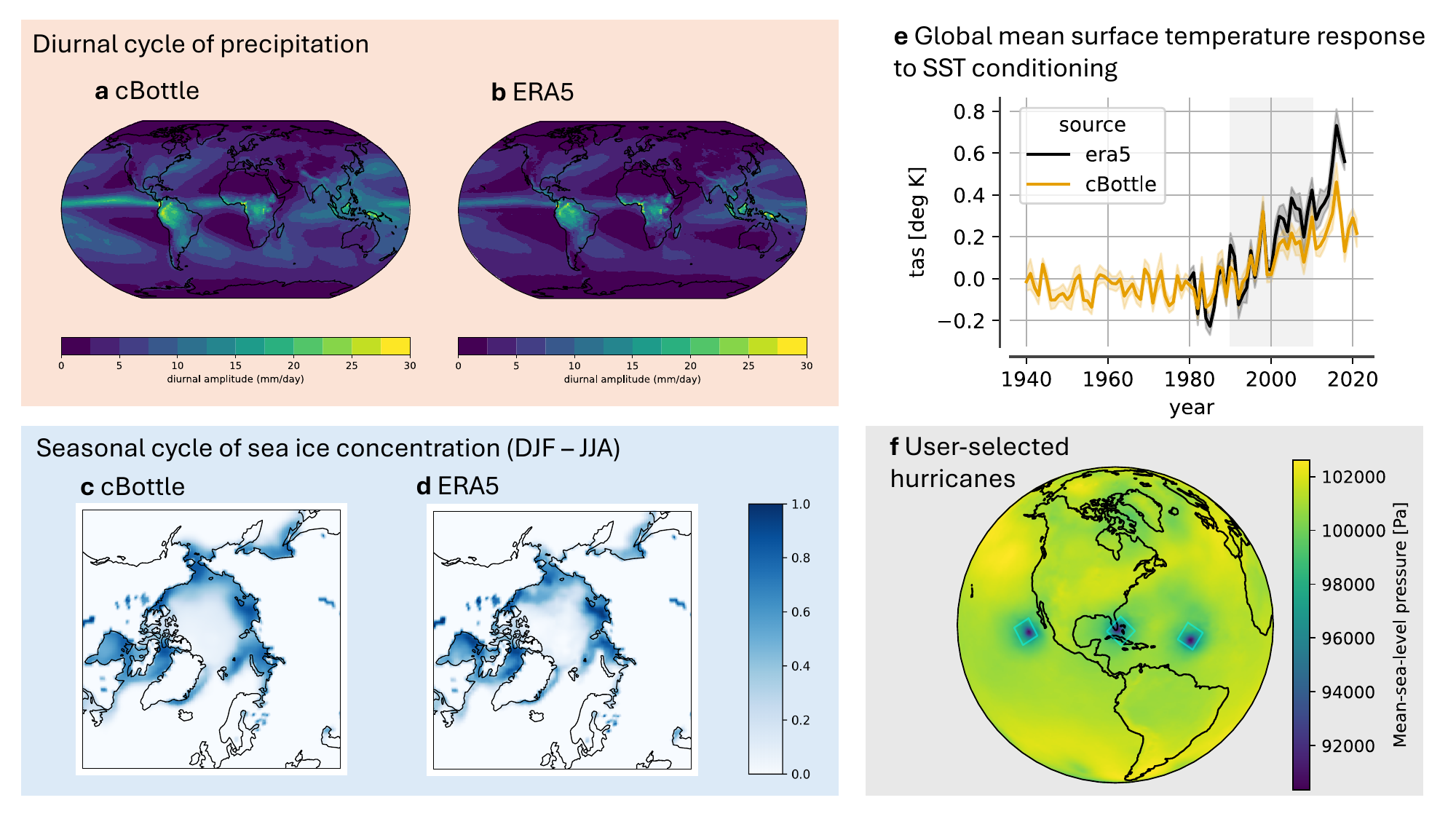}
    \caption{Fidelity to conditioning. Amplitude of the diurnal cycle of precipitation for (a) cBottle and (b) ERA5. DJF-JJA Arctic sea ice concentration computed over the test period for (c) cBottle and (d) ERA5. (e) \textit{cBottle}'s secular temperature trend during the AMIP period compared to ERA5. (f) User-provided hurricane conditioning. Minima of pressure shown in blue are linked to TCs generated by cBottle using classifier guidance. The input to the inference is the location of the cyclones, shown as turquoise boxes.}
    \label{fig:user-control}
\end{figure}

\paragraph{Diurnal Cycle}
In Figure \ref{fig:user-control}a,b) we compare the amplitude of the diurnal cycle of precipitation in \textit{cBottle} with ERA5. Although the amplitude is too large in cBottle, there is a realistic spatial structure of the precipitation diurnal cycle amplitude including spatial details interior to tropical continents -- and even the halo that surrounds the Maritime Continent from the emanating diurnally phased convection propagation there \cite{yang2001diurnal}. Section \ref{sec:methods:diurnal} provides additional analysis of the diurnal band including the signal-to-noise ratio which confirms \textit{cBottle} reproduces even small-amplitude signals like the locally detectable subtropical stratocumulus drizzle diurnal cycles (cf. Fig. \ref{fig:si:diurnal:pr}(d)) and the appropriate land-sea contrast in the surface temperature diurnal cycle (Fig. \ref{fig:si:diurnal:tas}). 

\paragraph{Seasonal cycle and Mean Climate} Figure \ref{fig:user-control}c) shows that \textit{cBottle} correctly generates the seasonal evolution of Arctic sea ice concentration compared to ERA5 in d) (See Fig.~\ref{fig:cycles} for the concentrations by season). During boreal summer (JJA) the sea ice margin retreats and its concentration is reduced locally over the Beaufort Sea, Chukchi Sea and Bering Sea -- poleward of Alaska and Siberia. These results are evidence that the time-of-year and/or SST-conditioning in \textit{cBottle} work as intended. The seasonal cycle is similarly realistic for other fields (see Figure \ref{fig:seasons}). The annual mean climate of the \textit{cBottle} generated precipitation samples also matches the ERA5 and ICON data (see Figure \ref{fig:pr-zonal}).

\begin{figure}
    \centering
\newlength{\imgheight}
\setlength{\imgheight}{1in}
    \includegraphics[width=\linewidth]{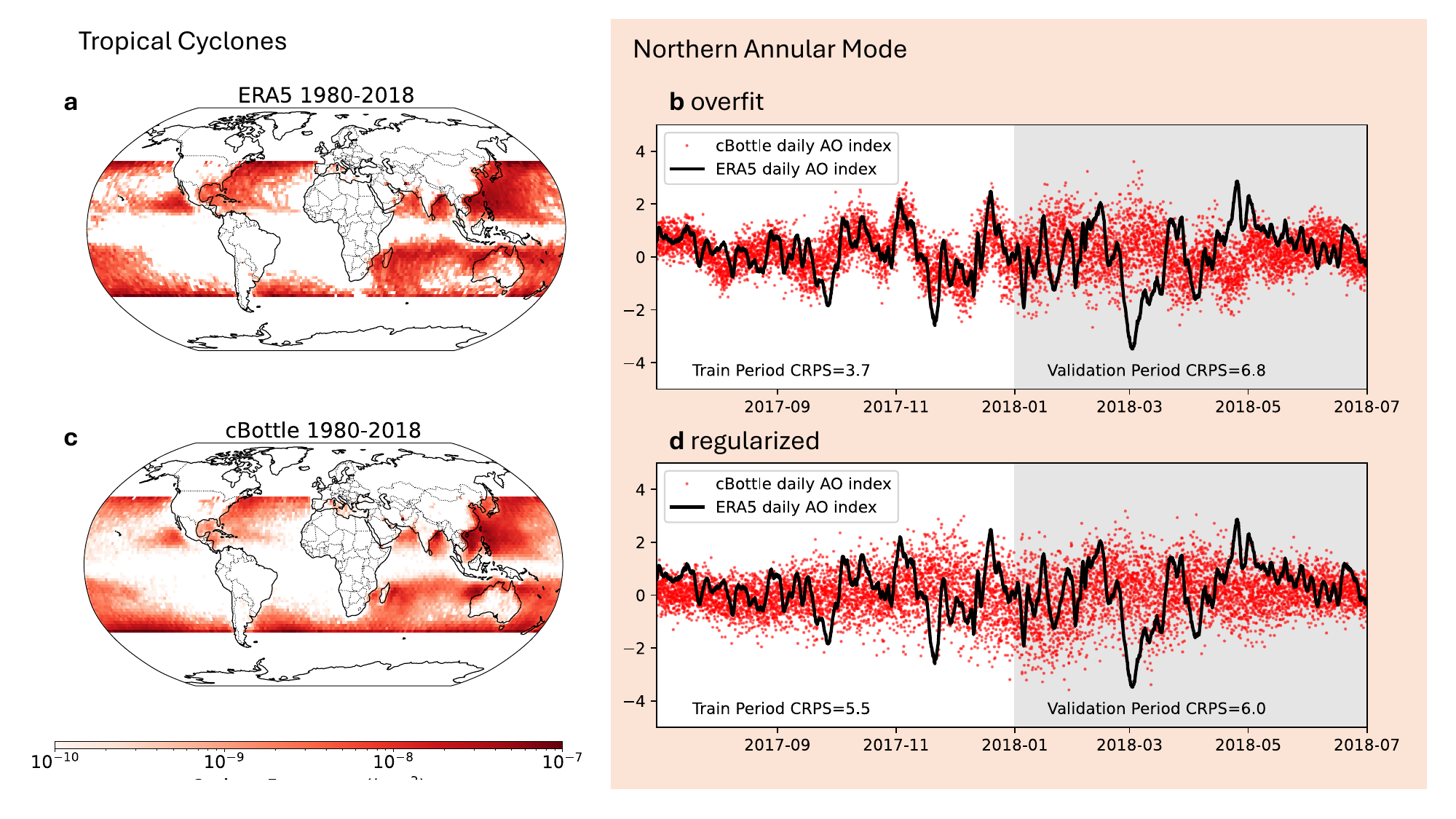}
    \caption{Internal variability. (a,c) Tropical cyclone occurrence probability derived from ERA5 and \textit{cBottle} (1980–2018), presented as heatmaps normalized per $km^2$. (b,d) The Northern Annular Mode (NAM) index for cBottle samples with overfitting (b) and large-noise-level regularization (d). The NAM indices are computed by projecting the daily Z1000 anomalies onto the NAM loading pattern (Fig \ref{fig:si:nam}. For convenience, the CRPS is computed by comparing the 0 UTC from the ground truth index to the 24-member ensemble of \texttt{cBottle} samples taken from ever hour of the same day.}
    \label{fig:ao-simple}
\end{figure}

\paragraph{Northern Annular Mode and SST conditioning}
We next assess a planetary scale mode of internal atmospheric variability. In the northern hemispheric midlatitudes, the Northern Annular Mode (NAM) is the leading empirical orthogonal function (EOF), which accounts for one-fifth of overall variance in the monthly anomalies of 1000 hPa geopotential height computed from November to April \cite{thompson2000annular}.
Figure \ref{fig:ao-simple}d shows that the daily variability of \textit{cBottle}'s NAM index is plausible though slightly weak---with a root-mean-square amplitude in 2018 of 0.84 versus 1.02 in ERA5. Moreover, its low frequency envelope is in phase with inter-annual fluctuations in the ERA5 NAM state, which is more variable during the winter months, suggesting learnt associations between the generated NAM state and the SST and day of year inputs to \textit{cBottle}. Appropriate conditioning on SST is further confirmed by analyzing the El Nino Southern Oscillation (ENSO; see SI Section \ref{sec:enso} and Fig. \ref{fig:nino-pr}). We note that this success in NAM generation depended on using our proposed regularized ensemble of experts approach (see Section \ref{sec:overfit}) to avoid severe over-fitting to the training data that is otherwise encountered when a single checkpoint trained to 9.8M images is naively used instead. Figure \ref{fig:ao-simple}b demonstrates such overfitting as unrealistic temporal coherence between ERA5 and cBottle in the test set; in contrast, our regularized model has a better fair continuous-ranked-probability score (CRPS) \citep{Zamo2018-xs} in the validation set and a smaller discrepancy between training and validation scores. 

\subsection{Interactivity} 
\textit{cBottle} being a diffusion model enables the unique capability of oversampling parts of the distribution that are particularly relevant, such as extreme events. We show this using the example of Tropical Cyclones (TCs), which are generated with a realistic climatology in vast AMIP samples (Fig. \ref{fig:ao-simple}a,b) but nonetheless rare in individual generations. This relates to a principal motivation of \textit{cBottle} to enable new interactive experiences in climate informatics, such as sidestepping the need for a user to sift through vast datasets to find e.g. extreme events of interest over locations of concern. 

To demonstrate user-informed sampling from the distribution of TCs, we train a simple classifier on historical observations from the International Best Track Archive for Climate Stewardship (IBTrACS) \cite{ibtracs}. We use the classifier for guidance during inference  \cite{classifierguidance} (see section~\ref{sec:classifier-guidance-method} for details), enabling us to steer the output state to be compatible with an input map of desired Tropical Cyclone locations. Fig.~\ref{fig:user-control}f shows a sample generated using this technique, with three TCs successfully generated in the locations requested in the conditioning input. This demonstrates that an AI climate state sampler can generate extreme events on-demand.

In our experiments, imposing such conditions in certain locations did not necessarily lead to reduced variability elsewhere, and channels are correlated in a realistic way. Unphysical conditioning, such as guiding towards a TC on the equator, is possible but is somewhat impeded by the model (see SI section~\ref{sec:si:guidance-results}).

\subsection{Multi-modal inference}

We demonstrate how the separate modalities and generative vs. super-resolution modules of \textit{cBottle} can be composed into novel workflows, providing the beginnings of a foundational versatility. We show three examples tested on a real-world atmospheric-river that occurred during the validation period \citep{Viale2018-pu}.

\begin{figure}
    \centering
    \includegraphics[width=\linewidth]{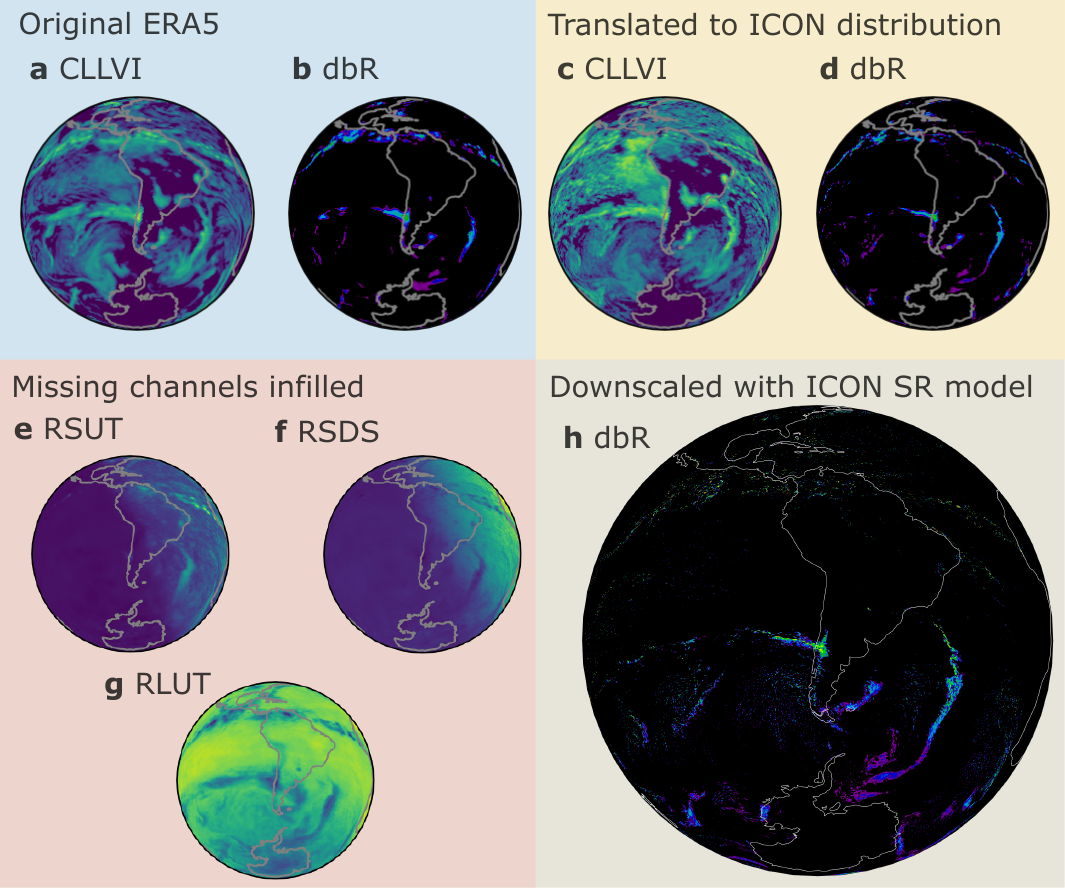}
    \caption{Multi-modal inference methods. The original ERA5 data for vertically integrated liquid water (CLLVI) (a) and $dbR=10log(precip)$ (b). (c-d) These fields translated to the ICON distribution. (e-g) The missing radiation channels infilled by cBottle-3d: outgoing shortwave (RSUT), surface downwelling shortwave (RSDS), and outgoing longwave (RLUT) (h) The downscaled dbR for this same example. This is computed by applying cBottle-SR after translation from ERA5 to ICON. These are shown for an atmospheric river example from 2006-07-11 12UTC.
    \label{fig:multimodal}}
\end{figure}

\paragraph{Channel infilling}\label{sec:channel_fill}
When the channel of interest is not only biased, but damaged, or not available at all in one of the two datasets, we can use \textit{cBottle} to infill it based on the other dataset. In Fig. \ref{fig:multimodal} e--g) we show infilled radiation channels for the ERA5 sample in a, b), channels which in the original ERA5 data show streaking artifacts and were excluded from the training data. Compared to the ERA5 ground truths in Fig. \ref{fig:multimodal} a, b), \textit{cBottle} successfully reconstructs plausible ICON fields. This feature can be used to diagnose any given field given arbitrary subsets of the other fields so it removes the need to train combinatorially many \emph{diagnostic} models to predict certain fields (e.g. precipitation) given others (e.g. free-tropospheric states).

\paragraph{Translating between modalities}
The ICON model prefers a liquid-heavy partitioning between cloud water and ice compared to ERA5 (cf Fig. \ref{fig:multimodal}a, c). It has nearly double the liquid water as ERA5. \textit{cBottle} can translate from such an ERA5 sample (a, b) to an ICON sample (c, d) (see Methods). Overall, the location of the macroscale weather phenomena in the translated data is similar, though there are some apparent differences. This ability to translate between datasets with a single foundation model is an exciting potential application, and we leave more detailed analysis to future studies.

\paragraph{Downscaling} To downscale ERA5 data to higher resolution, we first translate the ERA5 sample to the ICON distribution using the coarse-resolution model, and then apply the super-resolution model. The results in Fig.\ref{fig:multimodal}h shows that the precipitation is concentrated in the same regions as Fig. \ref{fig:multimodal}a,b.

In summary, these results demonstrate that \textit{cBottle} can plausibly reconstruct physical variables and unresolved features throughout the globe. This underscores the potential of \textit{cBottle} as a valuable tool for data recovery in atmospheric science, particularly in scenarios where observational records are incomplete, corrupted, or degraded by systematic artifacts or high-resolution dynamical-downscaling simulations have not been performed.

\section{Conclusions}

We have introduced a generative AI model capable of synthesizing full global atmospheric states at 5 km resolution—over 12.5 million pixels per sample—without requiring autoregressive rollout. This milestone in resolution and sampling fidelity is achieved through a two-stage diffusion framework: a coarse-resolution generator conditioned on minimal, climate-controlling inputs (monthly SST, time of year/day), followed by a patch-based super-resolution module. Together, they enable low-latency, high-fidelity generation of multivariate climate states at the same ambitious resolution of the world's next-generation operational climate simulation technology.

In addition to resolution, \textit{cBottle} demonstrates foundational versatility. Trained jointly on reanalysis (ERA5) and global cloud-resolving simulations (ICON), it permits multimodal translation between datasets, channel in-filling for corrupted or missing variables, bias correction and downscaling and zero-shot generalization of its global super-resolution module. This is made possible by a shared latent representation and masked training strategy (see Methods), allowing cBottle to flexibly bridge data modalities and tasks.

Perhaps most importantly, cBottle proposes a new form of interactive climate modeling. By integrating classifier guidance into the diffusion process, users can conditionally generate physically plausible extreme events—such as tropical cyclones—at specified locations. This demonstrates that an AI climate model can sample rare events on-demand, opening the door to targeted scenario generation and new forms of climate risk analysis.

Methodologically, we have introduced diffusion modeling on the HEALPix grid and proposed a noise schedule tuned to the spectral characteristics of climate data. In particular we show (see Methods) that the defaults most often used in natural image generation literature are insufficient for generating extremely large-scale modes like the seasonal and diurnal cycles. Fixing this required two main innovations: i) increasing the range of noise levels sampled during training and inference; and ii) adding appropriate time and position embedding. Specifically, we propose tuning the noise schedule based on spectral analysis of the data instead of expensive hyper-parameter tuning. We have also identified and mitigated over-fitting at high noise levels using an ensemble-of-experts approach; it is logical to speculate that similar issues may explain the lack of calibration in regional down-scaling models \citep{Mardani2025-eq}. By comparison our super-resolution models trained on large global storm-resolving models simulations are much less prone to over-fitting both owing to the quantity of the training data and the enhanced locality of the super-resolution task compared to global state generation. These perspectives should be helpful to anyone training diffusion models on scientific data.

Additional limitations of our work are important to note. Non-stationary trends are not optimally represented in the current version of cBottle, with its long-term trend only half as big as observed. Similarly, the amplitude of NAM index of the generated samples is 20\% too small, suggesting the model under-represents the variance in the largest scales. We attribute both of these weaknesses to the heavy-handed regularization  required to prevent the coarse model from over-fitting at large noise levels. Nonetheless the modularity, scalability, and interactivity of cBottle represent a step towards a truly interactive Earth digital twin.

\section{Methods}
\label{Sec:methods}


\subsection{Overview}

CBottle is a cascaded diffusion model formulated on the \healpix discretization of the sphere. It is trained on atmospheric simulations and reanalysis and consists of the following components:
\begin{itemize}
    \item \texttt{cBottle-3d} generates hourly atmospheric conditions at approximately 1 degree resolution given monthly sea surface temperatures (SST), the day of year,  data-set label, and the time of day as conditions. It is trained on ERA5 and ICON.
    \item \texttt{cBottle-3d-tc} is a variant of cBottle-3d capable of generating tropical cyclones in user-specified locations. It is trained with tropical cyclone track information data from the International Best Track Archive for Climate Stewardship (IBTrACS) \cite{ibtracs}.  
    \item \texttt{cBottle-SR} is a super-resolution model trained to generate 5-km atmospheric states given 1 degree coarse resolution inputs. This is trained only on the ICON simulation data.
\end{itemize}
The following sections provide a high-level overview of the data sources and our modeling approach. Detailed information about selected inputs and targets, architecture details, and hyper-parameter settings are shown in Section \ref{sec:architectures}.

\subsection{\healpix}
\label{sec:healpix}

\paragraph{Background}

Using AI for global data poses a geometric challenge of appropriately representing data on a spherical surface. To this end, \healpix \citep{gorski2005healpix} grids offer an attractive solution with an equal area grid structure that also has iso-latitude properties and a hierarchical indexing scheme for multi-resolution workflows. The \healpix grid is constructed of 12 faces with each face containing a square $2^l$ x $2^l$ grid of pixels where $l$ represents the resolution level of the grid.
This structure is well-suited for deep-learning algorithms, especially for convolutional architectures and has been used in \citep{krachmalnicoff2019convolutional}, \citep{perraudin2019deepsphere} and \citep{Karlbauer2024-kr}; the equal area property is also useful. Due to these desirable properties, we use the \healpix grid as the spherical discretization method in this work. A common convention for \healpix grids refers to the length of each face $N_{side}=2^l$. We refer to two main global grids: HPX64 ($N_{side}=64$) corresponding to 100km resolution (49152 pixels) and HPX1024 ($N_{side}=1024$) at 6km resolution (12,582,912 pixels).

\paragraph{Padding and convolution\label{sec:hpxpad}} For coarse resolution applications (below \healpix level 8), we use architectures similar to \citep{Mardani2023-cd} that were designed for 2D rectangular domains. This is done by adapting the convolution operations in the architecture to perform convolutions on each \healpix face independently.
However, this approach does not account for the connected topology of the \healpix grid and prevents the flow of information between adjacent faces. To address this, we apply padding by exchanging pixels from the neighboring \healpix faces using the protocol outlined in \citep{Karlbauer2024-kr} (see Section \ref{sec:si:hpx-conv}). This technique is commonly referred to as a ghost-cell or halo update in numerical methods.

\subsection{Datasets}
\label{sec:data}

The inputs and outputs we trained with are displayed in Tab. \ref{tab:fields}. Briefly, the models are trained to generate 3D states of the atmosphere and some key 2D variables relevant to climate impacts (precipitation, surface air temperature, downwelling short-wave) and observations (top of atmosphere radiative fluxes).Our models are trained on two global datasets: the ECMWF Reanalysis version 5 (ERA5)\citep{ERA5} which is observationally constrained, and a global storm resolving simulation with the ICON climate model \citep{Icon2023}, which is a free-running simulation unpaired with reality.

\paragraph{ERA5} ERA5 is a commonly used dataset that represents the best guess of the global atmospheric state at a coarse resolution of 25 km. Briefly, it is created by running a data assimilation algorithm that incorporates observations into ECMWF's numerical model. However, it is not available at a high enough resolution in space or time to resolve meso-scale motions on kilometer scale, and in particular deep convection. We train on hourly data from 1980--2017, leaving 2018 as validation. For consistency with the well-known AMIP protocol \citep{amip}, we use its forcing instead of ERA5 for the sea-surface temperature input.

The ERA5 shortwave and longwave radiation fields are corrupted by a known, but undocumented, bug which appears as clear stripe-like artifacts that point North-West (see Fig \ref{fig:si:infill}). This is caused be a random seed in the radiation scheme being set to $lat+lon$ in every atmospheric column \citep{Chantry2024}. Therefore, we do not use the fields from ERA5 for training our models.

\paragraph{ICON Cycle 3} The ICON data is a free-running simulation with the ICON atmospheric model coupled to a dynamic ocean and land \citep{nextGEMS}. The atmospheric component solves the nonhydrostatic fluid mechanics equations on a global ico-sahedral mesh with a resolution of around 5 km. Unlike ERA5 the model explicitly resolves certain convective motions so a convection parameterization is not used. So even beyond the simple increase in resolution, we expect this to increase the fidelity of the precipitation and cloud fields in this dataset relative to ERA5 where these processes are parameterized. We train on 2020-01-20 03:00:00 through 2024-03-06 12:00:00 (inclusive) and validate on 2024-03-06 15:00:00 through 2025-07-22 00:00:00.

\paragraph{IBTrACS}
The International Best Track Archive for Climate Stewardship (IBTrACS) \cite{ibtracs} is used to train the TC classifier for guiding states towards TCs in a desired location. The dataset contains information collected from Regional Specialized Meteorological Centers (RSMCs) about position, intensity, and other parameters of TCs. Here, we use only the information about the storms' position, based on the best track from multiple sources. While data is available from 1840 to present day, we use the part that overlaps with our ERA5 training set from 1980--2017. IBTrACS thus provides the labels for locations of TCs in the ERA5 data at a given time.

See Section \ref{sec:si:methods:data} for detailed information about these data sources and our pre-processing.

\subsection{Denoising Diffusion}
\label{sec:diffusions}

\paragraph{Background} As mentioned in the introduction, cBottle is a de-noising diffusion model \citep{song2021scorebasedgenerativemodelingstochastic,Ho2020-pp}. This is a powerful class of generative models that learn to reverse a Gaussian stochastic process which starts from a sample of the data and ends in pure white noise. Specifically, we adopt the score-based diffusion framework \citep{song2021scorebasedgenerativemodelingstochastic, Mardani2023-cd}, which learns this de-noising process via a score-matching network. The training objective is given by
\begin{align}
\argmin_\theta \mathbb{E}_{\mathbf{x} \sim p_{\text{data}}} \mathbb{E}_{\sigma \sim p_\sigma} \mathbb{E}_{\mathbf{\epsilon} \sim \mathcal{N}(0, \sigma^2 \mathbf{I})} \left\| \mathcal{D}_\theta(\mathbf{x} + \mathbf{\epsilon}; \sigma) - \mathbf{x} \right\|^2,
\label{eq:loss}
\end{align}
where the learned denoiser $\mathcal{D}_\theta(\cdot; \sigma)$ approximates the score function according to $\nabla_{\mathbf{x}} \log p(\mathbf{x}; \sigma) \approx \frac{\mathcal{D}_\theta(\mathbf{x}; \sigma) - \mathbf{x}}{\sigma^2}$. Here, $p_\sigma$ denotes the noise level distribution used during training.

To sample from the model, we employ the Elucidated Diffusion Model (EDM) sampler proposed by \citep{Karras2022-edm}, which solves the corresponding reverse-time ordinary differential equation (ODE):
\begin{align}
\frac{d\mathbf{x}}{d\sigma} =\frac{\mathbf{x}- \mathcal{D}_{\theta}(\mathbf{x}, \sigma)}{\sigma} .\label{eq:ode}
\end{align}
Specifically, we use Algorithm 2 from \citet{Karras2022-edm}, with the $S_{churn}$ parameter set to 0. This ODE is sampled backwards from $\sigma=\sigma_{max}$ to $\sigma=\sigma_{min}$.

\paragraph{Diffusions on the sphere} To train diffusion models on the sphere, the noise $\epsilon_i \sim ~N(0,1)$ is drawn independently for every pixel $1\leq i \leq 12\cdot 4^l$. Because the \healpix pixels are all equal area, we conjecture that this is identical to the pixel-wise average of a finite-variance continuous Gaussian process on the sphere.

\paragraph{Noise schedule} The choice of noise levels seen during training ($p_\sigma$) and inference ($\sigma_{min}$ and $\sigma_{max}$) is crucial for the quality of a diffusion model. Compared to the typical settings used for natural image generation, we find that much larger noise levels must be sampled during training for a model to generate realistic seasonal and diurnal cycles. When the noise is too small, the model never learns to create these signal on its own---a phenomenom known as the signal-leak bias \citep{Everaert2023-bi}.  We therefore select a broadly distributed log-uniform noise distribution with $\sigma_{min}=0.02$ and $\sigma_{max}=200.0$. These values are chosen based on a spectral analysis to cover the dynamic range between the lowest and highest variances modes of the data---without a need for expensive hyperparameter tuning. Specifically, for data $\mathbf{}\in\mathbb{R}^n$, $\sigma_{max}$ should be chosen proportional to $\sqrt{n} s_1$ where $s_1^2$ is the fraction of variance explained by the first EOF of the data. See Section \ref{sec:si:tuning-noise} for a detailed derivation.

\paragraph{Ensemble of experts regularization} Similar spectral considerations reveal that diffusion models are prone to over-fitting at high noise levels when trained on a limited number of samples. This occurs because there are few independent observations of such large variance signal such as El Ni\~no or the NAM in the observed record, and a large neural networks is fully capable of memorizing these when given a suitably rich input like SST. We avoid this over-fitting by adapting the ensemble of experts approach of \citet{Yogesh2022-zv}. While they trained separate models for different noise levels in order to increase overall model capacity, the method can just as easily be adapted to regularize the large noise levels more than the small.  In our approach, separate denoisers (i.e experts) for small ($\sigma<10$), medium ($10<\sigma\leq)100$, and large ($100<\sigma\leq\sigma_{max}$) noise levels. For simplicity and to avoid extra training, we regularize via early stopping, and simply use less-trained checkpoints for the larger noise levels. Section \ref{sec:overfit} contains more details and sensitivity tests.

\subsection{Multi-diffusion}
\label{sec:multi-diffusion}

Emulating ICON at HPX1024 resolution poses substantial computational challenges due to the massive size of this km-scale global dataset. Directly applying a traditional diffusion model to the full-resolution domain is infeasible as it would require over 2,000 GB of GPU memory, assuming a linear relationship between spatial resolution and memory footprint. This far exceeds the capacity of current hardware, even on high-end multi-GPU systems.

To address the computational bottleneck and take advantage of the inherently local nature of super-resolution tasks, multi-diffusion \citep{bartal2023multidiffusionfusingdiffusionpaths} is adapted as a scalable and memory-efficient sampling strategy. It enables high-resolution generation without sacrificing global spatial coherence or requiring model architecture changes.
The central idea behind multi-diffusion is to replace the global denoiser used in standard sampling algorithms with an aggregated local denoiser. At each sampling step, the aggregated local denoiser operates on the global feature map in a sliding-window fashion, denoising it patch by patch. These locally denoised patches are then reassembled to reconstruct the global feature map. This approach is simple to adapt to the \healpix grid by exchanging neighboring information between the 12 faces via the padding procedure described above. See Section \ref{sec:si:multi-diffusion} for details.

\subsection{Multi-modal training and inference}

\paragraph{Masked Training} \label{sec:masking}

Since we intend our framework to extend to a variety of data sources, with different coverage in space, time, and the included fields, we use masked training.  Our approach is to simply train a diffusion model to generate the union of all channels present in all the datasets and mask any missing values for a given dataset---such as ERA5's corrupted radiation fluxes---with 0 after normalization.

More specifically, any masked values in the noisy image $x + n$ and target $x$ are filled with 0 before calling the network and computing the loss. We then normalize the loss by the fraction of masked out data points, so that the training equally weights modalities with different numbers of observables. Pseudo-code for this approach is shown in SI \ref{sec:si:masking}.

\paragraph{Channel infilling} Given the masked training described above, we can use the trained model to infill channels such as the ERA5 radiation fields given the other ERA5 channels. 

Following Appendix I.2 of \citet{Song2020-pj}, let $\Omega: \mathbb{R}^{C \times \mathbf{d}} \to \mathbb{R}^{C_{\mathcal{K}} \times \mathbf{d}}$ and $\bar{\Omega}: \mathbb{R}^{C \times \mathbf{d}} \to \mathbb{R}^{C_{\mathcal{U}} \times \mathbf{d}}$ denote the slicing operators that extract the known and unknown channels, respectively, where $\mathbf{d}$ denotes an arbitrary set of additional dimensions (e.g., spatial or temporal). The unknown subset at sampling time $t$ is then denoted by $\mathbf{z}_t = \bar{\Omega}(\mathbf{x}_t)$. The conditional score function required to impute the missing channels can be approximated as
\begin{align}
\nabla_{\mathbf{z}} \log p\big(\mathbf{z}_t \mid \Omega(\mathbf{x}_0), \sigma_t\big) 
\approx \nabla_{\mathbf{z}} \log p\big([\mathbf{z}_t;\, \Omega(\mathbf{x}_t)] \mid \sigma_t\big),
\end{align}
where $[\cdot\,;\cdot]$ denotes concatenation along the channel dimension.

\paragraph{Translating between modalities}
To map an ICON sample to the distribution of the ERA5 data (closer to observations), we take advantage of the latent representation of the diffusion model. Our diffusion model being trained with ICON and ERA5 data can be regarded as two different models (with shared weights, the only difference being whether we train or inference with an ICON or with an ERA5 flag passed to the model). With these two models, we can follow a Dual Diffusion Implicit Bridge approach \citep{DDIB}, which consists in evaluating the diffusion ODE forward (mapping an ICON sample to noise), switch the flag to ERA5 and evaluate the ODE backwards (mapping the noise to ERA5). It may seem surprising that this yields a realistic sample, but as \citet{Liu2022-eu} explains the uniqueness property of differential equations ensures that trajectories between the data and noise distributions cannot cross one another. Therefore using the deterministic mapping from data to noise for one domain and from noise to data for the other domain, allows us to discover mappings between the two domain with low transport cost.


Unlike e.g. rectified flow \cite{Liu2022-eu} our model is not trained explicitly to map from one dataset to the other. That approach would require training a separate model for each pair of datasets so does not scale well to many different data sources. By training one model on all potential datasets, we can use the denoiser to define a shared latent space, a desirable property of a foundation model.

\subsection{Classifier Guidance}\label{sec:classifier-guidance-method}
We demonstrate the capacity of cBottle to be used in a more interactive way than both auto-regressive climate emulators such as ACE2 \citep{ACE2} and numerical GCMs with the example of Tropical Cyclones (TCs). To do this, we implement classifier guidance \cite{classifierguidance} with an image segmentation model trained on historical cyclone data from IBTracs \cite{ibtracs}.

We initially experimented with explicitly conditioning on the training by randomly drawing a coarse-resolution HPX8 pixel and labeling it based on whether it contained TC or not. However, this approach failed to generate TCs in the desired location---likely because of the sparseness of this input. By comparison a classifier trained with cross-entropy loss to segment the image into TC and not-TC classes learns from every TC present in any given sample, enabling a more efficient use of the data. This classifier is trained with the noise image  $x(\sigma)$ and noise level $\sigma$ as an input so that it can be used in the sampler ODE (\ref{eq:ode}) where only $x(\sigma)$ is available, rather than clean state $x(0)$. During the inference phase, this classifier is used to guide the sampler based on a user-specified mask indicating the desired TC locations. See Section \ref{sec:si:classifier} for more details.

\section{Data Availability}

The ICON data is archived at the Max Planck Institute for Meteorology \citep{IconCycle3}. ERA5\citep{ERA5} is available for download via the Copernicus Climate Data Store and is mirrored at many sites around the globe. The monthly sea surface temperature inputs for the ERA5 model were downloaded from the inputs4MIPS section of Earth System Grid Federation.

Our trained model parameters are available on NVIDIA NGC at this URL: \\\texttt{https://catalog.ngc.nvidia.com/orgs/nvidia/teams/earth-2/models/cbottle}.

The code is available at \verb|https://github.com/NVlabs/cBottle|.

\section*{Acknowledgements}

This work would not have occurred had Bjorn Stevens and Daniel Klocke of the Max-Planck Institute for Meteorology not provided the ICON Cycle 3 simulation in a format perfect for training machine learning models, and we also owe David Pruitt of NVIDIA a debt for helping transmit this sizeable dataset onto our systems. We also thank Oliver Watt-Meyer and Christopher Bretherton of the Allen Institute of Artificial Intelligence for their critical observation that the released version cBottle was over-fitting to the NAM, and Karsten Kries (NVIDIA) for helpful discussions about the spectral perspective on diffusion models.

\bibliography{references}

\begin{thebibliography}{53}
\providecommand{\natexlab}[1]{#1}
\providecommand{\url}[1]{\texttt{#1}}
\expandafter\ifx\csname urlstyle\endcsname\relax
  \providecommand{\doi}[1]{doi: #1}\else
  \providecommand{\doi}{doi: \begingroup \urlstyle{rm}\Url}\fi

\bibitem[Bauer et~al.(2021)Bauer, Stevens, and Hazeleger]{DestinE}
Peter Bauer, Bjorn Stevens, and Wilco Hazeleger.
\newblock A digital twin of earth for the green transition.
\newblock \emph{Nature Climate Change}, 11\penalty0 (2):\penalty0 80--83, 2021.
\newblock \doi{10.1038/s41558-021-00986-y}.
\newblock URL \url{https://www.nature.com/articles/s41558-021-00986-y}.

\bibitem[Stevens et~al.(2024)Stevens, Adami, Ali, Anzt, Aslan, Attinger,
  B\"ack, Baehr, Bauer, Bernier, Bishop, Bockelmann, Bony, Brasseur, Bresch,
  Breyer, Brunet, Buttigieg, Cao, Castet, Cheng, Dey~Choudhury, Coen, Crewell,
  Dabholkar, Dai, Doblas-Reyes, Durran, El~Gaidi, Ewen, Exarchou, Eyring,
  Falkinhoff, Farrell, Forster, Frassoni, Frauen, Fuhrer, Gani, Gerber,
  Goldfarb, Grieger, Gruber, Hazeleger, Herken, Hewitt, Hoefler, Hsu, Jacob,
  Jahn, Jakob, Jung, Kadow, Kang, Kang, Kashinath, Kleinen-von K\"onigsl\"ow,
  Klocke, Kloenne, Kl\"ower, Kodama, Kollet, K\"olling, Kontkanen, Kopp, Koran,
  Kulmala, Lappalainen, Latifi, Lawrence, Lee, Lejeun, Lessig, Li, Lippert,
  Luterbacher, Manninen, Marotzke, Matsouoka, Merchant, Messmer, Michel,
  Michielsen, Miyakawa, M\"uller, Munir, Narayanasetti, Ndiaye, Nobre, Oberg,
  Oki, \"Ozkan-Haller, Palmer, Posey, Prein, Primus, Pritchard, Pullen,
  Putrasahan, Quaas, Raghavan, Ramaswamy, Rapp, Rauser, Reichstein, Revi,
  Saluja, Satoh, Schemann, Schemm, Schnadt~Poberaj, Schulthess, Senior, Shukla,
  Singh, Slingo, Sobel, Solman, Spitzer, Stier, Stocker, Strock, Su, Taalas,
  Taylor, Tegtmeier, Teutsch, Tompkins, Ulbrich, Vidale, Wu, Xu, Zaki, Zanna,
  Zhou, and Ziemen]{EVE}
B.~Stevens, S.~Adami, T.~Ali, H.~Anzt, Z.~Aslan, S.~Attinger, J.~B\"ack,
  J.~Baehr, P.~Bauer, N.~Bernier, B.~Bishop, H.~Bockelmann, S.~Bony,
  G.~Brasseur, D.~N. Bresch, S.~Breyer, G.~Brunet, P.~L. Buttigieg, J.~Cao,
  C.~Castet, Y.~Cheng, A.~Dey~Choudhury, D.~Coen, S.~Crewell, A.~Dabholkar,
  Q.~Dai, F.~Doblas-Reyes, D.~Durran, A.~El~Gaidi, C.~Ewen, E.~Exarchou,
  V.~Eyring, F.~Falkinhoff, D.~Farrell, P.~M. Forster, A.~Frassoni, C.~Frauen,
  O.~Fuhrer, S.~Gani, E.~Gerber, D.~Goldfarb, J.~Grieger, N.~Gruber,
  W.~Hazeleger, R.~Herken, C.~Hewitt, T.~Hoefler, H.-H. Hsu, D.~Jacob, A.~Jahn,
  C.~Jakob, T.~Jung, C.~Kadow, I.-S. Kang, S.~Kang, K.~Kashinath,
  K.~Kleinen-von K\"onigsl\"ow, D.~Klocke, U.~Kloenne, M.~Kl\"ower, C.~Kodama,
  S.~Kollet, T.~K\"olling, J.~Kontkanen, S.~Kopp, M.~Koran, M.~Kulmala,
  H.~Lappalainen, F.~Latifi, B.~Lawrence, J.~Y. Lee, Q.~Lejeun, C.~Lessig,
  C.~Li, T.~Lippert, J.~Luterbacher, P.~Manninen, J.~Marotzke, S.~Matsouoka,
  C.~Merchant, P.~Messmer, G.~Michel, K.~Michielsen, T.~Miyakawa, J.~M\"uller,
  R.~Munir, S.~Narayanasetti, O.~Ndiaye, C.~Nobre, A.~Oberg, R.~Oki,
  T.~\"Ozkan-Haller, T.~Palmer, S.~Posey, A.~Prein, O.~Primus, M.~Pritchard,
  J.~Pullen, D.~Putrasahan, J.~Quaas, K.~Raghavan, V.~Ramaswamy, M.~Rapp,
  F.~Rauser, M.~Reichstein, A.~Revi, S.~Saluja, M.~Satoh, V.~Schemann,
  S.~Schemm, C.~Schnadt~Poberaj, T.~Schulthess, C.~Senior, J.~Shukla, M.~Singh,
  J.~Slingo, A.~Sobel, S.~Solman, J.~Spitzer, P.~Stier, T.~Stocker, S.~Strock,
  H.~Su, P.~Taalas, J.~Taylor, S.~Tegtmeier, G.~Teutsch, A.~Tompkins,
  U.~Ulbrich, P.-L. Vidale, C.-M. Wu, H.~Xu, N.~Zaki, L.~Zanna, T.~Zhou, and
  F.~Ziemen.
\newblock Earth virtualization engines (eve).
\newblock \emph{Earth System Science Data}, 16\penalty0 (4):\penalty0
  2113--2122, 2024.
\newblock \doi{10.5194/essd-16-2113-2024}.
\newblock URL \url{https://essd.copernicus.org/articles/16/2113/2024/}.

\bibitem[Watt-Meyer et~al.(2024)Watt-Meyer, Henn, McGibbon, Clark, Kwa,
  Perkins, Wu, Harris, and Bretherton]{ACE2}
Oliver Watt-Meyer, Brian Henn, Jeremy McGibbon, Spencer~K Clark, Anna Kwa,
  W~Andre Perkins, Elynn Wu, Lucas Harris, and Christopher~S Bretherton.
\newblock {ACE2}: Accurately learning subseasonal to decadal atmospheric
  variability and forced responses.
\newblock \emph{arXiv [physics.ao-ph]}, November 2024.

\bibitem[Chapman et~al.(2025)Chapman, Schreck, Sha, Gagne, Kimpara, Zanna,
  Mayer, and Berner]{CAMulator}
William~E Chapman, John~S Schreck, Yingkai Sha, David~John Gagne, II, Dhamma
  Kimpara, Laure Zanna, Kirsten~J Mayer, and Judith Berner.
\newblock {CAMulator}: Fast emulation of the community atmosphere model.
\newblock \emph{arXiv [physics.ao-ph]}, April 2025.

\bibitem[Cresswell-Clay et~al.(2024)Cresswell-Clay, Liu, Durran, Liu, Espinosa,
  Moreno, and Karlbauer]{Cresswell2024}
Nathaniel Cresswell-Clay, Bowen Liu, Dale Durran, Zihui Liu, Zachary~I
  Espinosa, Raul Moreno, and Matthias Karlbauer.
\newblock A deep learning earth system model for efficient simulation of the
  observed climate.
\newblock \emph{arXiv [physics.ao-ph]}, September 2024.

\bibitem[Watson-Parris et~al.(2022)Watson-Parris, Rao, Olivié, Seland, Nowack,
  Camps-Valls, Stier, Bouabid, Dewey, Fons, Gonzalez, Harder, Jeggle, Lenhardt,
  Manshausen, Novitasari, Ricard, and Roesch]{ClimBench}
D~Watson-Parris, Y~Rao, D~Olivié, Ø~Seland, P~Nowack, G~Camps-Valls, P~Stier,
  S~Bouabid, M~Dewey, E~Fons, J~Gonzalez, P~Harder, K~Jeggle, J~Lenhardt,
  P~Manshausen, M~Novitasari, L~Ricard, and C~Roesch.
\newblock {ClimateBench} {v1}.0: A benchmark for data‐driven climate
  projections.
\newblock \emph{J. Adv. Model. Earth Syst.}, 14\penalty0 (10), October 2022.
\newblock ISSN 1942-2466.
\newblock \doi{10.1029/2021ms002954}.

\bibitem[Ho et~al.(2020{\natexlab{a}})Ho, Jain, and Abbeel]{Ho2020}
Jonathan Ho, Ajay Jain, and Pieter Abbeel.
\newblock Denoising diffusion probabilistic models.
\newblock \emph{arXiv [cs.LG]}, June 2020{\natexlab{a}}.

\bibitem[Song et~al.(2020{\natexlab{a}})Song, Sohl-Dickstein, Kingma, Kumar,
  Ermon, and Poole]{Song2020}
Yang Song, Jascha Sohl-Dickstein, Diederik~P Kingma, Abhishek Kumar, Stefano
  Ermon, and Ben Poole.
\newblock Score-based generative modeling through stochastic differential
  equations.
\newblock \emph{arXiv [cs.LG]}, November 2020{\natexlab{a}}.

\bibitem[Karras et~al.(2022)Karras, Aittala, Aila, and Laine]{Karras2022-edm}
Tero Karras, Miika Aittala, Timo Aila, and Samuli Laine.
\newblock Elucidating the design space of diffusion-based generative models,
  2022.

\bibitem[Ho et~al.(2021)Ho, Saharia, Chan, Fleet, Norouzi, and
  Salimans]{Ho2021}
Jonathan Ho, Chitwan Saharia, William Chan, David~J Fleet, Mohammad Norouzi,
  and Tim Salimans.
\newblock Cascaded diffusion models for high fidelity image generation.
\newblock \emph{J. Mach. Learn. Res.}, 23\penalty0 (47):\penalty0 47:1--47:33,
  May 2021.
\newblock ISSN 1532-4435,1533-7928.

\bibitem[Li et~al.(2024)Li, Carver, Lopez-Gomez, Sha, and Anderson]{Li2024-fs}
Lizao Li, Robert Carver, Ignacio Lopez-Gomez, Fei Sha, and John Anderson.
\newblock Generative emulation of weather forecast ensembles with diffusion
  models.
\newblock \emph{Sci Adv}, 10\penalty0 (13):\penalty0 eadk4489, March 2024.
\newblock ISSN 2375-2548.
\newblock \doi{10.1126/sciadv.adk4489}.

\bibitem[Price et~al.(2024)Price, Sanchez-Gonzalez, Alet, Andersson, El-Kadi,
  Masters, Ewalds, Stott, Mohamed, Battaglia, Lam, and Willson]{Gencast}
Ilan Price, Alvaro Sanchez-Gonzalez, Ferran Alet, Tom~R Andersson, Andrew
  El-Kadi, Dominic Masters, Timo Ewalds, Jacklynn Stott, Shakir Mohamed, Peter
  Battaglia, Remi Lam, and Matthew Willson.
\newblock Probabilistic weather forecasting with machine learning.
\newblock \emph{Nature}, pages 1--7, December 2024.
\newblock ISSN 1476-4687,1476-4687.
\newblock \doi{10.1038/s41586-024-08252-9}.

\bibitem[Mardani et~al.(2025{\natexlab{a}})Mardani, Brenowitz, Cohen, Pathak,
  Chen, Liu, Vahdat, Nabian, Ge, Subramaniam, et~al.]{Mardani2023-cd}
Morteza Mardani, Noah Brenowitz, Yair Cohen, Jaideep Pathak, Chieh-Yu Chen,
  Cheng-Chin Liu, Arash Vahdat, Mohammad~Amin Nabian, Tao Ge, Akshay
  Subramaniam, et~al.
\newblock Residual corrective diffusion modeling for km-scale atmospheric
  downscaling.
\newblock \emph{Communications Earth \& Environment}, 6\penalty0 (1):\penalty0
  124, 2025{\natexlab{a}}.

\bibitem[Rozet and Louppe(2023)]{Rozet2023}
Fran\c{c}ois Rozet and Gilles Louppe.
\newblock Score-based data assimilation.
\newblock In A.~Oh, T.~Naumann, A.~Globerson, K.~Saenko, M.~Hardt, and
  S.~Levine, editors, \emph{Advances in Neural Information Processing Systems},
  volume~36, pages 40521--40541. Curran Associates, Inc., 2023.
\newblock URL
  \url{https://proceedings.neurips.cc/paper_files/paper/2023/file/7f7fa581cc8a1970a4332920cdf87395-Paper-Conference.pdf}.

\bibitem[Manshausen et~al.(2024)Manshausen, Cohen, Pathak, Pritchard, Garg,
  Mardani, Kashinath, Byrne, and Brenowitz]{Manshausen2024}
Peter Manshausen, Yair Cohen, Jaideep Pathak, Mike Pritchard, Piyush Garg,
  Morteza Mardani, Karthik Kashinath, Simon Byrne, and Noah Brenowitz.
\newblock Generative data assimilation of sparse weather station observations
  at kilometer scales.
\newblock \emph{arXiv [cs.LG]}, June 2024.

\bibitem[Bassetti et~al.(2024)Bassetti, Hutchinson, Tebaldi, and
  Kravitz]{bassetti2024diffesm}
Seth Bassetti, Brian Hutchinson, Claudia Tebaldi, and Ben Kravitz.
\newblock Diffesm: Conditional emulation of temperature and precipitation in
  earth system models with 3d diffusion models.
\newblock \emph{Journal of Advances in Modeling Earth Systems}, 16\penalty0
  (10):\penalty0 e2023MS004194, 2024.

\bibitem[Quilcaille et~al.(2022)Quilcaille, Gudmundsson, Beusch, Hauser, and
  Seneviratne]{quilcaille2022showcasing}
Yann Quilcaille, Lukas Gudmundsson, Lea Beusch, Mathias Hauser, and Sonia~I
  Seneviratne.
\newblock Showcasing mesmer-x: Spatially resolved emulation of annual maximum
  temperatures of earth system models.
\newblock \emph{Geophysical Research Letters}, 49\penalty0 (17):\penalty0
  e2022GL099012, 2022.

\bibitem[Wang et~al.(2025)Wang, Souza, Ferrari, and Sapsis]{wang2025stochastic}
Mengze Wang, Andre~Nogueira Souza, Raffaele Ferrari, and Themistoklis Sapsis.
\newblock Stochastic emulators of spatially resolved extreme temperatures of
  earth system models.
\newblock \emph{Authorea Preprints}, 2025.
\newblock \doi{10.1029/2023ms004021}.

\bibitem[Hersbach et~al.(2020)Hersbach, Bell, Berrisford, Hirahara, Horányi,
  Muñoz-Sabater, Nicolas, Peubey, Radu, Schepers, Simmons, Soci, Abdalla,
  Abellan, Balsamo, Bechtold, Biavati, Bidlot, Bonavita, Chiara, Dahlgren, Dee,
  Diamantakis, Dragani, Flemming, Forbes, Fuentes, Geer, Haimberger, Healy,
  Hogan, Hólm, Janisková, Keeley, Laloyaux, Lopez, Lupu, Radnoti, Rosnay,
  Rozum, Vamborg, Villaume, and {Jean-Noël Thépaut}]{ERA5}
Hans Hersbach, Bill Bell, Paul Berrisford, Shoji Hirahara, András Horányi,
  Joaquín Muñoz-Sabater, Julien Nicolas, Carole Peubey, Raluca Radu, Dinand
  Schepers, Adrian Simmons, Cornel Soci, Saleh Abdalla, Xavier Abellan,
  Gianpaolo Balsamo, Peter Bechtold, Gionata Biavati, Jean Bidlot, Massimo
  Bonavita, Giovanna Chiara, Per Dahlgren, Dick Dee, Michail Diamantakis,
  Rossana Dragani, Johannes Flemming, Richard Forbes, Manuel Fuentes, Alan
  Geer, Leo Haimberger, Sean Healy, Robin~J Hogan, Elías Hólm, Marta
  Janisková, Sarah Keeley, Patrick Laloyaux, Philippe Lopez, Cristina Lupu,
  Gabor Radnoti, Patricia Rosnay, Iryna Rozum, Freja Vamborg, Sebastien
  Villaume, and {Jean-Noël Thépaut}.
\newblock The {ERA5} global reanalysis.
\newblock \emph{Quart. J. Roy. Meteor. Soc.}, 146\penalty0 (730):\penalty0
  1999--2049, July 2020.
\newblock ISSN 0035-9009,1477-870X.
\newblock \doi{10.1002/qj.3803}.

\bibitem[Hohenegger et~al.(2023)Hohenegger, Korn, Linardakis, Redler, Schnur,
  Adamidis, Bao, Bastin, Behravesh, Bergemann, Biercamp, Bockelmann, Brokopf,
  Br\"uggemann, Casaroli, Chegini, Datseris, Esch, George, Giorgetta, Gutjahr,
  Haak, Hanke, Ilyina, Jahns, Jungclaus, Kern, Klocke, Kluft, K\"olling,
  Kornblueh, Kosukhin, Kroll, Lee, Mauritsen, Mehlmann, Mieslinger, Naumann,
  Paccini, Peinado, Praturi, Putrasahan, Rast, Riddick, Roeber, Schmidt,
  Schulzweida, Sch\"utte, Segura, Shevchenko, Singh, Specht, Stephan, von
  Storch, Vogel, Wengel, Winkler, Ziemen, Marotzke, and Stevens]{Icon2023}
C.~Hohenegger, P.~Korn, L.~Linardakis, R.~Redler, R.~Schnur, P.~Adamidis,
  J.~Bao, S.~Bastin, M.~Behravesh, M.~Bergemann, J.~Biercamp, H.~Bockelmann,
  R.~Brokopf, N.~Br\"uggemann, L.~Casaroli, F.~Chegini, G.~Datseris, M.~Esch,
  G.~George, M.~Giorgetta, O.~Gutjahr, H.~Haak, M.~Hanke, T.~Ilyina, T.~Jahns,
  J.~Jungclaus, M.~Kern, D.~Klocke, L.~Kluft, T.~K\"olling, L.~Kornblueh,
  S.~Kosukhin, C.~Kroll, J.~Lee, T.~Mauritsen, C.~Mehlmann, T.~Mieslinger,
  A.~K. Naumann, L.~Paccini, A.~Peinado, D.~S. Praturi, D.~Putrasahan, S.~Rast,
  T.~Riddick, N.~Roeber, H.~Schmidt, U.~Schulzweida, F.~Sch\"utte, H.~Segura,
  R.~Shevchenko, V.~Singh, M.~Specht, C.~C. Stephan, J.-S. von Storch,
  R.~Vogel, C.~Wengel, M.~Winkler, F.~Ziemen, J.~Marotzke, and B.~Stevens.
\newblock Icon-sapphire: simulating the components of the earth system and
  their interactions at kilometer and subkilometer scales.
\newblock \emph{Geoscie ntific Model Development}, 16\penalty0 (2):\penalty0
  779--811, 2023.
\newblock \doi{10.5194/gmd-16-779-2023}.
\newblock URL \url{https://gmd.copernicus.org/articles/16/779/2023/}.

\bibitem[Eyring et~al.(2016)Eyring, Bony, Meehl, Senior, Stevens, Stouffer, and
  Taylor]{cmip6}
Veronika Eyring, Sandrine Bony, Gerald~A Meehl, Catherine~A Senior, Bjorn
  Stevens, Ronald~J Stouffer, and Karl~E Taylor.
\newblock Overview of the coupled model intercomparison project phase 6
  ({CMIP6}) experimental design and organization.
\newblock \emph{Geosci. Model Dev.}, 9\penalty0 (5):\penalty0 1937--1958, May
  2016.
\newblock ISSN 1991-959X,1991-9603.
\newblock \doi{10.5194/gmd-9-1937-2016}.

\bibitem[Gates et~al.(1999)Gates, Boyle, Covey, Dease, Doutriaux, Drach,
  Fiorino, Gleckler, Hnilo, Marlais, Phillips, Potter, Santer, Sperber, Taylor,
  and Williams]{amip}
W~Lawrence Gates, James~S Boyle, Curt Covey, Clyde~G Dease, Charles~M
  Doutriaux, Robert~S Drach, Michael Fiorino, Peter~J Gleckler, Justin~J Hnilo,
  Susan~M Marlais, Thomas~J Phillips, Gerald~L Potter, Benjamin~D Santer,
  Kenneth~R Sperber, Karl~E Taylor, and Dean~N Williams.
\newblock An overview of the results of the atmospheric model intercomparison
  project ({AMIP} {I}).
\newblock \emph{Bull. Am. Meteorol. Soc.}, 80\penalty0 (1):\penalty0 29--55,
  January 1999.
\newblock ISSN 0003-0007,1520-0477.
\newblock \doi{10.1175/1520-0477(1999)080<0029:aootro>2.0.co;2}.

\bibitem[Yang and Slingo(2001)]{yang2001diurnal}
Gui-Ying Yang and Julia Slingo.
\newblock The diurnal cycle in the tropics.
\newblock \emph{Monthly Weather Review}, 129\penalty0 (4):\penalty0 784--801,
  2001.

\bibitem[Thompson and Wallace(2000)]{thompson2000annular}
David~WJ Thompson and John~M Wallace.
\newblock Annular modes in the extratropical circulation. part i:
  Month-to-month variability.
\newblock \emph{Journal of climate}, 13\penalty0 (5):\penalty0 1000--1016,
  2000.

\bibitem[Zamo and Naveau(2018)]{Zamo2018-xs}
Michaël Zamo and Philippe Naveau.
\newblock Estimation of the continuous ranked probability score with limited
  information and applications to ensemble weather forecasts.
\newblock \emph{Math. Geosci.}, 50\penalty0 (2):\penalty0 209--234, February
  2018.
\newblock ISSN 1874-8961,1874-8953.
\newblock \doi{10.1007/s11004-017-9709-7}.

\bibitem[Knapp et~al.(2010)Knapp, Kruk, Levinson, Diamond, and
  Neumann]{ibtracs}
Kenneth~R Knapp, Michael~C Kruk, David~H Levinson, Howard~J Diamond, and
  Charles~J Neumann.
\newblock The international best track archive for climate stewardship
  (ibtracs) unifying tropical cyclone data.
\newblock \emph{Bulletin of the American Meteorological Society}, 91\penalty0
  (3):\penalty0 363--376, 2010.

\bibitem[Dhariwal and Nichol(2021)]{classifierguidance}
Prafulla Dhariwal and Alexander Nichol.
\newblock Diffusion models beat gans on image synthesis.
\newblock \emph{Advances in neural information processing systems},
  34:\penalty0 8780--8794, 2021.

\bibitem[Viale et~al.(2018)Viale, Valenzuela, Garreaud, and
  Ralph]{Viale2018-pu}
Maximiliano Viale, Raúl Valenzuela, René~D Garreaud, and F~Martin Ralph.
\newblock Impacts of atmospheric rivers on precipitation in southern south
  america.
\newblock \emph{J. Hydrometeorol.}, 19\penalty0 (10):\penalty0 1671--1687,
  October 2018.
\newblock ISSN 1525-7541,1525-755X.
\newblock \doi{10.1175/jhm-d-18-0006.1}.

\bibitem[Mardani et~al.(2025{\natexlab{b}})Mardani, Brenowitz, Cohen, Pathak,
  Chen, Liu, Vahdat, Nabian, Ge, Subramaniam, Kashinath, Kautz, and
  Pritchard]{Mardani2025-eq}
Morteza Mardani, Noah Brenowitz, Yair Cohen, Jaideep Pathak, Chieh-Yu Chen,
  Cheng-Chin Liu, Arash Vahdat, Mohammad~Amin Nabian, Tao Ge, Akshay
  Subramaniam, Karthik Kashinath, Jan Kautz, and Mike Pritchard.
\newblock Residual corrective diffusion modeling for km-scale atmospheric
  downscaling.
\newblock \emph{Commun. Earth Environ.}, 6\penalty0 (1), February
  2025{\natexlab{b}}.
\newblock ISSN 2662-4435.
\newblock \doi{10.1038/s43247-025-02042-5}.

\bibitem[Gorski et~al.(2005)Gorski, Hivon, Banday, Wandelt, Hansen, Reinecke,
  and Bartelmann]{gorski2005healpix}
Krzysztof~M Gorski, Eric Hivon, Anthony~J Banday, Benjamin~D Wandelt, Frode~K
  Hansen, Mstvos Reinecke, and Matthia Bartelmann.
\newblock Healpix: A framework for high-resolution discretization and fast
  analysis of data distributed on the sphere.
\newblock \emph{The Astrophysical Journal}, 622\penalty0 (2):\penalty0 759,
  2005.

\bibitem[Krachmalnicoff and Tomasi(2019)]{krachmalnicoff2019convolutional}
Nicoletta Krachmalnicoff and Maurizio Tomasi.
\newblock Convolutional neural networks on the healpix sphere: a pixel-based
  algorithm and its application to cmb data analysis.
\newblock \emph{Astronomy \& Astrophysics}, 628:\penalty0 A129, 2019.

\bibitem[Perraudin et~al.(2019)Perraudin, Defferrard, Kacprzak, and
  Sgier]{perraudin2019deepsphere}
Nathana{\"e}l Perraudin, Micha{\"e}l Defferrard, Tomasz Kacprzak, and Raphael
  Sgier.
\newblock Deepsphere: Efficient spherical convolutional neural network with
  healpix sampling for cosmological applications.
\newblock \emph{Astronomy and Computing}, 27:\penalty0 130--146, 2019.

\bibitem[Karlbauer et~al.(2024)Karlbauer, Cresswell-Clay, Durran, Moreno,
  Kurth, Bonev, Brenowitz, and Butz]{Karlbauer2024-kr}
Matthias Karlbauer, Nathaniel Cresswell-Clay, Dale~R Durran, Raul~A Moreno,
  Thorsten Kurth, Boris Bonev, Noah Brenowitz, and Martin~V Butz.
\newblock Advancing parsimonious deep learning weather prediction using the
  {HEALPix} mesh.
\newblock \emph{J. Adv. Model. Earth Syst.}, 16\penalty0 (8), August 2024.
\newblock ISSN 1942-2466.
\newblock \doi{10.1029/2023ms004021}.

\bibitem[Chantry(2024)]{Chantry2024}
Matthew Chantry.
\newblock Personal communication.
\newblock Personal communication, 2024.
\newblock Conversation on September 11, 2024.

\bibitem[Koldunov et~al.(2023{\natexlab{a}})Koldunov, K\"{o}lling,
  Pedruzo-Bagazgoitia, Rackow, Redler, Sidorenko, Wieners, and
  Ziemen]{nextGEMS}
Nikolay Koldunov, Tobias K\"{o}lling, Xabier Pedruzo-Bagazgoitia, Thomas
  Rackow, Ren\'{e} Redler, Dmitry Sidorenko, Karl-Hermann Wieners, and
  Florian~Andreas Ziemen.
\newblock nextgems: output of the model development cycle 3 simulations for
  icon and ifs, 2023{\natexlab{a}}.
\newblock URL \url{https://doi.org/10.26050/WDCC/nextGEMS\_cyc3}.

\bibitem[Song et~al.(2021)Song, Sohl-Dickstein, Kingma, Kumar, Ermon, and
  Poole]{song2021scorebasedgenerativemodelingstochastic}
Yang Song, Jascha Sohl-Dickstein, Diederik~P. Kingma, Abhishek Kumar, Stefano
  Ermon, and Ben Poole.
\newblock Score-based generative modeling through stochastic differential
  equations, 2021.
\newblock URL \url{https://arxiv.org/abs/2011.13456}.

\bibitem[Ho et~al.(2020{\natexlab{b}})Ho, Jain, and Abbeel]{Ho2020-pp}
Jonathan Ho, Ajay Jain, and Pieter Abbeel.
\newblock Denoising diffusion probabilistic models.
\newblock \emph{arXiv [cs.LG]}, June 2020{\natexlab{b}}.

\bibitem[Everaert et~al.(2023)Everaert, Fitsios, Bocchio, Arpa, Süsstrunk, and
  Achanta]{Everaert2023-bi}
Martin~Nicolas Everaert, Athanasios Fitsios, Marco Bocchio, Sami Arpa, Sabine
  Süsstrunk, and Radhakrishna Achanta.
\newblock Exploiting the signal-leak bias in diffusion models.
\newblock \emph{arXiv [cs.CV]}, September 2023.

\bibitem[Yogesh et~al.(2022)Yogesh, Seungjun, Xun, Arash, Jiaming, Qinsheng,
  Karsten, Miika, Timo, Samuli, Bryan, Tero, and Ming-Yu]{Yogesh2022-zv}
Balaji Yogesh, Nah Seungjun, Huang Xun, Vahdat Arash, Song Jiaming, Zhang
  Qinsheng, Kreis Karsten, Aittala Miika, Aila Timo, Laine Samuli, Catanzaro
  Bryan, Karras Tero, and Liu Ming-Yu.
\newblock {EDiff}-{I}: Text-to-image diffusion models with an ensemble of
  expert denoisers.
\newblock \emph{arXiv [cs.CV]}, November 2022.

\bibitem[Bar-Tal et~al.(2023)Bar-Tal, Yariv, Lipman, and
  Dekel]{bartal2023multidiffusionfusingdiffusionpaths}
Omer Bar-Tal, Lior Yariv, Yaron Lipman, and Tali Dekel.
\newblock Multidiffusion: Fusing diffusion paths for controlled image
  generation, 2023.
\newblock URL \url{https://arxiv.org/abs/2302.08113}.

\bibitem[Song et~al.(2020{\natexlab{b}})Song, Sohl-Dickstein, Kingma, Kumar,
  Ermon, and Poole]{Song2020-pj}
Yang Song, Jascha Sohl-Dickstein, Diederik~P Kingma, Abhishek Kumar, Stefano
  Ermon, and Ben Poole.
\newblock Score-based generative modeling through stochastic differential
  equations.
\newblock \emph{arXiv [cs.LG]}, November 2020{\natexlab{b}}.

\bibitem[Su et~al.(2022)Su, Song, Meng, and Ermon]{DDIB}
Xuan Su, Jiaming Song, Chenlin Meng, and Stefano Ermon.
\newblock Dual diffusion implicit bridges for image-to-image translation.
\newblock \emph{arXiv preprint arXiv:2203.08382}, 2022.

\bibitem[Liu et~al.(2022)Liu, Gong, and Liu]{Liu2022-eu}
Xingchao Liu, Chengyue Gong, and Qiang Liu.
\newblock Flow straight and fast: Learning to generate and transfer data with
  rectified flow.
\newblock \emph{arXiv [cs.LG]}, September 2022.

\bibitem[Koldunov et~al.(2023{\natexlab{b}})Koldunov, K\"{o}lling,
  Pedruzo-Bagazgoitia, Rackow, Redler, Sidorenko, Wieners, and
  Ziemen]{IconCycle3}
Nikolay Koldunov, Tobias K\"{o}lling, Xabier Pedruzo-Bagazgoitia, Thomas
  Rackow, Ren\'{e} Redler, Dmitry Sidorenko, Karl-Hermann Wieners, and
  Florian~Andreas Ziemen.
\newblock nextgems: output of the model development cycle 3 simulations for
  icon and ifs, 2023{\natexlab{b}}.
\newblock URL \url{https://doi.org/10.26050/WDCC/nextGEMS\_cyc3}.

\bibitem[Chen(2023)]{Chen2023}
Ting Chen.
\newblock On the importance of noise scheduling for diffusion models.
\newblock \emph{arXiv [cs.CV]}, January 2023.

\bibitem[Wang et~al.(2023)Wang, Cheng, Sun, Wang, Qi, Liao, Wang, and
  Liu]{Wang2023-ha}
Huazheng Wang, Daixuan Cheng, Haifeng Sun, Jingyu Wang, Qi~Qi, Jianxin Liao,
  Jing Wang, and Cong Liu.
\newblock How does diffusion influence pretrained language models on
  out-of-distribution data?
\newblock \emph{arXiv [cs.CL]}, July 2023.

\bibitem[Favero et~al.(2025)Favero, Sclocchi, and Wyart]{Favero2025-sa}
Alessandro Favero, Antonio Sclocchi, and Matthieu Wyart.
\newblock Bigger isn't always memorizing: Early stopping overparameterized
  diffusion models.
\newblock \emph{arXiv [cs.LG]}, May 2025.

\bibitem[Kingma and Ba(2014)]{Adam}
Diederik~P Kingma and Jimmy Ba.
\newblock Adam: A method for stochastic optimization.
\newblock \emph{arXiv [cs.LG]}, December 2014.

\bibitem[Ho et~al.(2022)Ho, Salimans, Gritsenko, Chan, Norouzi, and
  Fleet]{ho2022-vd}
Jonathan Ho, Tim Salimans, Alexey Gritsenko, William Chan, Mohammad Norouzi,
  and David~J. Fleet.
\newblock Video diffusion models, 2022.
\newblock URL \url{https://arxiv.org/abs/2204.03458}.

\bibitem[Blattmann et~al.(2023)Blattmann, Rombach, Ling, Dockhorn, Kim, Fidler,
  and Kreis]{blattmann2023-ldm}
Andreas Blattmann, Robin Rombach, Huan Ling, Tim Dockhorn, Seung~Wook Kim,
  Sanja Fidler, and Karsten Kreis.
\newblock Align your latents: High-resolution video synthesis with latent
  diffusion models, 2023.
\newblock URL \url{https://arxiv.org/abs/2304.08818}.

\bibitem[Voleti et~al.(2022)Voleti, Jolicoeur-Martineau, and
  Pal]{voleti2022-mcvd}
Vikram Voleti, Alexia Jolicoeur-Martineau, and Christopher Pal.
\newblock Mcvd: Masked conditional video diffusion for prediction, generation,
  and interpolation, 2022.
\newblock URL \url{https://arxiv.org/abs/2205.09853}.

\bibitem[Rao et~al.(2002)Rao, do~Carmo, and Franchito]{Rao2002}
V.~Brahmananda Rao, A.~M.~C. do~Carmo, and Sergio~H. Franchito.
\newblock Seasonal variations in the southern hemisphere storm tracks and
  associated wave propagation.
\newblock \emph{Journal of the Atmospheric Sciences}, 59\penalty0 (6):\penalty0
  1029 -- 1040, 2002.
\newblock \doi{10.1175/1520-0469(2002)059<1029:SVITSH>2.0.CO;2}.
\newblock URL
  \url{https://journals.ametsoc.org/view/journals/atsc/59/6/1520-0469_2002_059_1029_svitsh_2.0.co_2.xml}.

\bibitem[Sobel et~al.(2001)Sobel, Nilsson, and Polvani]{sobel2001weak}
Adam~H Sobel, Johan Nilsson, and Lorenzo~M Polvani.
\newblock The weak temperature gradient approximation and balanced tropical
  moisture waves.
\newblock \emph{Journal of the atmospheric sciences}, 58\penalty0
  (23):\penalty0 3650--3665, 2001.

\end{thebibliography}
\bibliographystyle{unsrtnat}

\clearpage
\appendix

\section{Supplemental Methods}
\subsection{\healpix convolutions\label{sec:si:hpx-conv}}

\begin{figure}
    \centering
    \includegraphics[width=0.49\textwidth]{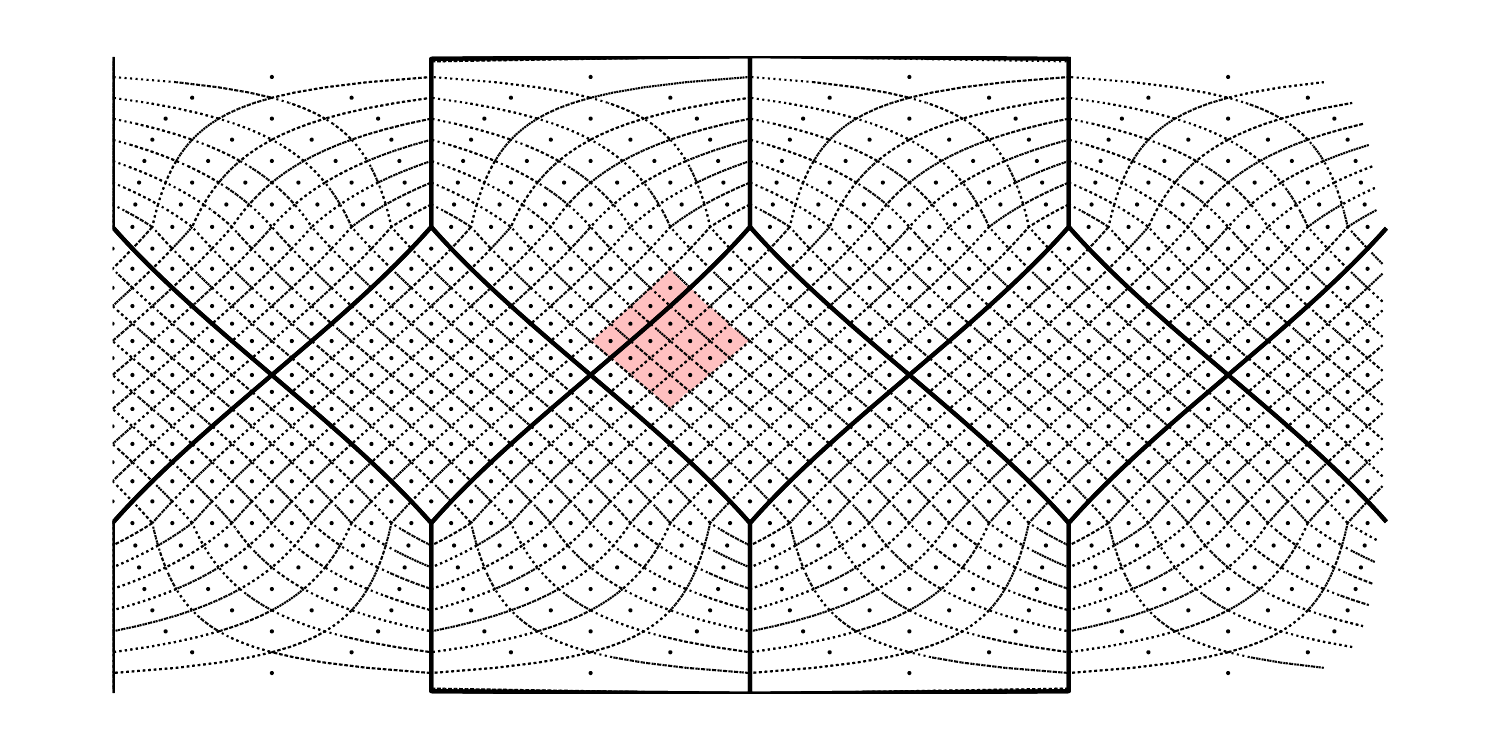}
    \includegraphics[width=0.49\textwidth]{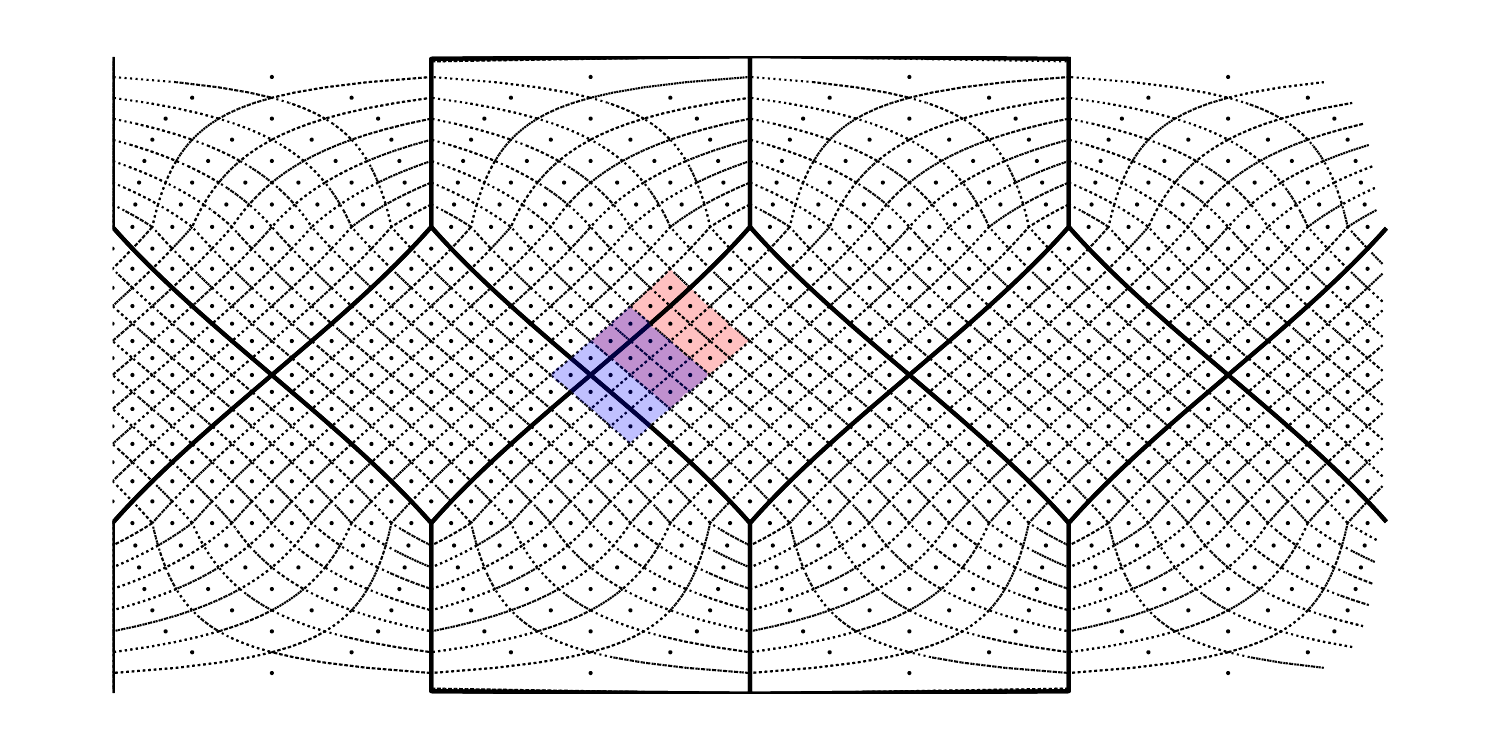}
    \includegraphics[width=0.49\textwidth]{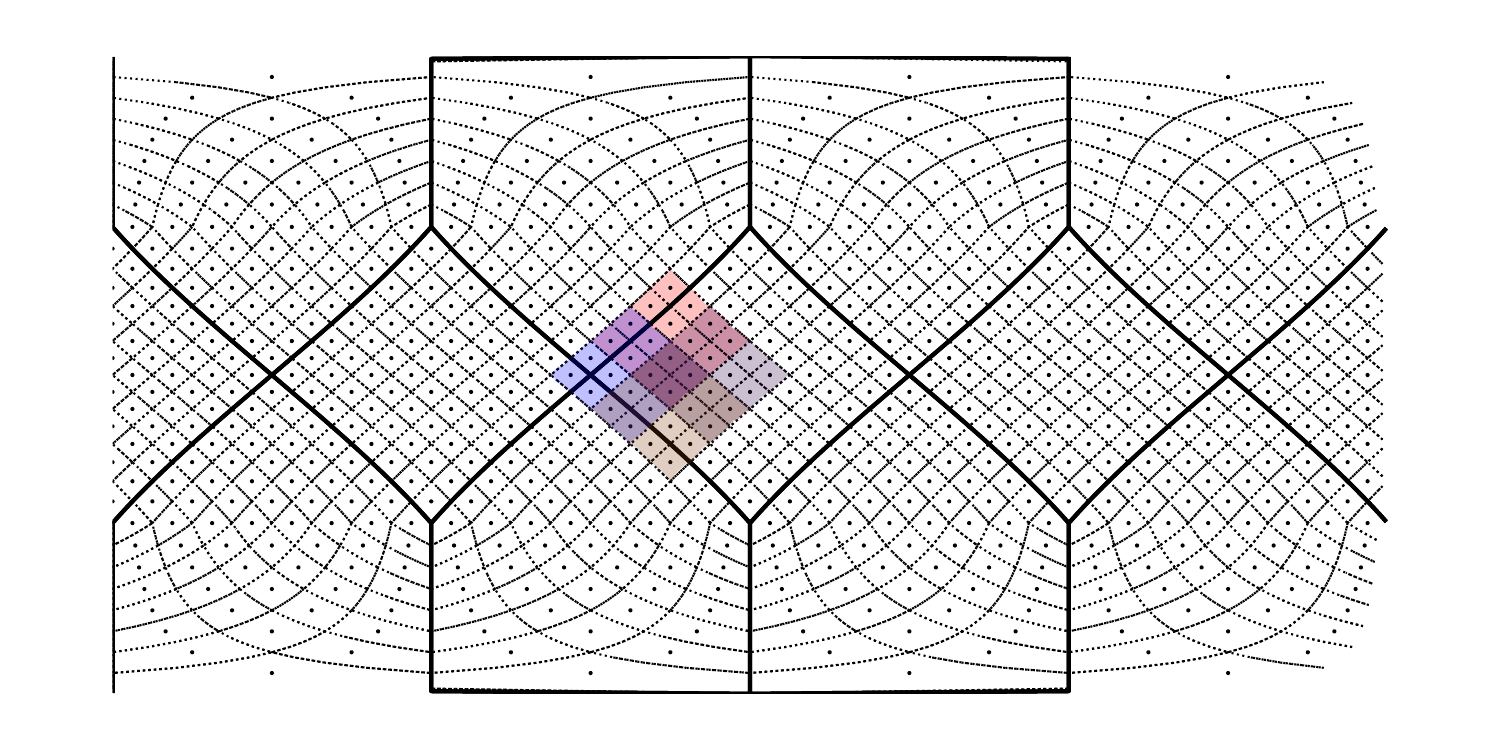}
    \includegraphics[width=0.49\textwidth]{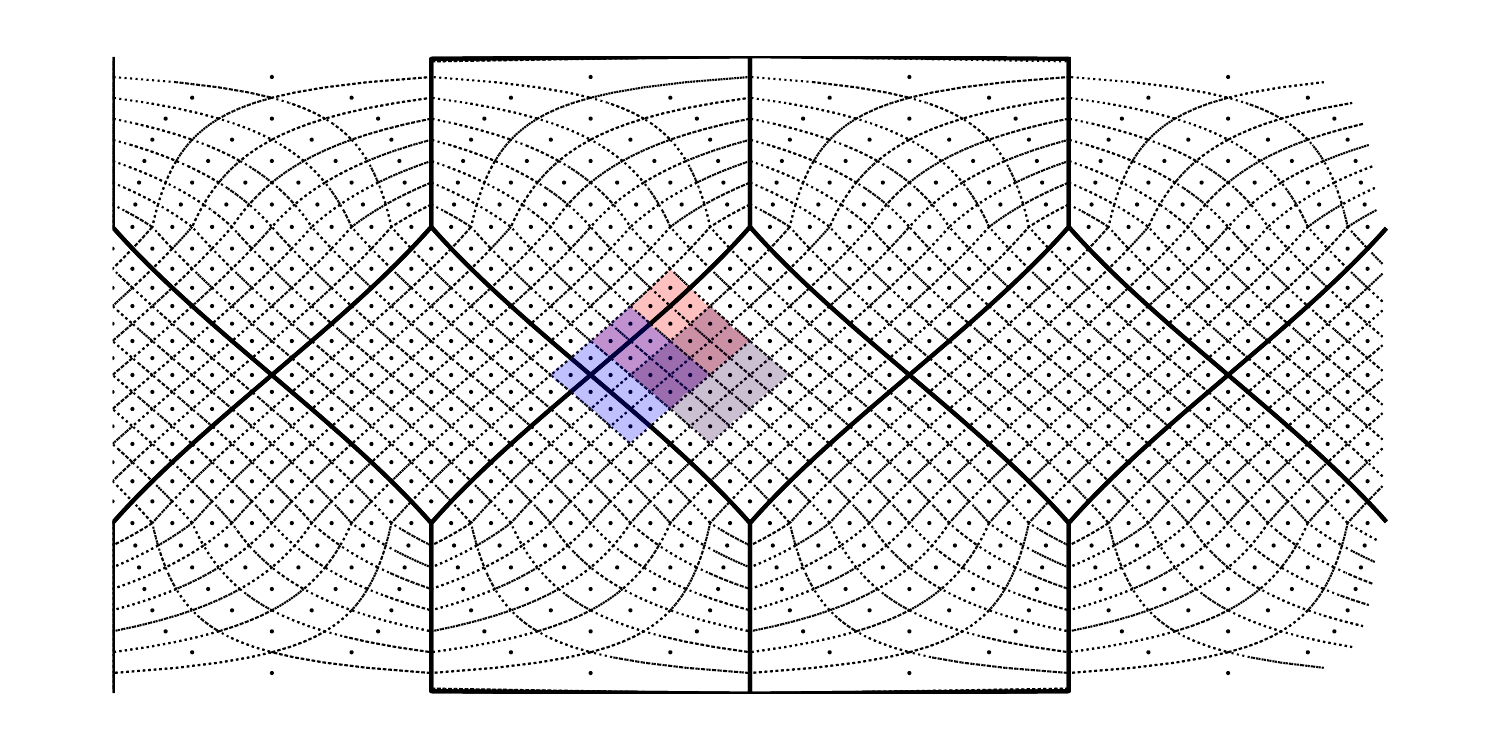}
    \caption{Schematic showing the multi-diffusion approach for spherical geometry with \healpix grids. A sequence of the overlapping patch sampling protocol is shown going clockwise from the top left. Each colored region indicates a single patch and overlap between the patches minimizes boundary artifacts by combining patch representations into a global representation following \ref{eq:patch_denoising}.}
    \label{fig:hpx_multidiffusion_across_face}
\end{figure}

\begin{figure}
    \centering
    \includegraphics[width=0.99\textwidth]{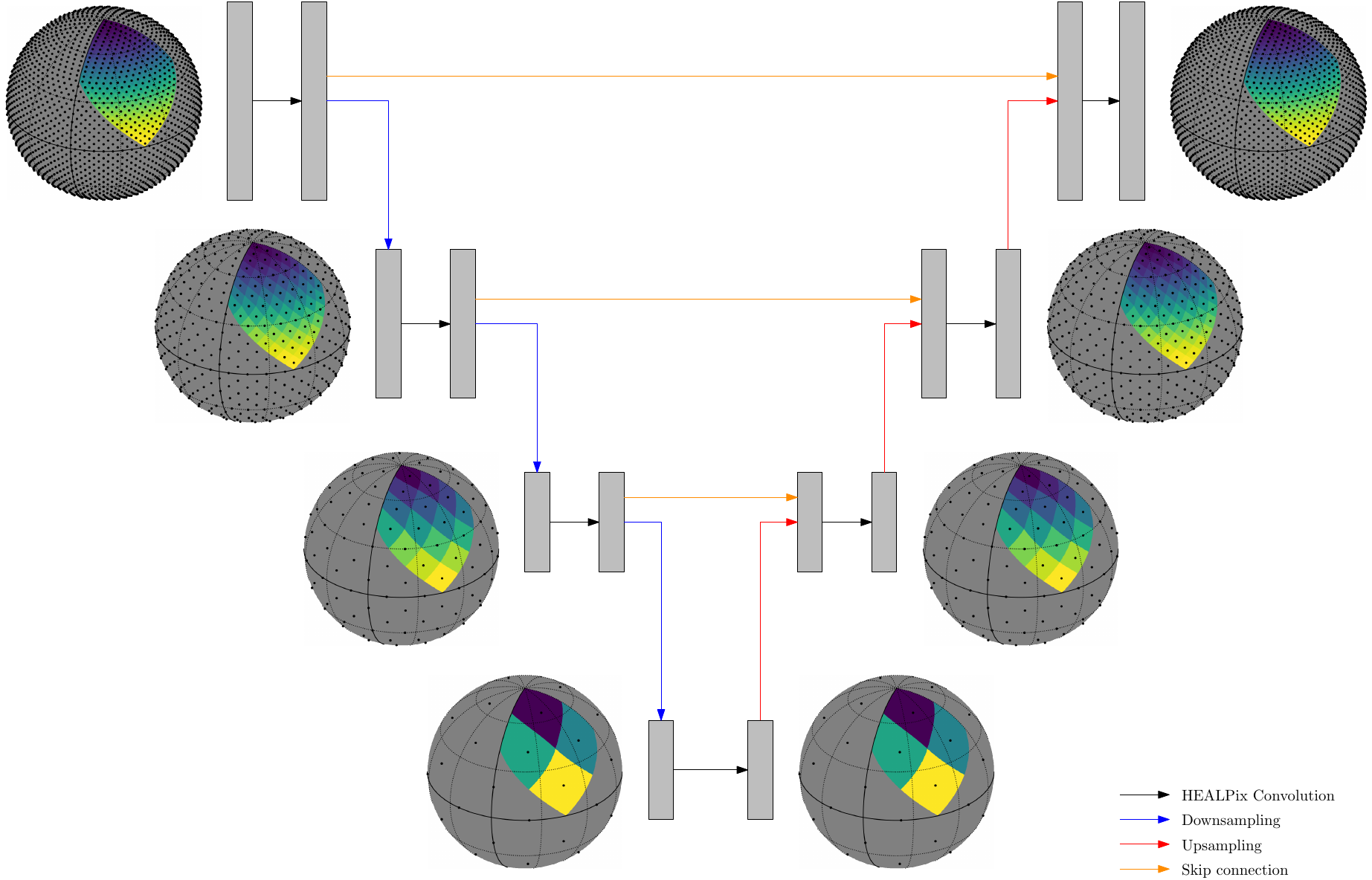}
    \caption{Schematic showing the global UNet architecture for spherical geometry with \healpix grids. Each convolution is applied on \healpix faces padded as in \citep{Karlbauer2024-kr}.}
    \label{fig:hpx_global_UNet}
\end{figure}

Consider a field represented on a level $l$ of the \healpix grid, $x \in \mathbb{R}^{c_{in} \times 12 \times 2^l\times 2^l}$. One may apply a discrete convolution kernel $\mathcal{K}_{conv} \in \mathbb{R}^{c_{out} \times c_{in} \times 2k+1\times 2k+1}$ as follows:
\begin{align}
    x &= \left( x_1, x_2, \cdots x_{12} \right), & \quad x_f \in \mathbb{R}^{c_{in} \times 2^l\times 2^l} \\
    x^p &= \mathcal{P}\left(x; k\right) = \left( x^p_1, x^p_2, \cdots x^p_{12} \right), & \quad x^p_f \in \mathbb{R}^{c_{in} \times (2^l+2k)\times (2^l+2k)} \\
    y_f &= \mathcal{K}_{conv} \ast x^p_f, & \quad y_f \in \mathbb{R}^{c_{out} \times 2^l\times 2^l} \\
    y &= \left( y_1, y_2, \cdots y_{12} \right), & \quad y \in \mathbb{R}^{c_{out} \times 12 \times 2^l\times 2^l}
\end{align}
where $x_f$ and $y_f$ are the pixels of $x$ and $y$ respectively on the \healpix face $f$, $\mathcal{P}(.; k)$ is the padding operation of \citep{Karlbauer2024-kr} and $\ast$ indicates the convolution operation.
Figure~\ref{fig:hpx_global_UNet} shows a schematic of how the UNet architecture of \citep{song2021scorebasedgenerativemodelingstochastic} is adapted to work on global \healpix fields.
\subsection{Data sources and pre-processing \label{sec:si:methods:data}}

\paragraph{ERA5} For pre-processing, we retrieved hourly ERA5 data for 1980--2018 (inclusive) via the data lake at the National Energy Research Computing Center (NERSC) at $0.25^{\circ}$ resolution on the lat-lon grid. The 3d atmospheric states are available on pressure levels. We then regridded this using bilinear interpolation to the \healpix grid with $N_{side}=256$. For training the coarse-resolution model, this coarsened by averaging pooling to an $N_{side}=64$.
We train on years 1980--2017 (inclusive) and validate on 2018, though many of our climate diagnostics below are also computed on the validation set (as is common in generative modeling).

\paragraph{AMIP SST Forcing} Our models take a monthly time-scale SST as an input. For the ERA5 data which is paired with reality, we linearly interpolate the adjusted monthly SST time series following the AMIP protocol \citep{amip}.

\paragraph{ICON Cycle 3} We obtained O(PB) of ICON data (ngc3028) of the nextGEMS cycle 3 simulations \citep{nextGEMS} in zarr format from DKRZ's JupyterHub server. 
This dataset was stored on the \healpix grid with $N_{side}=1024$, corresponding roughly to a resolution of $\sqrt{A / (12 N_{side}^2)}=6.3 \text{km}$. Prior to our handling of the data it was interpolated using nearest neighbors from the native icosahedral grid. Five years of this data are available at 3-hour (3D) and 30-minute (2D) resolution in time. Conveniently, this dataset featured pre-coarsened data (again using averaging pooling) at our coarse-resolution of $N_{side}=64$.

We interpolated this coarse data to fixed pressure levels using linear interpolation in the vertical direction.
To fill in values at pressure level locations that are below the surface, we use extrapolation based on hydrostatic balance constraints and an assumption of a constant temperature lapse rate of $6.5$ K/km for temperature and geo-potential..
For all other variables, we use constant extrapolation down from the surface.
This procedure approximately reproduces ERA5's undocumented procedure for filling-in below-surface levels.

We compute the SST input for ICON with a rolling window average in time with a window size of 30 days based on the HPX64 spatial and daily temporal resolution data in the ICON dataset.

\paragraph{AMIP SST Forcing} We downloaded the file: 
input4MIPs.CMIP6Plus.CMIP.PCMDI.PCMDI-AMIP-1-1-9.ocean.mon.tosbcs.gn from the Earth System Grid Federation, and use bilinear interpolation in space to map it to HPX64 resolution.

\subsection{Diffusion Training}
\subsubsection{Tuning noise schedules based on the data distribution}
\label{sec:si:tuning-noise}

The choice of noise levels seen during training ($p_\sigma$) and inference ($\sigma_{min}$ and $\sigma_{max}$) is crucial for the quality of a diffusion models.
Unfortunately, our models failed to produce seasonal or diurnal cycles when using the standard settings from \citep{Karras2022-edm}, suggesting that these hyper-parameters settings tuned for image generation are not suitable for generating atmospheric states (see Figure \ref{fig:diurnal_conditioning_rsut}). The basic issue is known as the \emph{signal leak bias} when a model is not trained with enough noise to entirely swamp the largest modes present in a dataset \citep{Everaert2023-bi}.
Without enough noise, it is easy to recover high-amplitude and large-scale signals like the seasonal or diurnal cycles---for example by spatial averaging---so the model never learns to create these signals on its own.
In this section, we propose a criterion for adjusting this settings based only on a spectral analysis of the data, without a need for expensive hyper-parameter tuning.

Specifically, the upper and lower bound of the noise ranges spanned during training $p_{\sigma}$ and the initial and final times for \eqref{eq:ode},  $\sigma_{min,max}$, should be determined by
\begin{align}
\sigma^2_{min} & \lesssim \inf_{||v||=1} E|v'x|^2 \label{eq:sigmin:def} \\
\sigma^2_{max} & \gtrsim \sup_{||v||=1} E|v'x|^2. \label{eq:sigmax:def}
\end{align}
The right hand sides are the minimal and maximal variances along any direction of the data spaces---and are equivalent to the maximum and minimum singular values of the covariance matrix $E|xx'|$. The \emph{signal leak bias} occurs when $\sigma_{max}^2$ is too small, such that the highest variances mode of the data is not covered up. When $\sigma_{min}$ is too large, the diffusion model will not denoise the lower-amplitude modes of the data---typically the highest spatial frequencies.  Choosing  $\sigma_{min}$ too large manifests as an incorrectly flat spectra for the highest wave numbers (see e.g., the Z500 power spectra presented by \citet{Gencast}).
 
To derive this, let $\textbf{v}\in \mathbb{R}^n$ be an arbitrary vector and for convenience assume that $|v|=1$. The data projected onto this vector should overwhelm the noise for $\sigma_{min}$  so that 
\[
|v'x| \gg \sigma_{min} |v'\epsilon|
\]
Likewise for high noise levels we demand that the signal is lost in the noise which means that
\[
|v'x| \ll \sigma_{max} |v'\epsilon|.
\]
\eqref{eq:sigmax:def}  are recovered from this equations by taking the expectation of these inequalities and noting that $E|v'\epsilon|^2 = |v|^2 E|\epsilon|^2 = 1$.

Note that \eqref{eq:sigmax:def} is equal to the variance explained by the maximum principal component of the data.  Recall that the overall variance can be decomposed $n=Tr Cov(x)=\sum_i s^2_i$  where $n$ is the dimension of $x$ (pixels and channels). To understand the behavior for changing $n$ it is useful to discuss the fraction of variance explained by a singular value $\tilde{s}^2_i=s_i^2 / n $, and expect this fraction to converge as we refine the resolution. This implies then, that $\sigma_{max}^2 \gtrsim n \tilde{s}_1^2 = O(n)$. 

In atmospheric science, the seasonal cycle and diurnal cycles are the dominant modes of variability since they are: 1) planetary in spatial scale; and 2) large in amplitude.  In mid-latitude land locations, the daily high and low temperature are separated by  $\sim 5-10^{\circ}$C, while the seasonal cycle is $\sim 10-20^{\circ}$C. Other variables don't have as strong a seasonal cycle over as large an area so we take the conservative estimate that the seasonal cycle explains 5\% of the global variance across our whole variable set. The HPX64 model has about 30 channels in total and 49152 pixels. So this would imply a $\sigma_{max}^2 \gtrsim 300$.  This is far larger than the range of the hyper parameters used for this resolution in typical image generation \citep{Karras2022-edm}. Similar analysis can explain the observation that generating  higher-resolution images requires larger noise levels \citep{Chen2023}.

While one can tune the log-normal distribution to fully cover the dynamic range of the data $[\sigma_{min},\sigma_{max}]$, we have found that sample quality is sensitive to shifts in the mean of the log-normal distribution. As a simpler approach, we propose the following noise schedule 
\begin{equation}
    p_\sigma \propto \sigma^{-1},  \sigma_{min} \leq \sigma \leq \sigma_{max}.
\end{equation}
Because this is uniform in $\log \sigma$ we call it a ``log-uniform'' noise distribution.
In earlier tests on the 2D data alone without SST conditioning, we achieved similar distortions of the power law distribution as reported by \citet{Gencast}, i.e. incorrectly flat spectra for the highest wave numbers. 
In practice, we set $\sigma_{min}=0.02$ and $\sigma_{max}=200$, settings which span the dynamic range of several of our variables (see Figure \ref{fig:spectra}; in hindsight we should have decreased $\sigma_{min}$ even further for the smoothest fields like Z500). For inference, we use the same sampler as \citet{Karras2022-edm} but with these same settings.

\subsubsection{Avoiding over-fitting at large noise levels\label{sec:overfit}}

Similar spectral considerations reveal that diffusion models are prone to over-fitting at high noise levels when trained on a limited number of samples. This occurs because there are very few independent observations of such large variance signals in the training data due to these conspiring factors:
\begin{itemize}
    \item Training at large noise-levels reduces the number of degrees of freedom per sample. This can be illustrated by assuming that $\sigma$ is sufficiently large such that only 10 singular values of the data are larger than $\sqrt{n} \sigma$ and that corresponding principal components are independent random variables. The same reasoning would apply if we chose the 10 most energetic Fourier modes. In either case, only 10 independent degrees of freedom per sample can be recovered at this noise level.
    \item Time series data is auto-correlated in time. Moreover, in many physical systems like the atmosphere, the auto-correlation of the largest modes of the data tend to be the slowest. For example, the highest-variance modes of the global ocean and extra-tropical atmosphere, respectively, are ENSO, which has an inter-annual timescale, and the annular modes with a multi-week timescale. So in 40-year duration of ERA5 there are only a few dozen ESNO events and few hundred oscillations of the annular modes.
    \item Conditional inputs that are easy to memorize. For example, the network can easily learn to map SST input to the timestamp of the sample and then use this timestamp to recover the desired outputs.
\end{itemize}

Since the large noise levels are exponentially less likely to be sampled during training, they comprise a small portion of the overall loss so this over-fitting can be easy to miss. Therefore, it is critical to monitor the relationship between the training and validation loss and the noise level $\sigma$. Indeed, we see that this is a problem for the coarse-resolution \texttt{cBottle-3d} model especially for $\sigma \geq 100$ (Fig. \ref{fig:si:overfit}). The small noise levels, by comparison, show no signs of over-fitting to the validation set.

We control this over-fitting with an ensemble of experts \citep{Yogesh2022-zv} approach where separate models are used for small $(0, 10]$, medium $(10, 100]$, and large $(100, \infty)$ noise levels $\sigma$ during inference. This allows us to regularize the large noise level models more than the small. For simplicity and to avoid extra training, we regularize via early stopping. Specifically, we select the small, medium, and large models to be the checkpoints after 9.856M, 2.0248M, and 512K samples of training, respectively. Experiments with activation and condition dropout yielded better validation losses, but the gains were marginal and required extensive tuning.

Other studies have also noted the tendency of diffusion models to over-fit to their training data. Similar to the present setting, \citet{Wang2023-ha} note that the generalization gap between training and out of distribution data for diffusion language models is larger for increasing noise levels. \citet{Favero2025-sa} also propose a criteria for early-stopping based regularization that depends on the dataset size, but do not explicitly connect this with the need to regularize high levels more than small or spectral analysis.

\subsection{Multi-diffusion: details \label{sec:si:multi-diffusion}}

\begin{figure}
    \centering
    \includegraphics[width=0.95\textwidth]{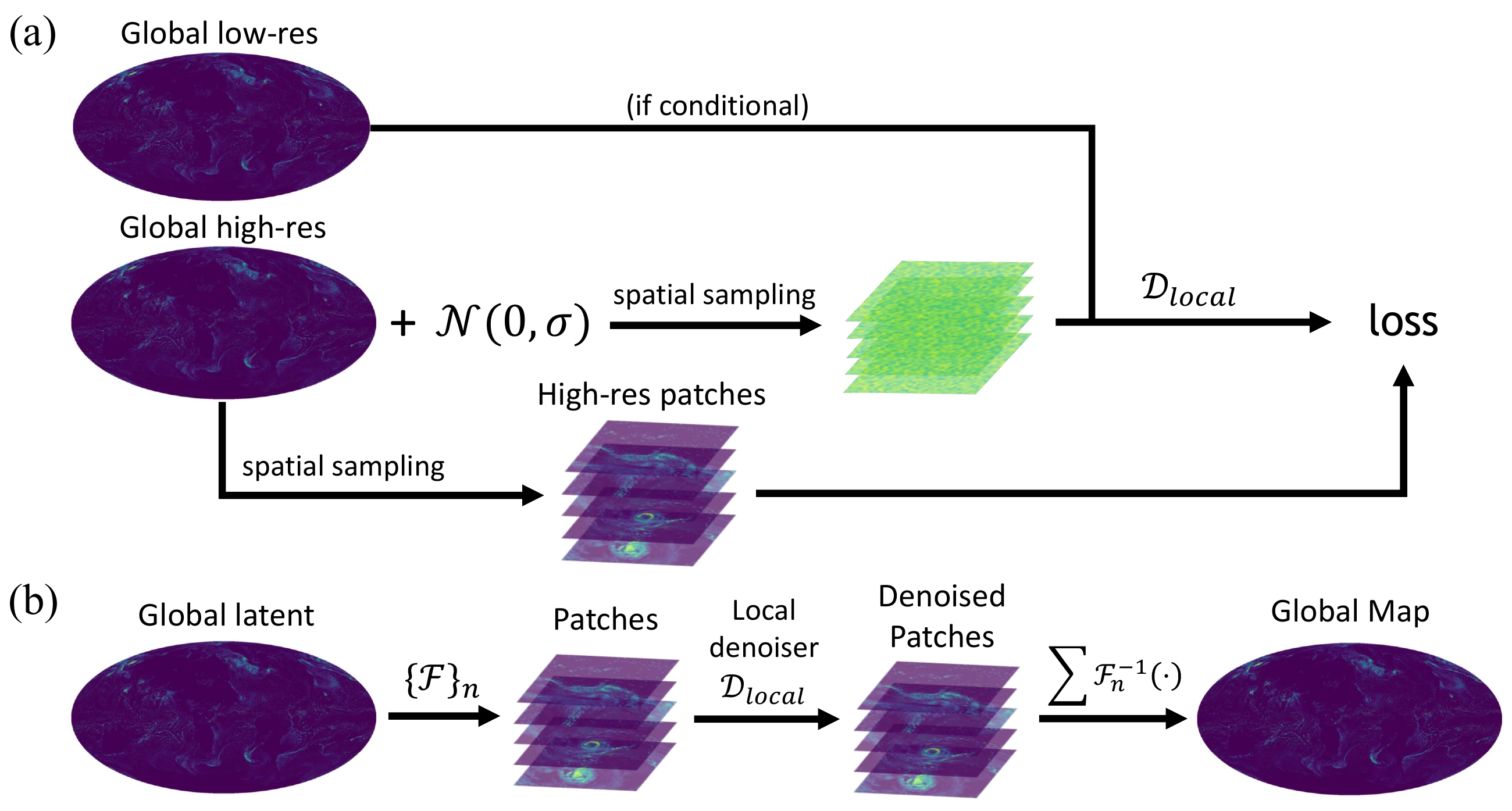}
    \caption{(a) Training workflow of the multi-diffusion framework. Global maps are decomposed into patches for parallel training of local denoisers conditioned on spatially aligned context. (b) Illustration of the aggregated local denoiser replacing the global denoiser in the sampling algorithm.}
    \label{fig:multidiffusion_flowchart}
\end{figure}

This procedure is illustrated in Figure~\ref{fig:multidiffusion_flowchart}(b).

More concretely, given a pretrained local denoiser $\Phi_\theta$ and a set of spatial sampling operators ${\mathcal{F}_i}: \mathcal{X} \rightarrow \mathcal{X}_{\text{local}, i}$, the global denoiser can be represented as an aggregation of local denoisers: 
\begin{align}
\mathcal{D}_\theta(\textbf{x}_t) = 
\frac{
    \sum_i \mathcal{F}_i^{-1}\left( \Phi_\theta \left( \mathcal{F}_i(\textbf{x}_t) \,\middle|\, \mathcal{F}_i(y), \sigma \right) \right)
}{
    \sum_i \mathcal{F}_i^{-1}(J_N)
},
\label{eq:patch_denoising}
\end{align}
where $\mathcal{F}_i^{-1}: \mathcal{X}_{\text{local}, i} \rightarrow \mathcal{X}$ denotes the inverse operation that maps local patches back to the global space, $y\in \mathcal{Y}$ denote the spatial conditions, $\mathcal{X}_{local} = \mathbb{R}^{c_{out}\times N\times N}$ denotes the local target space, $\mathcal{Y} = \mathbb{R}^{c_{in}\times N\times N}$ denotes the local condition space, $N$ is the patch size, and $J \in \mathcal{X}_{\text{local}}$ is an all-one matrix of size $N \times N$ used to account for the number of overlapping patches at each location in the global map.

Multi-diffusion can easily be adapted to the \healpix grid. Define a new spatial sampling operator ${\mathcal{G}_i}: \mathcal{X} \rightarrow \mathcal{X}_{\text{local}, i}$ as a composition of the padding operator $\mathcal{P}(.; N/2)$ and the 2D sampling operator $\mathcal{F}_i$ defined in Section~\ref{sec:multi-diffusion}. Here $N$ represents the patch size used in the multi-diffusion approach. The global denoiser can then be represented as
\begin{align}
\mathcal{D}_\theta(\textbf{x}_t) &= 
\frac{
    \sum_i \mathcal{G}_i^{-1}\left( \Phi_\theta \left( \mathcal{G}_i(\textbf{x}_t) \,\middle|\, \mathcal{G}_i(y), \sigma \right) \right)
}{
    \sum_i \mathcal{G}_i^{-1}(J_N)} \\
\mathcal{G}_i(\textbf{x}_t) &= \mathcal{F}_i(\mathcal{P}(\textbf{x}_t; N/2)),\label{eq:patch_hpx_denoising}
\end{align}
Figure \ref{fig:hpx_multidiffusion_across_face} shows a schematic of this approach.

A fundamental design principle of multi-diffusion is the global sharing of the latent representation $X$ across all local patches. The spatial sampling operators ${\mathcal{F}_i}$ define the patch layout, including the extent of overlap between adjacent regions and the total number of patches required to span the full domain. Introducing overlap is essential for reducing boundary artifacts and promoting consistency across patches. However, increasing the overlap also raises the computational cost, as it leads to a larger number of denoiser evaluations during sampling. As demonstrated in the ablation study in Subsection~\ref{sec:ablations}, strong performance can be maintained even with minimal overlap, enabling a favorable trade-off between quality and efficiency.
 
\subsection{Pseudo-code for masking in the loss function}
\label{sec:si:masking}

\begin{verbatim}
D_yn = net(where(mask, y + n, 0), sigma)
loss = weight * ((D_yn - where(mask, y, 0)) ** 2)
loss = where(mask, loss, 0) / mask.mean()
\end{verbatim}

\subsection{Classifier guidance details\label{sec:si:classifier}}

The classifier outputs a $12\times8\times8$ (HPX8) map of the probability that a given pixel contains a TC. For the architecture, we exploit the HPX8 resolution feature maps learned by the coarse-resolution denoiser (see Sec. \ref{sec:architectures}), and obtain the probability map by taking the lowest-resolution state of the UNet and feeding it through two convolutional layers (with 50\% dropout in between). During training, we minimize binary cross-entropy between the TC labels from IBTracs and the outputs of the classifier.

In the inference phase, guidance is implemented by adding a term to the ODE in eq.~\ref{eq:ode}: 
\begin{align}
\frac{d\mathbf{x}}{d\sigma} =\frac{\mathbf{x}- \mathcal{D}_{\theta}(\mathbf{x}, \sigma)}{\sigma} + \gamma \nabla_{\mathbf{x}} \mathcal{L}_{BCE}(\mathcal{C}(\mathbf{x}), y).\label{eq:cguidance}
\end{align}
Here, $\gamma$ represents a guidance scale hyperparameter, and $\mathcal{L}_{BCE}(\mathcal{C}(\mathbf{x}), \mathbf{y})$ the binary cross-entropy loss of the classifier output $\mathcal{C}(\mathbf{x})$ and the guidance input $\mathbf{y}$. In practice, we found that a constant guidance scale did not produce the desired effect, and so used an adaptive guidance scale $\gamma = \hat{\gamma} \rho$, where $\hat{\gamma}$ is constant and 
\begin{align}
\rho = \frac{||\frac{\mathbf{x}- \mathcal{D}_{\theta}(\mathbf{x}, \sigma)}{\sigma}||^2}{||\nabla_{\mathbf{x}} \mathcal{L}_{BCE}(\mathcal{C}(\mathbf{x}), y)||^2}
\end{align}
with the Frobenius norm $||\cdot ||^2$. This ensures that at each step, the fraction of the norm of the update in the score direction and of that in the classifier direction is constant. Without this normalization, we found the classifier gradient to be small compared to the score in particular at early denoising times.

\subsection{Architectures and hyperparameter settings}
\label{sec:architectures}

\begin{figure}
    \centering
    \includegraphics{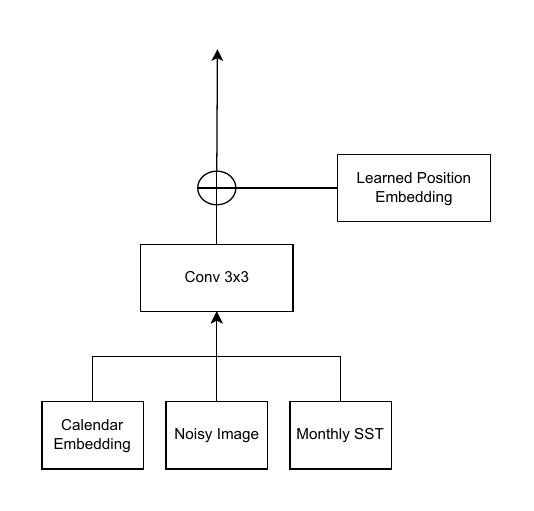}
    \caption{Position and calendar embedding.}
    \label{fig:embedding}
\end{figure}

\begin{table}[h]
\centering
\begin{tabular}{lp{1in}p{1in}p{1in}}
\toprule
                             & \textbf{cBottle-3d}       & \textbf{cBottle-video}         & \textbf{cBottle-SR}                \\
\midrule
Number of parameters         & 149M                      & 282M                          & 330M                                \\
Target channels               & 3D                       & 3D                             & 2D                                \\
Architecture                 & Song UNet                 & Song UNet                      & Song UNet                                \\
Learned Position Encoding    & \cmark                    & \cmark                         & \cmark \\
Calendar embedding           & \cmark                    & \cmark                         &  \xmark \\
Trained on                   & ICON+ERA5                 & ICON+ERA5                      & ICON                                     \\
Grid                         & HPX64 (49152)             & HPX64 (49152)                  & 128x128 crops of HPX1024                 \\
Training domain              & global                    & global                         & crops                                    \\
Total examples trained       & 9,856,000                 & 2,160,000                      & 95,520 samples / 191M patches            \\
Batch size                   & 64 (3M images) 256 (3-10M) & 32                            & 960                                       \\
Optimizer                    & Adam                      & Adam                           & SGD                                      \\
Learning rate                & 0.0001                    & 0.0001                         & 0.0001                                   \\
Noise distribution ($p_{\sigma}$) & log-uniform 0.02–200      & log-uniform 0.02–1000             & log-normal \newline $P_{\text{mean}} = -1.2$ $P_{\text{std}} = 1.2$ \\
\\
\multicolumn{4}{l}{\bf Computational Profile} \\
\midrule
Hardware trained on          & 16xH100                   & 32xH100                           & 64xH100                                     \\
Examples/GPU-second (train)  & 6.7                       & 0.32                   & 35.4 patches                         \\
\bottomrule
\end{tabular}
\caption{Comparison of \textit{cBottle} model variants.\label{tab:architectures}. A training example is defined a single HPX64 image for the coarse models but is a single 128x128 crop for cBottle-superres.}
\end{table}

We mostly reuse the standard UNet architecture used in diffusion modeling (the so-called Song-UNet in \citet{Karras2022-edm,Mardani2023-cd} and others). However, it was necessary to add position and temporal embedding (Figure \ref{fig:embedding} to allow the diffusion model to learn non-stationary signals in space and time, and temporal attention layers to support video diffusion. Table \ref{tab:architectures} summarizes our three model variants.

\paragraph{Calendar Embedding} Producing an accurate seasonal and diurnal cycle required a calendar and hour embedding strategy. We take care to use a embedding that respects the periodicity of these inputs. This is given by
\begin{align}
f_{4k} &= \cos \left(2 \pi k \frac{\text{doy}}{365.25}\right)\\
f_{4k+1} &= \sin \left(2 \pi k \frac{\text{doy}}{365.25}\right)\\
f_{4k+2} &= \cos \left(k t\right)\\
f_{4k+3} &= \sin \left(k t\right).
\end{align}
Here, the day of year (doy) is a scalar representing the time since the beginning of the year, and the time of day ($t$) is the local solar time in units radians $t = 2\pi( \frac{\text{second of day}}{86400} + \frac{\text{lon}}{360})$\footnote{Due to an implementation error we actually used $second - lon$.}. In the architectures below we use $k=1,\ldots,8$, and concatenate these inputs along with other image-like conditioning (e.g., monthly-averaged SST) and noisy target image. 

\paragraph{Learned position embedding} To allow the diffusion model to learn non-stationary spatial structures like climatology, we add a learned position embedding. After the initial convolution embedding of the image-like inputs (calendar, noisy image and monthly-averaged SST), a learned positional embedding for each pixel $49152 \times c$ is added.  The position embedding is initialized with independent standard normal random numbers.

\subsubsection{Model Settings}\label{sec:model-settings}
\paragraph{cBottle-3D}

The coarse image model generates HPX64 fields.
The coarse resolution image generator uses the standard settings for the Song-UNet presented by \citet{Karras2022-edm}, but we found the 3D training was made more stable by increasing base model embedding dimension from 128 to 192.
The source dataset is identified with a label (ERA5 or ICON) which is randomly dropped for 1/4 samples to encourage the model to learn features across both datasets. This label is provided to the UNet in the same manner as the imagenet class label in \citet{Karras2022-edm}.
Self-attention is only used on the coarsest resolution stage (8x8). All convolutional layers are implemented on the HPX grid using the padding approach described in Section \ref{sec:hpxpad}. 32 channels are devoted to calendar embedding.
The model was trained from scratch on a 50-50 mix of ICON and ERA5 data on 32 H100 GPUs. Even (odd) ranks were trained with ERA5 (ICON).
The model was trained to 3 million samples (with repeats) with a batch size of 64 ($\sim 50000$ steps) when the log probability loss started to plateau, and then the batch size was increased to 256 until 10 million samples. The Adam optimizer \citep{Adam} was used with a learning rate of $10^{-4}$.
During training the model uses the log-uniform noise distribution ($p_{\sigma}$) with $\sigma_{min}=0.02$ and $\sigma_{min}=200$. For inference, we use the power law sampler proposed by \citet{Karras2022-edm} with these same settings.



\paragraph{cBottle-SR}
\label{sec:superresmodel}

The super-resolution model, which performs upscaling from HPX64 to HPX1024 resolution (a factor of 16), is trained within a HEALPix multi-diffusion framework described in \ref{sec:multi-diffusion}. In this setting, the global input data—including latent variables, conditioning fields, and high-resolution targets—are partitioned into spatial patches. Each patch is then independently processed by a local denoising model, which is trained to reconstruct high-resolution content based on temporally and spatially aligned inputs. This approach enables efficient learning of fine-scale features while preserving global coherence through patch-wise training across the full spherical domain.

As illustrated in Figure \ref{fig:multidiffusion_flowchart}(a), this training strategy adapts the standard denoising score matching loss (see Eq. \ref{eq:loss}) to operate at the patch level. Specifically, the modified objective is defined as
\begin{align}
\min_\theta \mathbb{E}_{\mathbf{x} \sim p_{\text{data}}} \mathbb{E}_{\sigma \sim p_\sigma} \mathbb{E}_{\mathbf{\epsilon} \sim \mathcal{N}(0, \sigma^2 \mathbf{I})} \mathbb{E}_{i\sim \mathcal{U}}\Big|\Big|\Phi_\theta\Big(\mathcal{F}_i(\mathbf{x})+\mathbf{\epsilon};\mathcal{F}_i(y),\sigma\Big) - \mathcal{F}_i(\mathbf{x})\Big|\Big|^2,
\end{align}

Like \texttt{cBottle-3d} we use the Song-UNet with 128 hidden channels.
Since it is a local model, we do not need to use the HEALPix padding shown in \ref{fig:hpx_global_UNet}.

A patch size of 128$\times$128 is used. This is chosen to be much larger than 16x super-resolution factor but is still affordable.
To cope with non-stationary spatial features like topography, 20 learnable positional embeddings with a spatial size of HPX1024 grid are introduced and initialized using sinusoidal functions.
The inputs of the network consist of (1) the HPX64 data re-gridded bilinearly to the HPX1024 128$\times$128 patch in question, (2) the global position embeddings for this patch, and (4) the global HPX64 data interpolated to the 128×128 lat-lon grid.
These are concatenated with the latent before further processing by the encoder.
The model was trained to 95520 ICON samples (corresponding to 191 million patches, with repeats) with a batch size of 960.
The SGD optimizer was used with a learning rate of $10^{-4}$. During training the model uses the log-normal noise distribution of \citet{Karras2022-edm} ($p_{\sigma}$) with $P_{mean}=-1.2$ and $P_{std}=1.2$.
For inference, the super-resolution model uses a maximum noise level ($\sigma_{max}$) of 800, 18 sampling steps, and $S_{churn}=0$.

\subsection{Video Generation extension}
\label{sec:video}

Let $x \in \mathbb{R}^{T \times C \times X}$ denote an atmospheric trajectory of $T$ evenly spaced frames with $X$ pixels each. While the coarse image \textit{cBottle} produces realistic atmospheric snapshots, $T$ successive samples in time produce a sequence that lacks temporal coherence due to the independence of each sample.

\subsubsection{Architecture Modifications}

\paragraph{Temporal Attention}
We extend \textit{cBottle} to video generation by augmenting the 2D UNet architecture with temporal self-attention layers, following factorized space-time approaches \citep{ho2022-vd, blattmann2023-ldm}. The architecture preserves the UNet backbone of \textit{cBottle}, but introduces temporal layers at the end of spatial blocks. Within each UNet block, we first apply spatial modules with the time dimension folded into the batch dimension. For blocks with temporal attention, we fold the spatial dimension into the batch dimension, and then apply bidirectional self-attention across frames per pixel, using the channels as the embedding dimension. A learned relative-time embedding $R_{\Delta t} \in \mathbb{R}^{2T-1}$ capturing frame-to-frame dependencies based on temporal distance $\Delta t$ is added to the pre-softmax logits. The temporal attention output is then residually added to the main branch.

\paragraph{Masked-Conditioning}
To enable flexible inference capabilities, we adopt a masked-conditioning training scheme inspired by MCVD \citep{voleti2022-mcvd}. During training, we randomly sample mask patterns $m \in \{0,1\}^T$ (where 1 indicates observed frames) from a categorical distribution $p_{mask}$ that blends several tasks: Bernoulli random frame dropout, masking all but the first or last few frames contiguously, masking middle frames while keeping endpoints visible, and full sequence masking. On top of the calendar embedding and monthly-mean SST conditioning that the image-model uses, we add this masked sequence (ground truth at unmasked positions and zeros at masked positions) and the binary mask indicator as conditioning concatenated channel-wise with the input. The video diffusion training objective is identical to the image diffusion objective, but with this additional per-sample masked-conditioning and the loss averaged across the $T$ frames.


This unified training approach enables the video model to perform multiple tasks at inference time: generate entire sequences unconditionally, predict future frames given initial states, or infill arbitrary temporal gaps.

\subsubsection{Model settings and training}
The video model generates 3-day sequences with 6-hour resolution ($T=12$ frames) at the same HPX64 resolution of the image-based \textit{cBottle}.
Because $T$ correlated frames carry more redundant signal, we keep the same log-uniform noise schedule (with $\sigma_{min}=0.02$) and increase $\sigma_{max}$ from 200 to 1000 in both training and inference to preempt \emph{signal leak bias} (see Subsection \ref{sec:si:tuning-noise}). We insert temporal‐attention layers into every encoder block and into two positions within each decoder resolution stage: the up‑sampling block and the final block. We weight each of the four masking tasks defined above as follows: random frame dropout (0.3), block-wise masking (0.25), endpoint interpolation (0.1), and full sequence masking (0.35).

Similar to the coarse image model, a base model embedding dimension of 128 sufficed to model the 2D dataset's 12 fields. Given the video model's expanded 124-channel input on the 3D dataset, we increased the UNet width to 256 channels, finding it to converge markedly faster and achieve higher log-likelihood than at 192 channels. The model was trained from scratch like the image model on 32 H100 GPUs, using the same Adam optimizer \citep{Adam} and learning rate of $10^{-4}$. It was trained to 2.16M video samples with a batch size of 32.

\subsubsection{Results}

\begin{figure}
    \centering
    \includegraphics[width=0.98\textwidth]{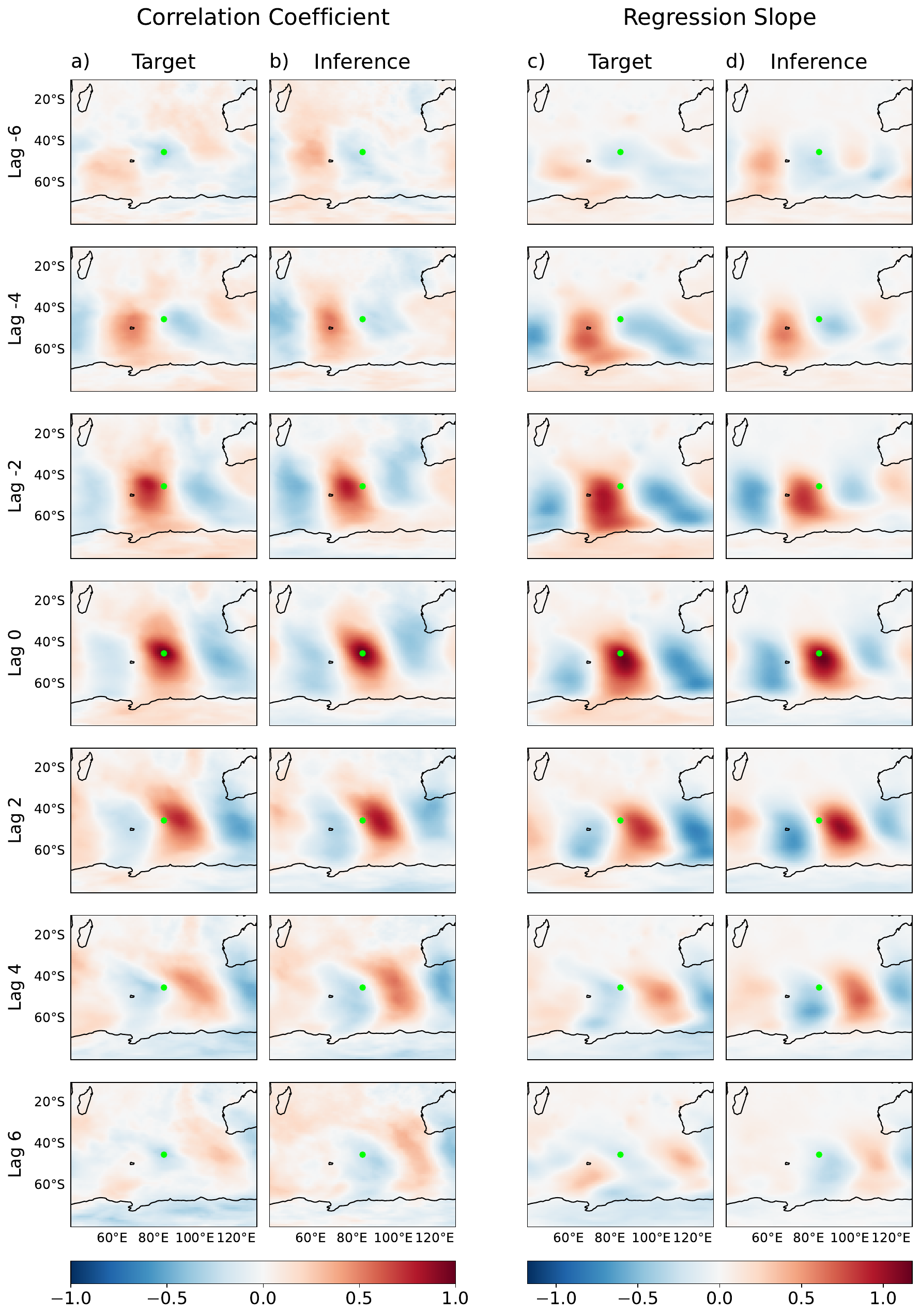}
    \caption{Lagged correlation (left) and regression slope (right) maps for mean sea level pressure computed on unconditional video model generations and ICON ground truth. Maps are computed relative to a seed point at 45$^{\circ}$S, 85$^{\circ}$E (green dot), and averaged over 60 non-overlapping 3-day sequences from December 2024 to May 2025.}
    \label{fig:video_spatio_temporal}
\end{figure}

To evaluate whether the video diffusion model captures realistic atmospheric dynamics through time---beyond just generating independent plausible snapshots---we analyze its ability to reproduce the patterns characteristic of mid-latitude storm track evolution in Figure \ref{fig:video_spatio_temporal}. Following the approach of \citep{Rao2002}, we compute one-point lag correlation maps on mean sea level pressure.
The video model generations successfully reproduce both the spatial structure of these patterns and their temporal evolution at comparable propagation speeds to the ICON ground truth. At negative lags, we observe a coherent wave train approaching the reference point from the west, and at positive lags, these correlation continue propagating eastward. The model's ability to replicate these spatio-temporal patterns indicates the temporal attention mechanism enables the model to capture the underlying physical dynamics and generate coherent trajectories of atmospheric state. 

\begin{table}[ht]
\centering
\begin{tabular}{p{1.5in}llcc}
\textbf{Targets} & \textbf{Shape} & \textbf{Datasets used for} & \textbf{2D} & \textbf{3D} \\
\hline
\\
\multicolumn{5}{l}{\bf Single level fields}\\
liquid water path & X & icon+era & \cmark & \cmark \\
ice water path & X & icon+era & \cmark & \cmark \\
2m air temperature & X & icon+era & \cmark & \cmark \\
surface zonal wind & X & icon+era & \cmark & \cmark \\
surface meridional wind & X & icon+era & \cmark & \cmark \\
Outgoing long wave radiation & X & icon & \cmark & \cmark \\
Outgoing shortwave radiation & X & icon & \cmark & \cmark \\
mean sea level pressure & X & icon+era & \cmark & \cmark \\
precipitation (total) & X & icon+era & \cmark & \cmark \\
downwelling SW at surface & X & icon & \cmark & \cmark \\
sea surface temperature (snapshot) & X & icon+era & \cmark & \cmark \\
sea ice concentration & X & icon+era & \cmark & \cmark \\
\\
\multicolumn{5}{l}{\bf Pressure level variables @  [1000, 850, 700, 500, 300, 200, 50, 10] mb}\\
$U $ & P, X & icon+era & \xmark & \cmark \\
$V $ & P, X & icon+era & \xmark & \cmark \\
$T $ & P, X & icon+era & \xmark & \cmark \\
$Z $ & P, X & icon+era & \xmark & \cmark \\
\\
\multicolumn{5}{l}{\textbf{Inputs}} \\
\hline
Monthly mean SST & X & \begin{tabular}[c]{@{}l@{}}era5 = AMIP Monthly SST\\ icon = rolling mean\end{tabular} & \cmark & \cmark \\
Day of year & & & \cmark & \cmark \\
Hour of day & & & \cmark & \cmark \\
Label & & & \cmark & \cmark \\
\hline
\end{tabular}
\caption{Input and target variables with dataset sources and dimensionality. X = space, P = levels. \label{tab:fields}}

\end{table}

\section{Supplemental information}

\renewcommand{\thefigure}{SI\arabic{figure}}
\setcounter{figure}{0}
\renewcommand{\thetable}{SI\arabic{table}}
\setcounter{table}{0}

\begin{figure}
    \centering
    \newlength{\w}
\setlength{\w}{3in}
\begin{tabular}{ll}
Mean Sea-Level Pressure Ground Truth & Generated \\
\includegraphics[width=\w]{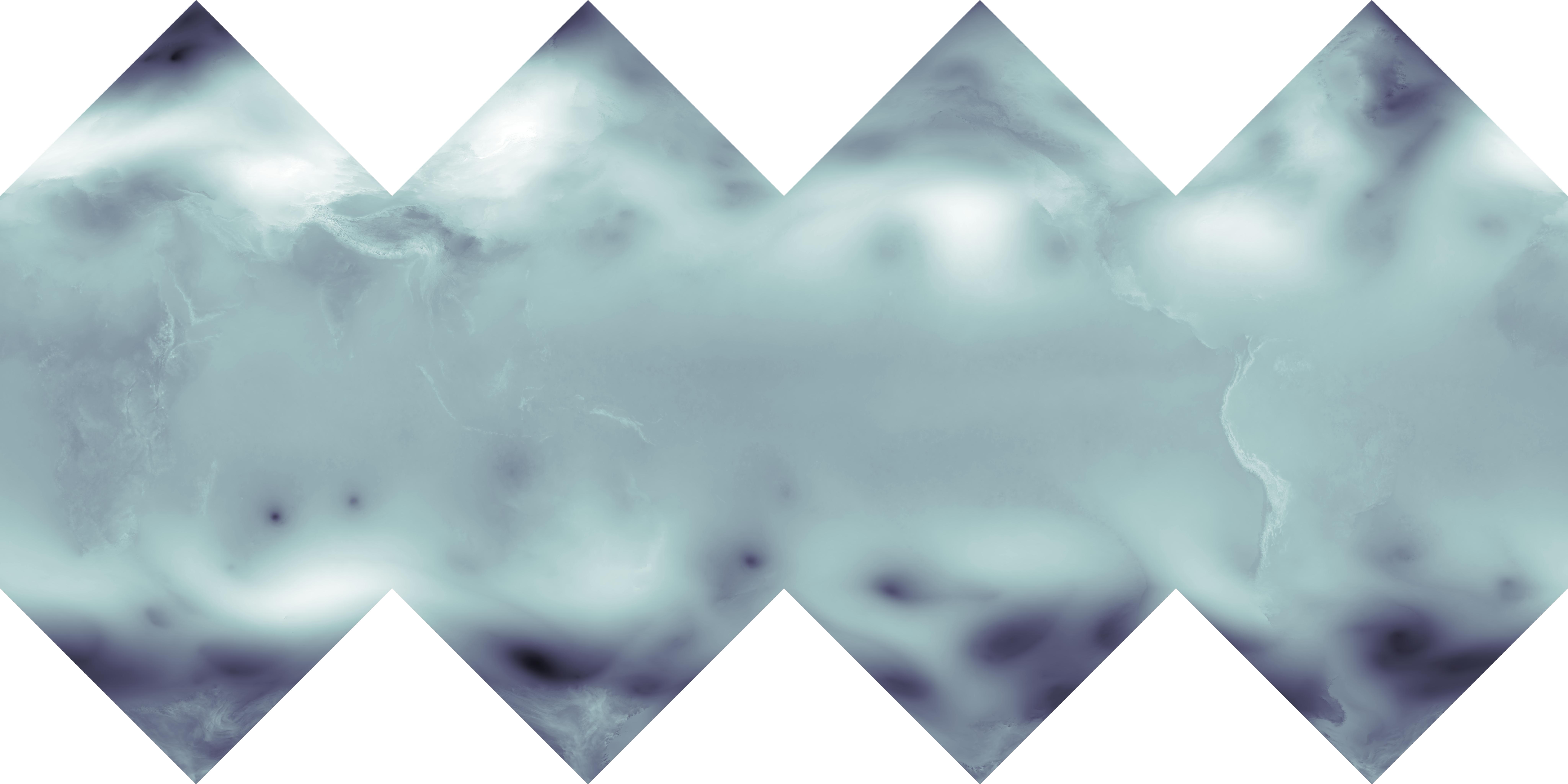} & \includegraphics[width=\w]{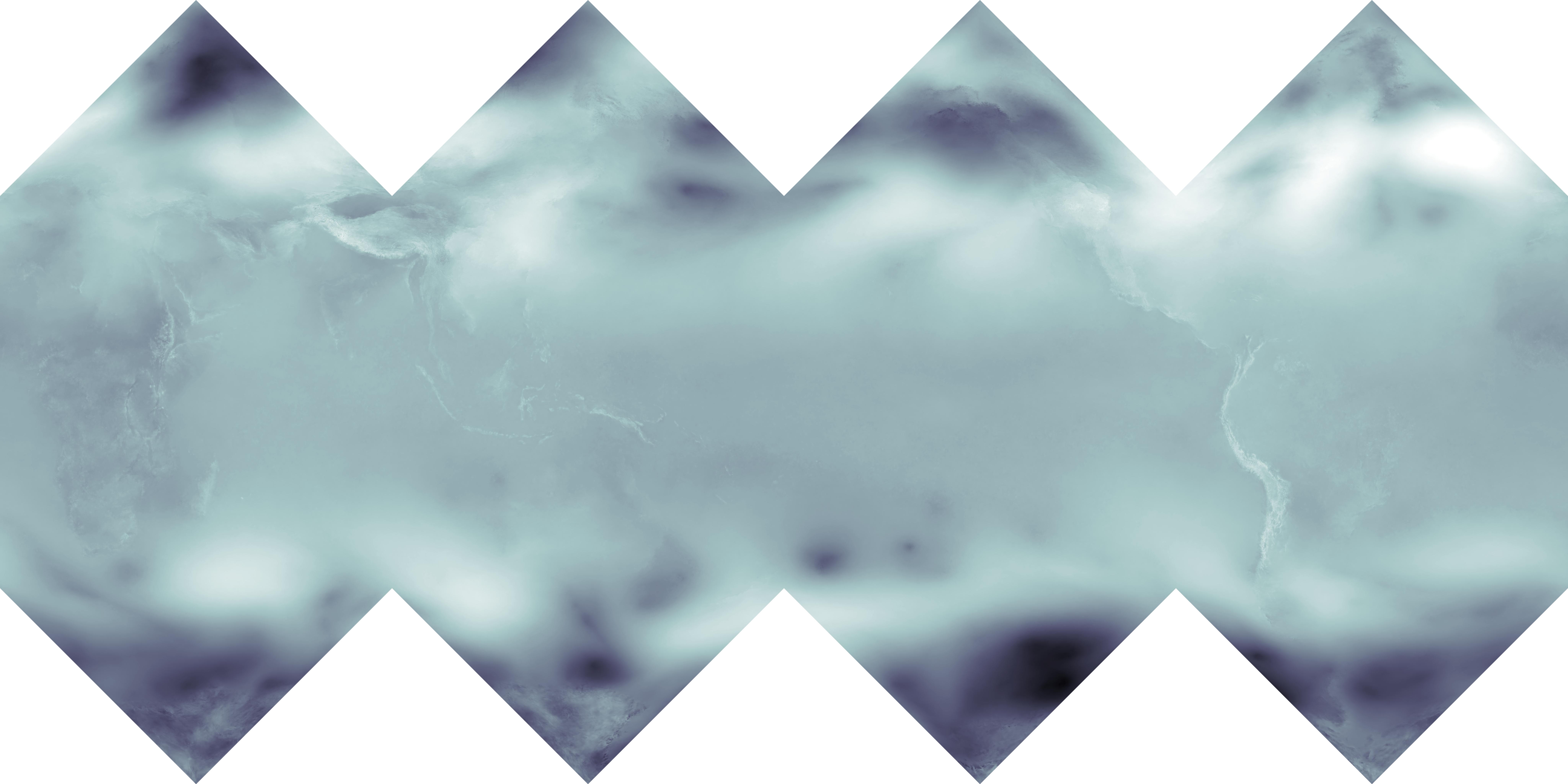} \\
Surface Wind Speed Ground Truth &  Generated \\
\includegraphics[width=\w]{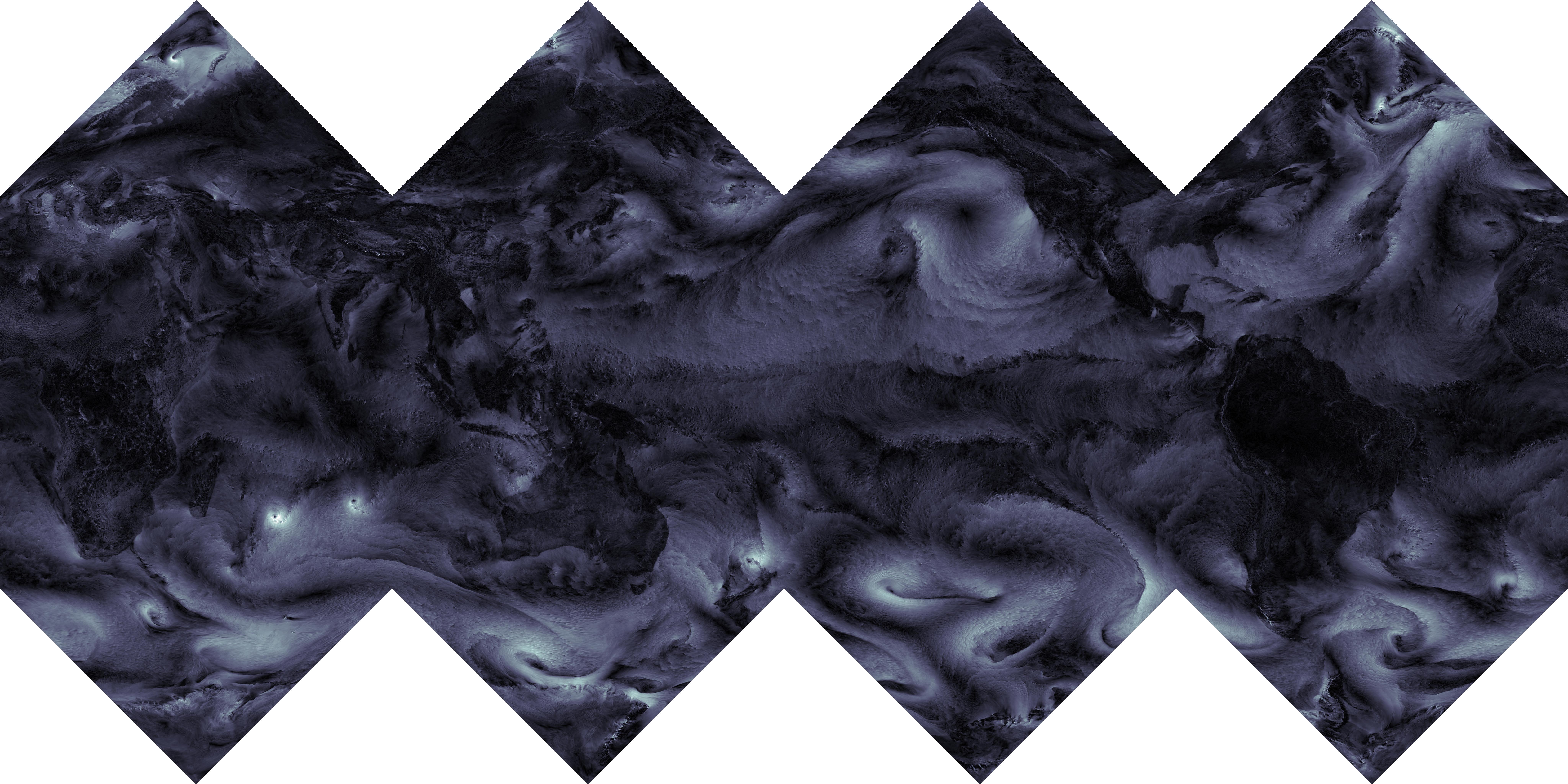} & \includegraphics[width=\w]{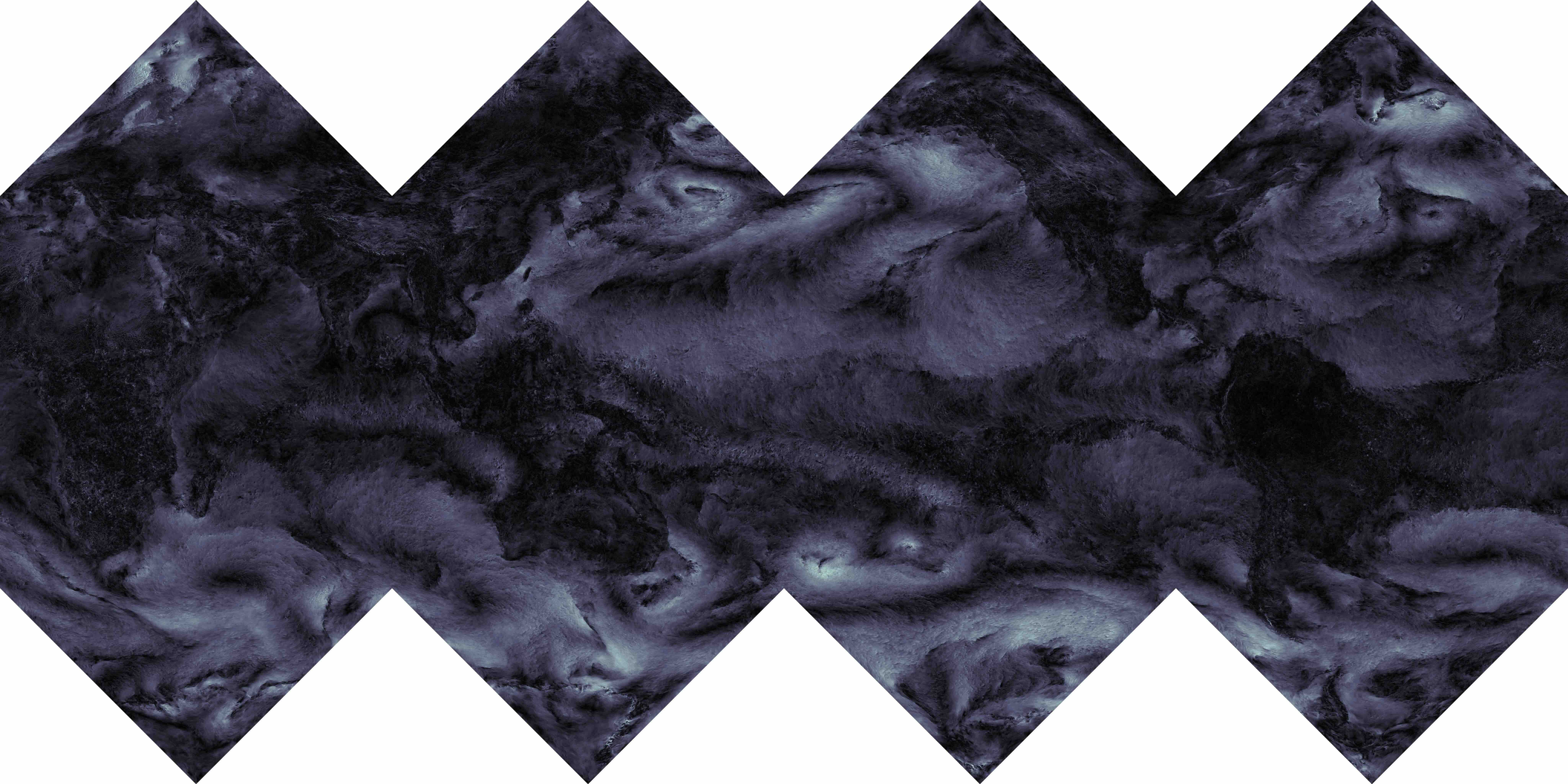} \\
Visible Light Ground Truth &  Generated \\
\includegraphics[width=\w]{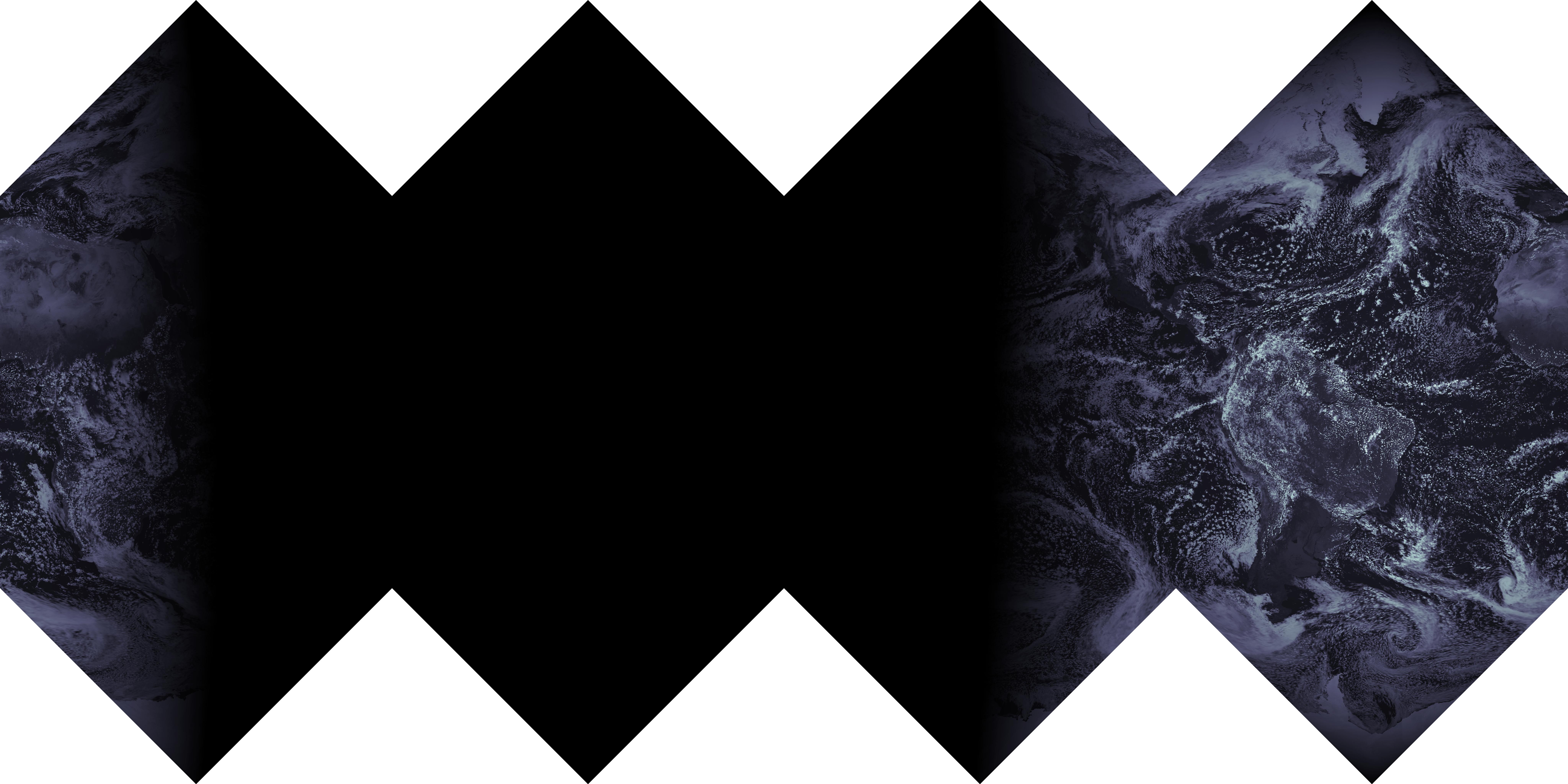} & \includegraphics[width=\w]{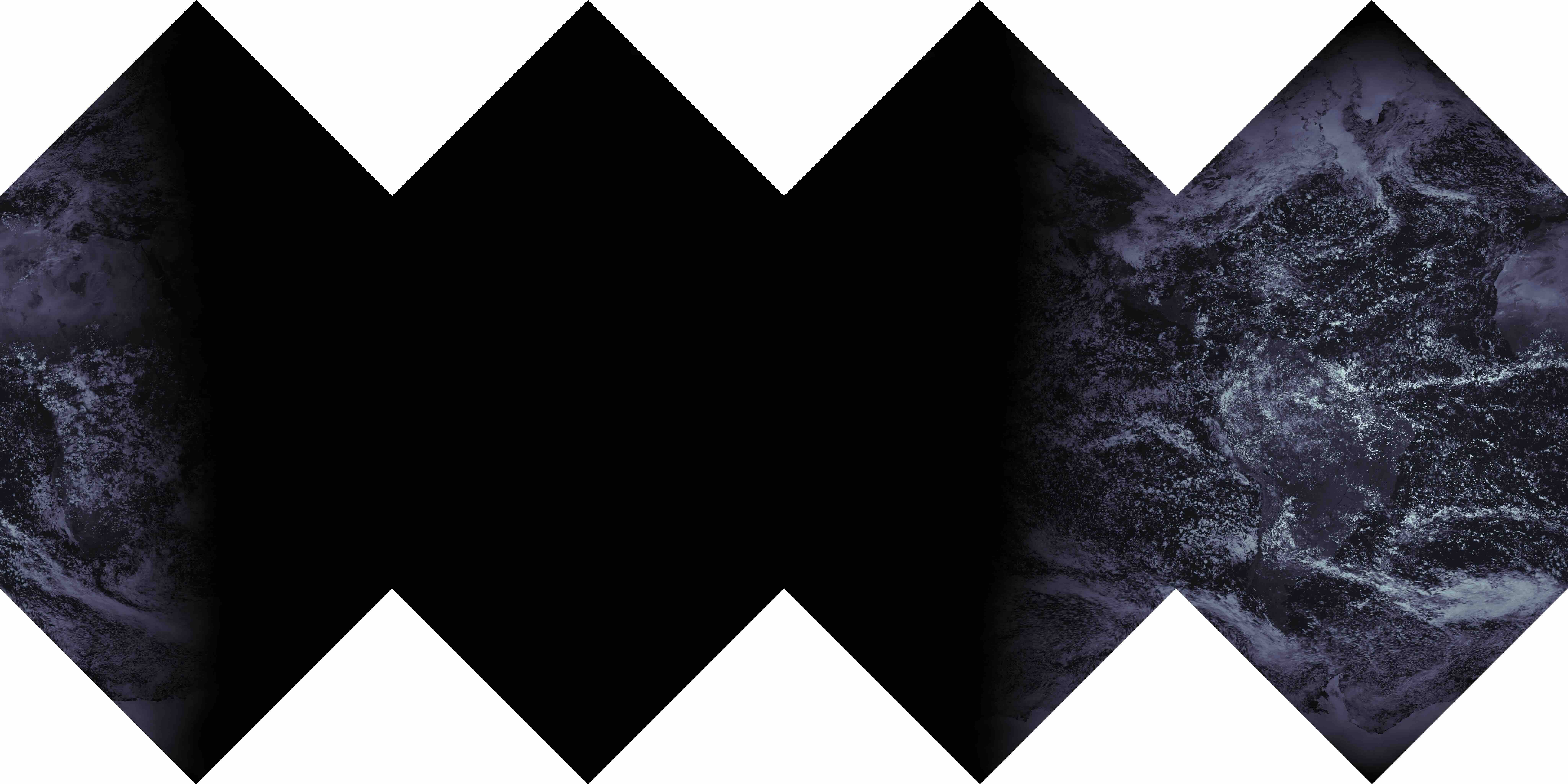} \\
Surface Precipitation Ground Truth & Generated \\
\includegraphics[width=\w]{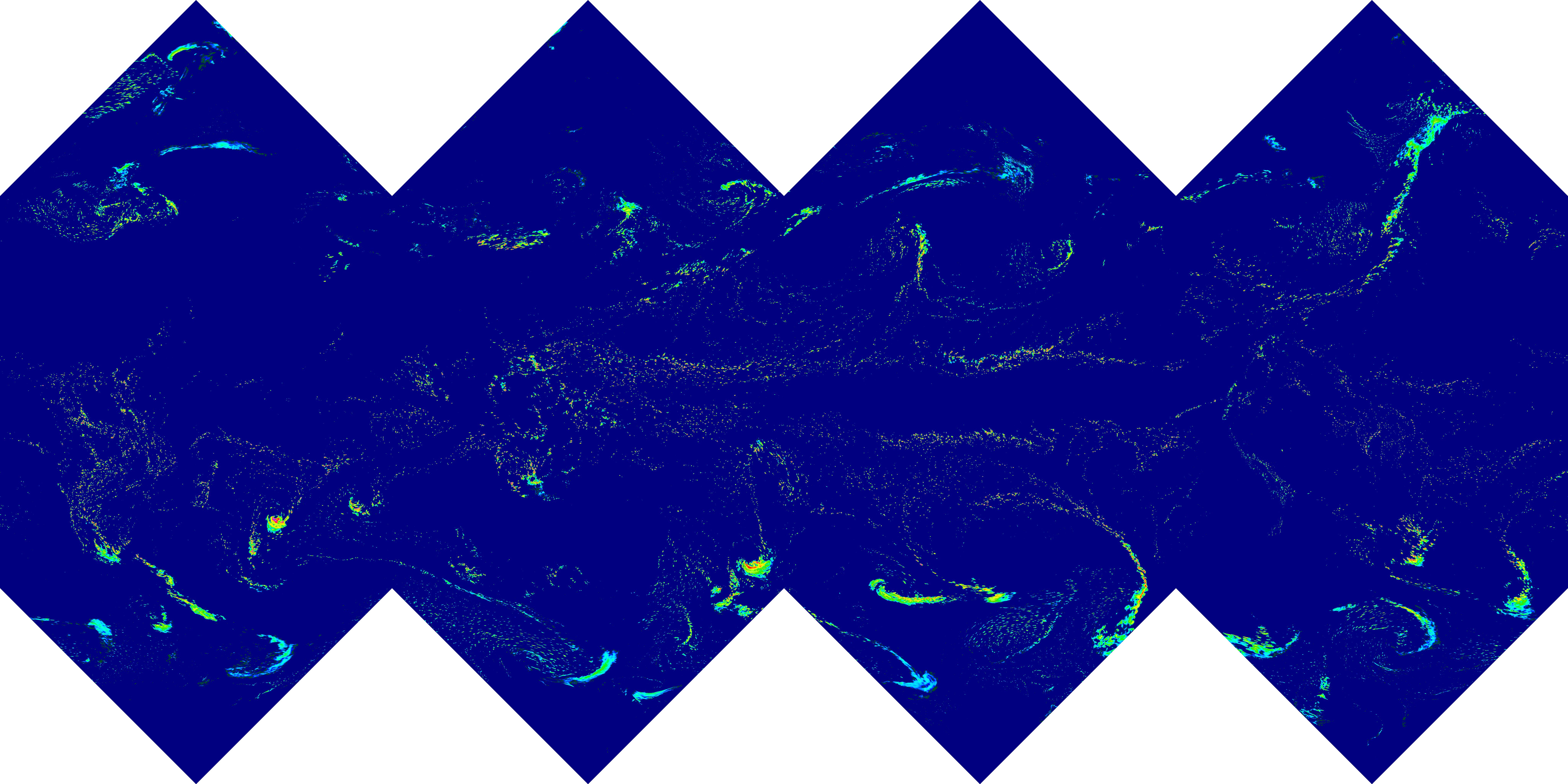} & \includegraphics[width=\w]{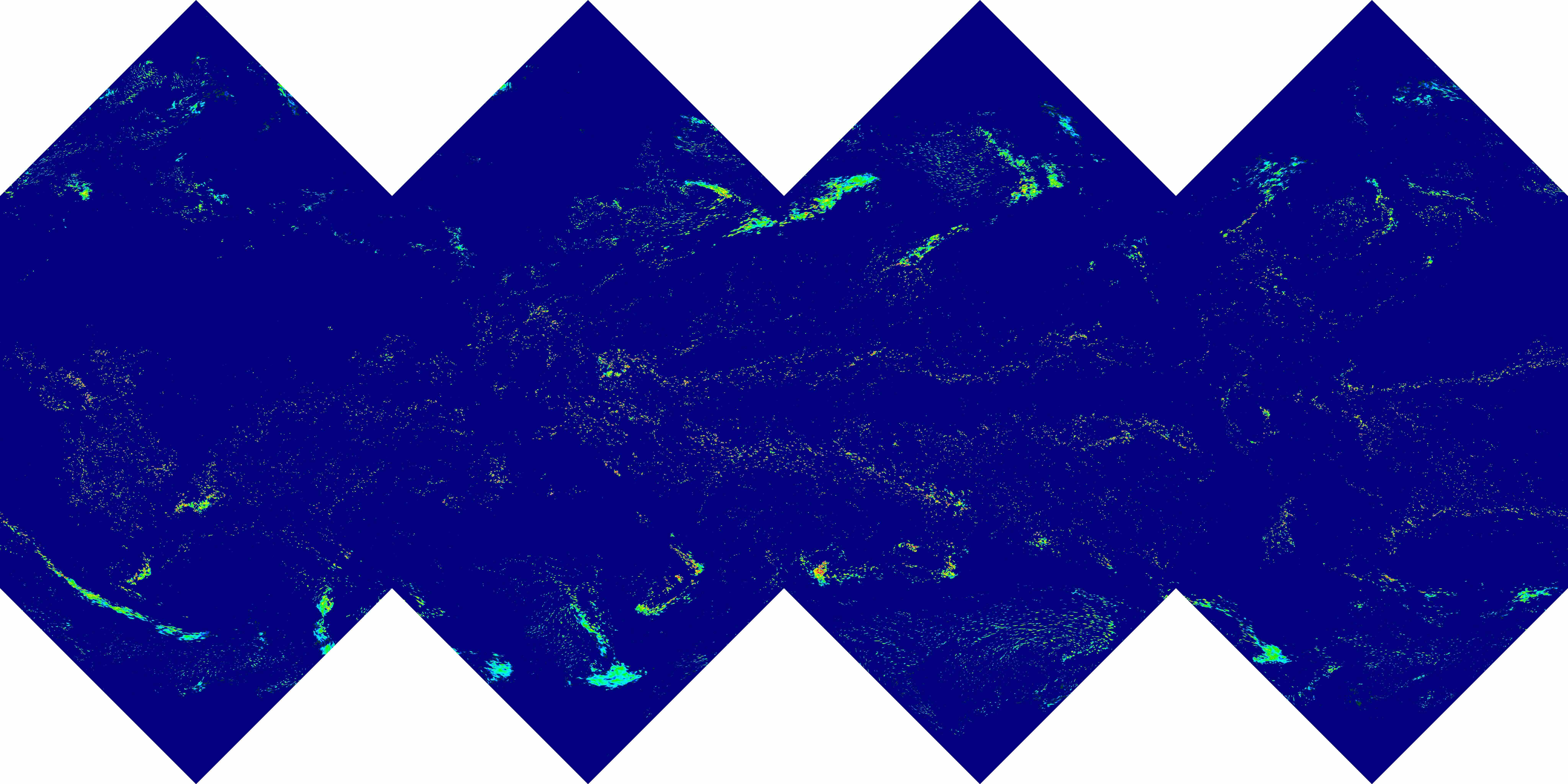} \\
\end{tabular}
    \caption{More fields for Figure \ref{fig:1}}.
    \label{fig:si:1}
\end{figure}

\begin{figure}
    \centering
    \includegraphics[width=1\textwidth]{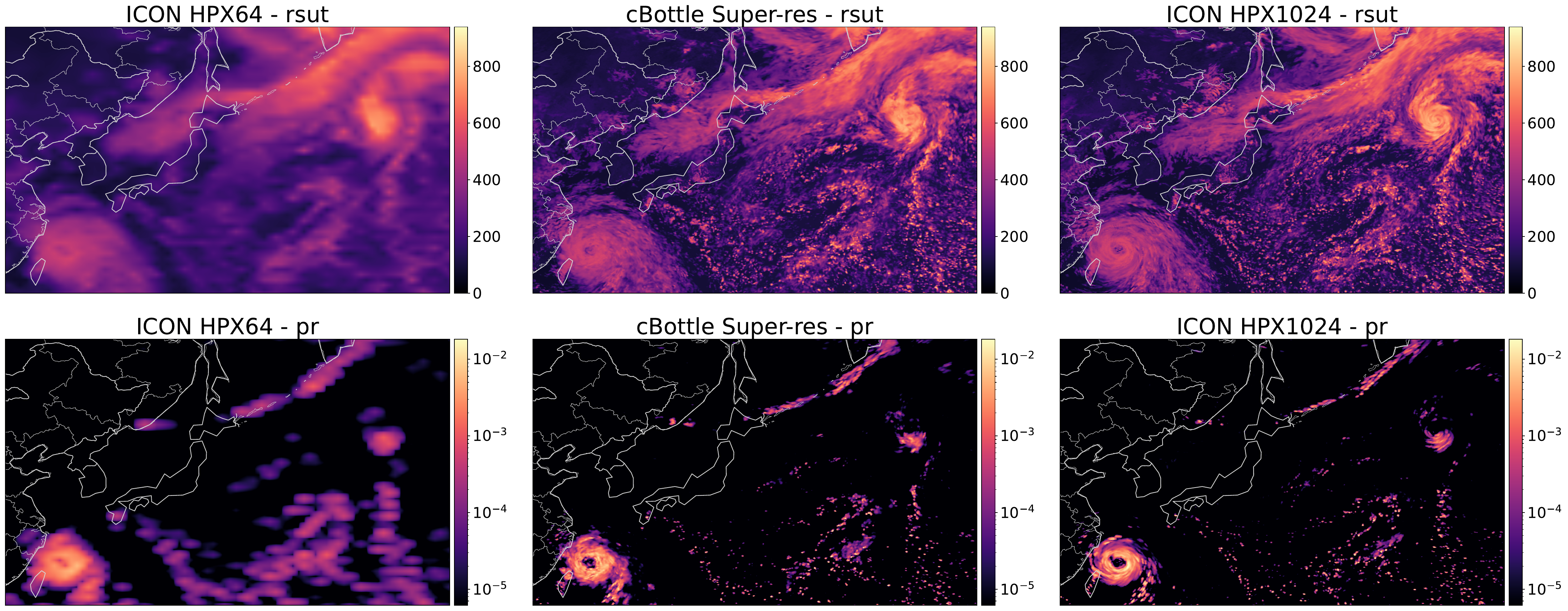}
    \caption{Demonstration of super-resolved atmospheric variables outgoing shortwave radiation, and precipitation flux of ICON HPX64 at a selected time during an extreme tropical cyclone event and adjacent midlatitude front. From left to right are the input ICON HPX64, HPX1024 super-resolved by cBottle, and ICON HPX1024. }
    \label{fig:superres}
\end{figure}

\begin{figure}
    \centering
    \includegraphics[width=\linewidth]{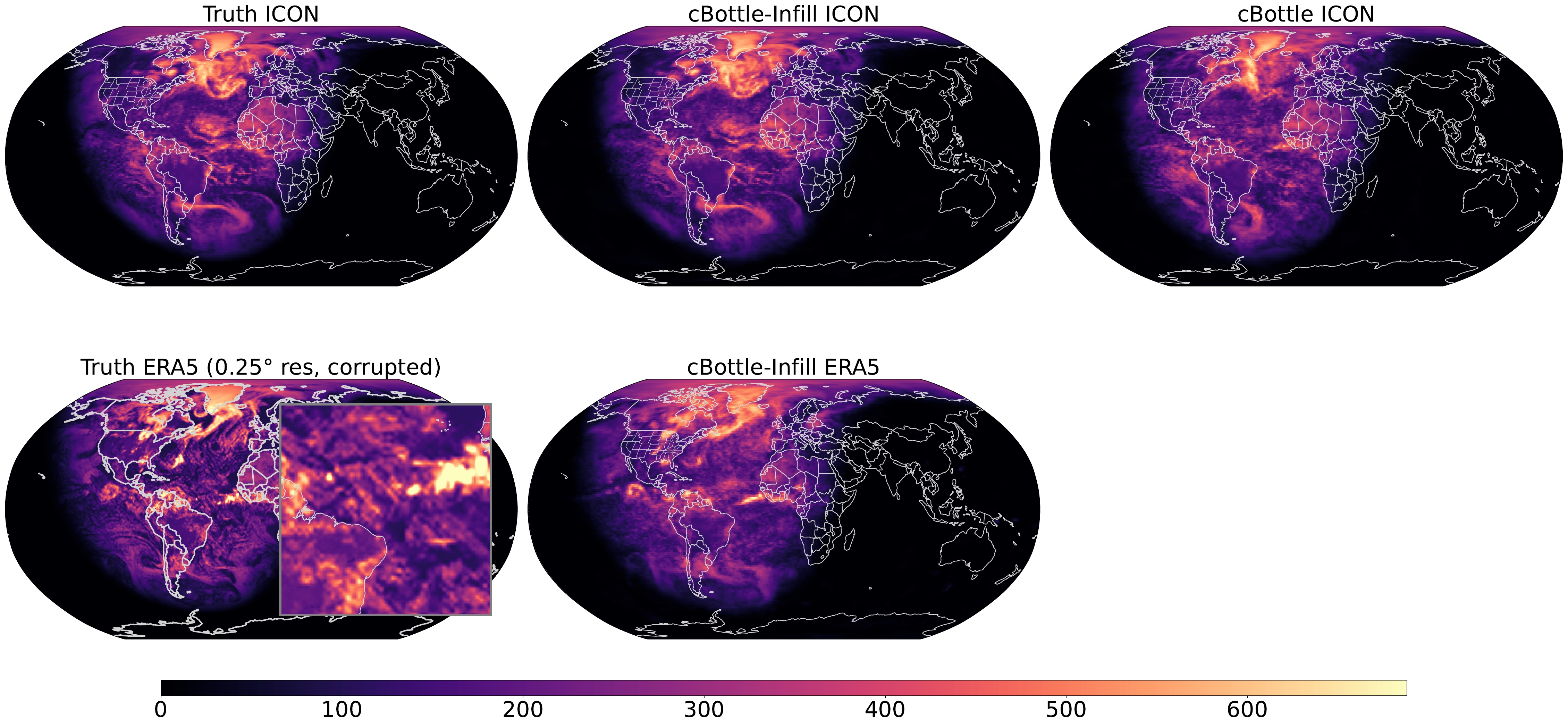} \\
    \caption{Imputation of outgoing shortwave radiation using \textit{cBottle}. \texttt{cBottle-Infill ICON}: ICON data with radiation channels masked and subsequently reconstructed via \textit{cBottle}. \texttt{cBottle ICON}: ICON data fully generated by \textit{cBottle} using only SST and timestamp as conditioning. \texttt{Target ERA5}: ERA5 reference with native spatial resolution and known streaking artifacts. \texttt{cBottle-Infill ERA5}: ERA5 data with radiation channels infilled by \textit{cBottle}. \label{fig:si:infill}}
\end{figure}

\subsection{Diurnal Cycle: Details}
\label{sec:methods:diurnal}
The diurnal cycle of rainfall should only be detectable above the noise of other modes of variability in certain subregions of the planet.

The diurnal amplitude is defined as

\[
    S := \frac{1}{366} \sum_d \max_{h} \overline{T}(d,h)  - \min_{h} \overline{T}(d, h).
\]
$\overline{T}(d, h)$ is the average of $T$ as a function of day of year $d$ and hour of day $h$. For the SNR computation, the noise is given by
\[
    N := \frac{1}{366} \sum_d \frac{1}{24} \sum_h s_T^2(d, h),
\]
where $s_T^2$ is the conditional sample variance. In this computation, $s_T^2$ and $\overline{T}$ are computed over 1990--2018 (inclusive). Then, the $SNR:=S/N$.

\begin{figure}
    \centering
    \begin{tabular}{cc}
        a. ERA5 &  b. cBottle \\
        \includegraphics[width=0.4\linewidth]{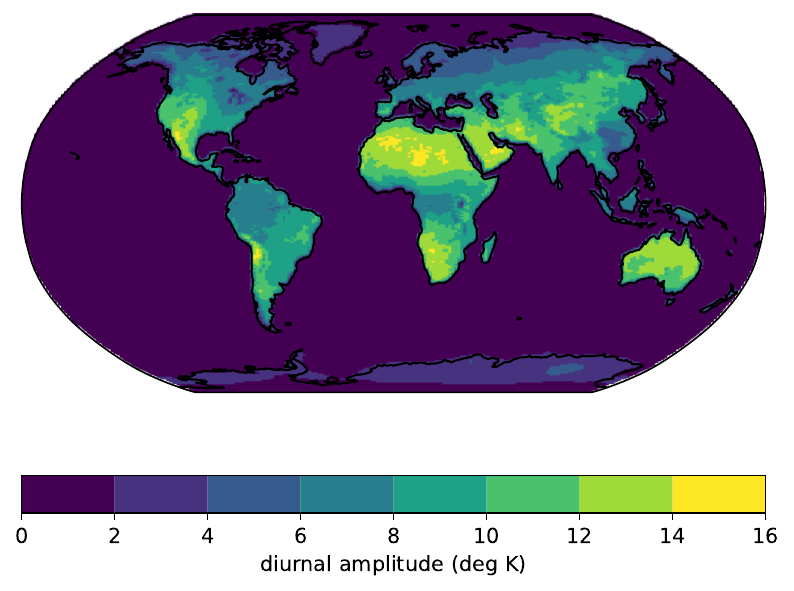} & 
        \includegraphics[width=0.4\linewidth]{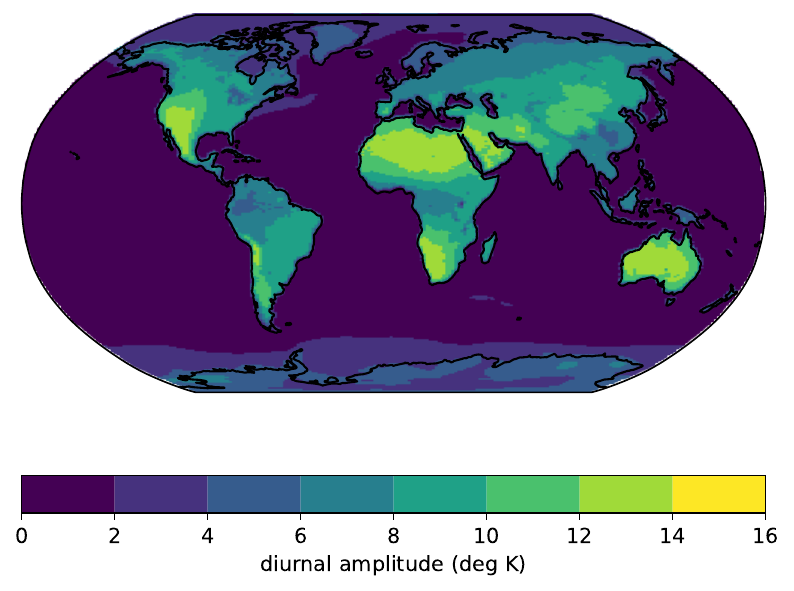} \\
        c. ERA5 &  d. cBottle \\
         \includegraphics[width=0.4\linewidth]{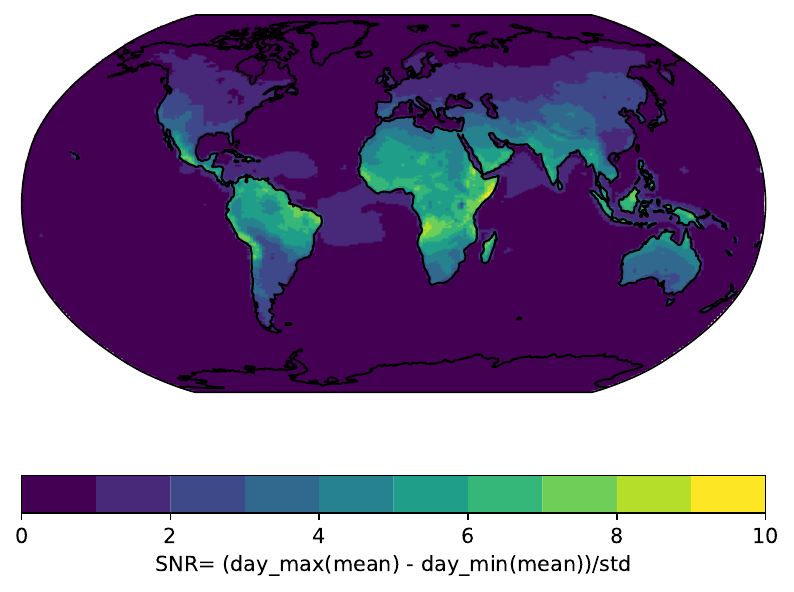} & 
        \includegraphics[width=0.4\linewidth]{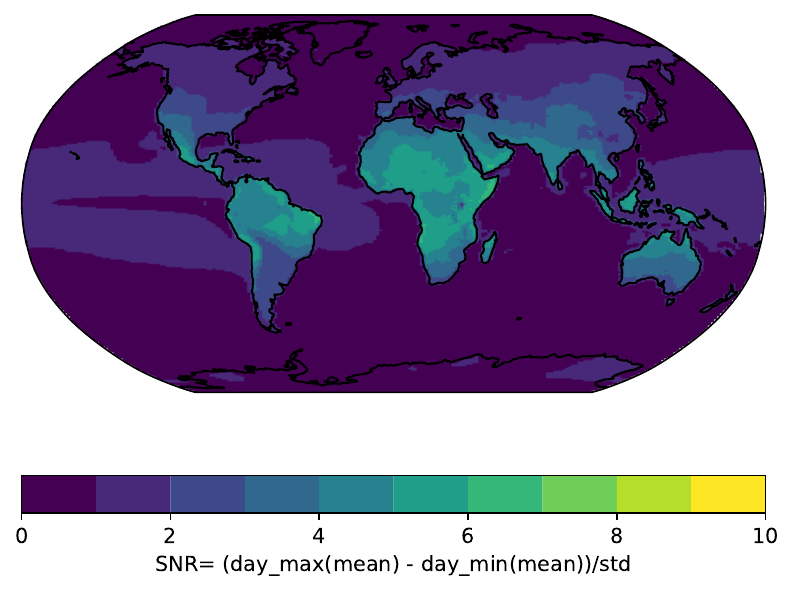} \\
    \end{tabular}
    \caption{Diurnal cycle diagnostics for surface air temperature. (top row) Amplitude of the diurnal cycle. (bottom row) signal to noise ratio.}
    \label{fig:si:diurnal:tas}
\end{figure}

\begin{figure}
    \centering
    \begin{tabular}{cc}
        a. ERA5 &  b. cBottle \\
        \includegraphics[width=0.4\linewidth]{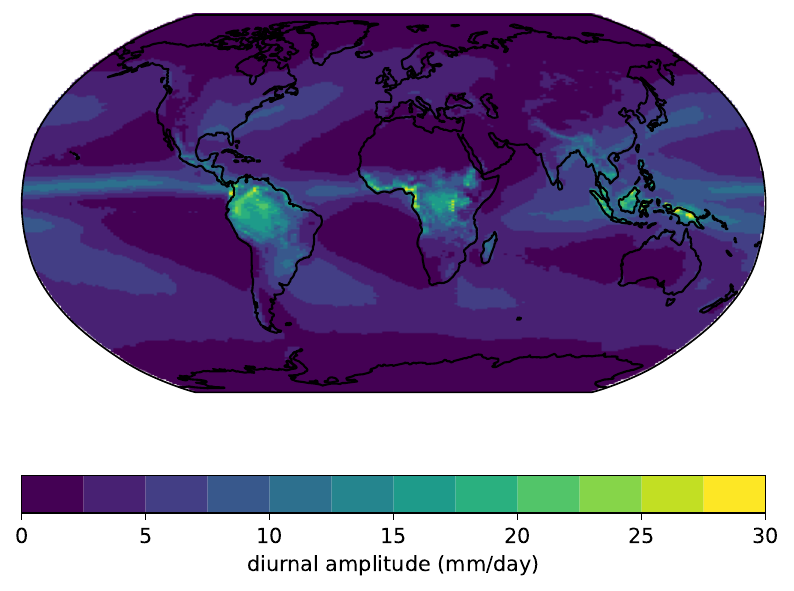} & 
        \includegraphics[width=0.4\linewidth]{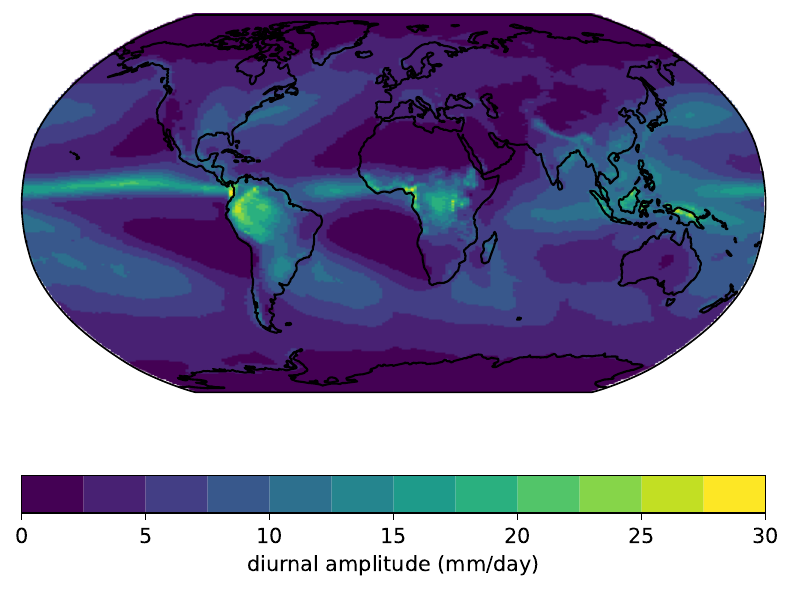} \\
        c. ERA5 &  d. cBottle \\
         \includegraphics[width=0.4\linewidth]{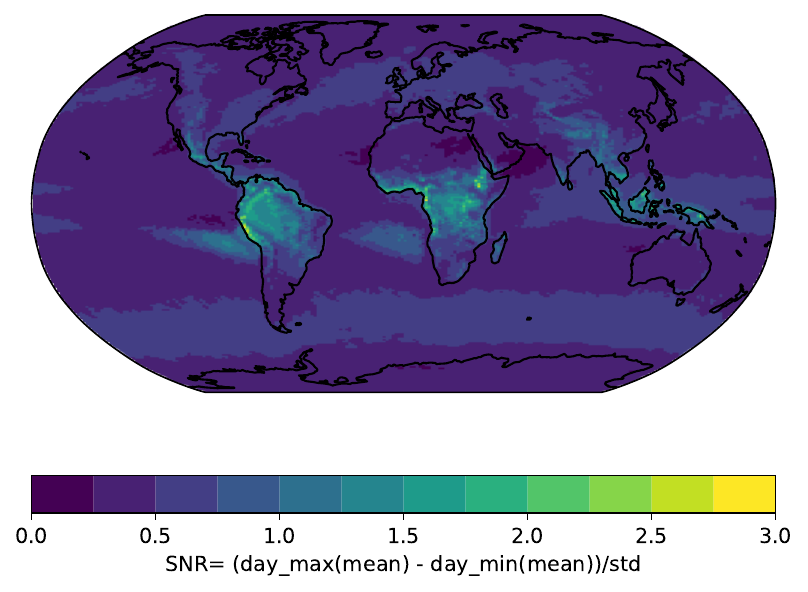} & 
        \includegraphics[width=0.4\linewidth]{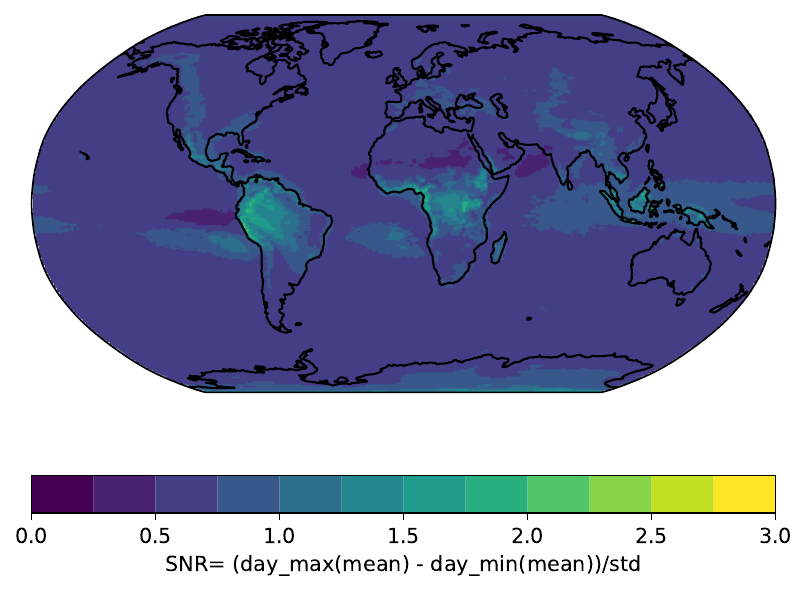} \\
    \end{tabular}
    \caption{Same as Figure \ref{fig:si:diurnal:tas} but for total surface precipitation.}
    \label{fig:si:diurnal:pr}
\end{figure}

\subsection{ENSO Response}
\label{sec:enso}

\begin{figure}
    \centering
    
    \includegraphics[width=\linewidth]{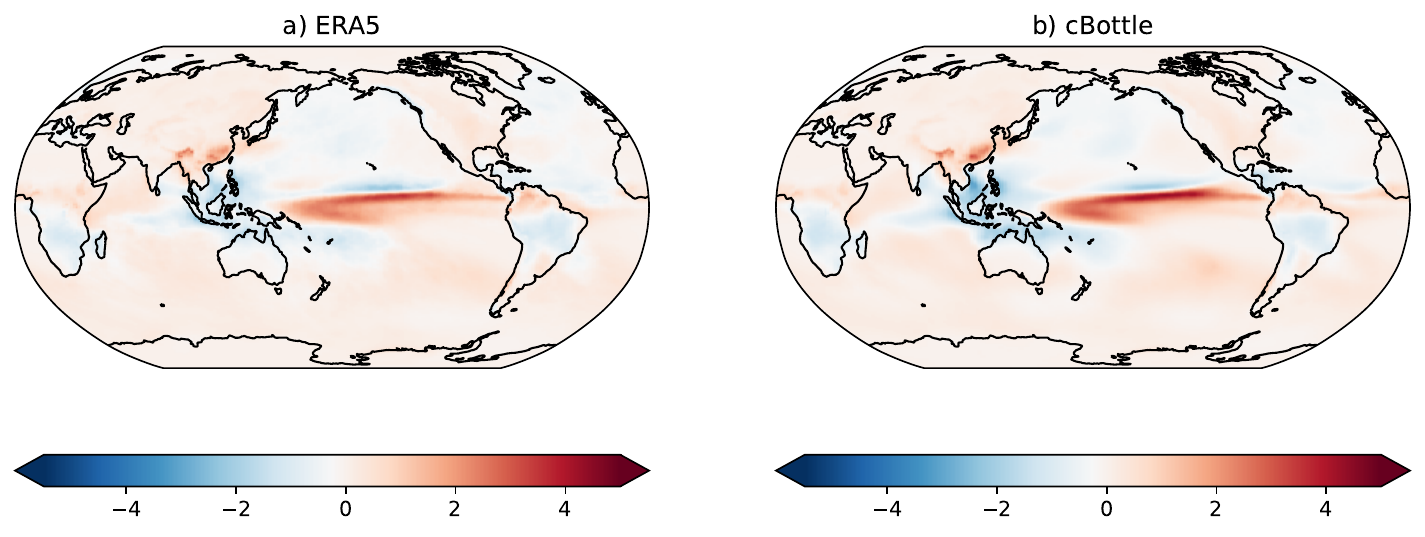}
    \caption{El Ni\~no-Southern Oscillation. Regression coefficient of the Nino3.4 index against monthly average precipitation. (left) ERA5. (right) Climate in a bottle. Computed over 1990-2020 inclusive.\label{fig:nino-pr}}

\end{figure}

Figure \ref{fig:nino-pr} demonstrates that \textit{cBottle} produces realistic ENSO precipitation variability, measured as the regression of its generated precipitation against the Nino3.4 index associated with its input SST. Within the tropics, increases of Central Pacific precipitation during the positive phase of ENSO associate with reductions of generated precipitation over tropical Africa, Indonesia and South America, consistent with Walker Cell modulation. 
 These results add to the evidence of \textit{cBottle}'s ability to condition generated variability appropriately on SST, in addition to time of year and time of day. 

\subsection{Interactivity: Details}\label{sec:si:guidance-results}
\begin{figure}
    \centering
    \begin{tabular}{cc}
        a. Unphysical samples &  b. Variability \\
        \includegraphics[width=0.37\linewidth]{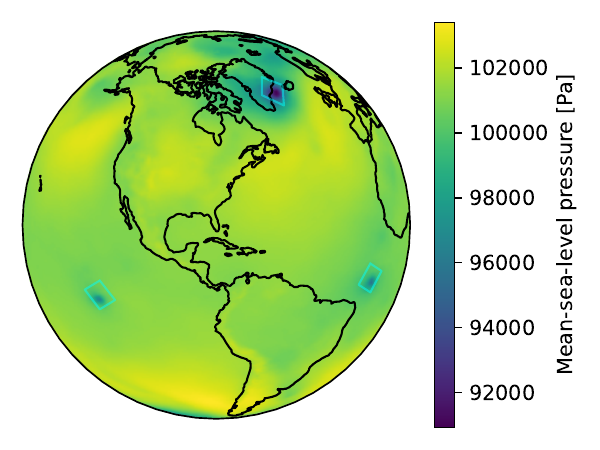} & 
        \includegraphics[width=0.4\linewidth]{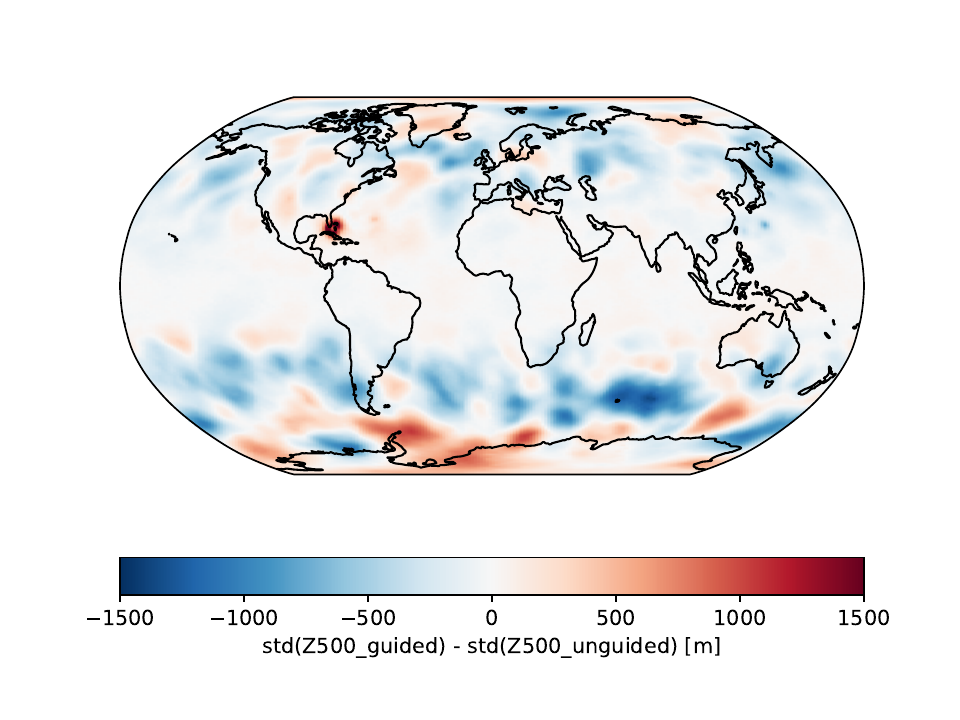}
    \end{tabular}
    \caption{Realism of generated TC states. a) The same Figure as Fig.~\ref{fig:user-control}f), but with TC guidance applied in different location. b) Difference between guided (TC guidance applied in one location near Florida) and unguided standard deviation in Z500 of a 10 member ensemble of states.}
    \label{fig:si:tc-guidance}
\end{figure}

Tropical Cyclones can in principle be requested in any location, and arbitrarily many can be requested in a single sample. Some such guidance inputs may be unphysical, such as TCs on the equator or at high latitudes. We find that while the guided model can produce such unphysical samples, TCs there tend to be more muted than physical ones, as demonstrated in Fig.~\ref{fig:si:tc-guidance}a). Here, we guide the same way as before, but with two locations on the equator and one off the coast of Greenland. All three locations show visible pressure minima, albeit somewhat weaker in the equator cases than in Fig.~\ref{fig:user-control}f). 

The guidance could also lead to a loss of variability in the output states. In the limit of very strong guidance and many locations, we cannot expect large variability in samples. However, in the moderate guidance setting described here, variability qualitatively seems to be preserved: In Fig~\ref{fig:si:tc-guidance}b) the ensemble standard deviations of a guided and unguided ensemble are similar (if the variability of the guided samples was very low, we would see a mostly blue map). 

\subsection{ICON in a bottle \label{sec:si:icon}}
\subsubsection{Super-Resolution\label{sec:si:icon-sr}}
Figure~\ref{fig:superres} provides a first qualitative assessment of the super-resolution on its own (starting from low-resolution ICON data). Here we choose a convenient time sample from the test set, in which a West Pacific tropical cyclone occurred adjacent to a large midlatitude weather system to its northeast. The performance of super-resolution is shown for two field variables -- upwelling solar radiation (first row) and precipitation rate (second row). Compared to the ground truth (right column), the results (middle column) are generally encouraging. For the tropical cyclone, appropriate orographic maximization of convection over the interior of Taiwan -- not present in the macroscale inputs (left column) -- is realistically generated. So too are details of the eyewall structure and surrounding rain bands, albeit with less spatial coherence than in the ground truth. Meanwhile, for the midlatitude weather system, a plausible fine-scale frontal band is generated, extending from the Kiril islands to just east of the Kamchatka peninsula, with consistent horizontal scale and position as the ground truth, across both field variables. 

\begin{figure}
    \centering
    \includegraphics[width=0.7\textwidth]{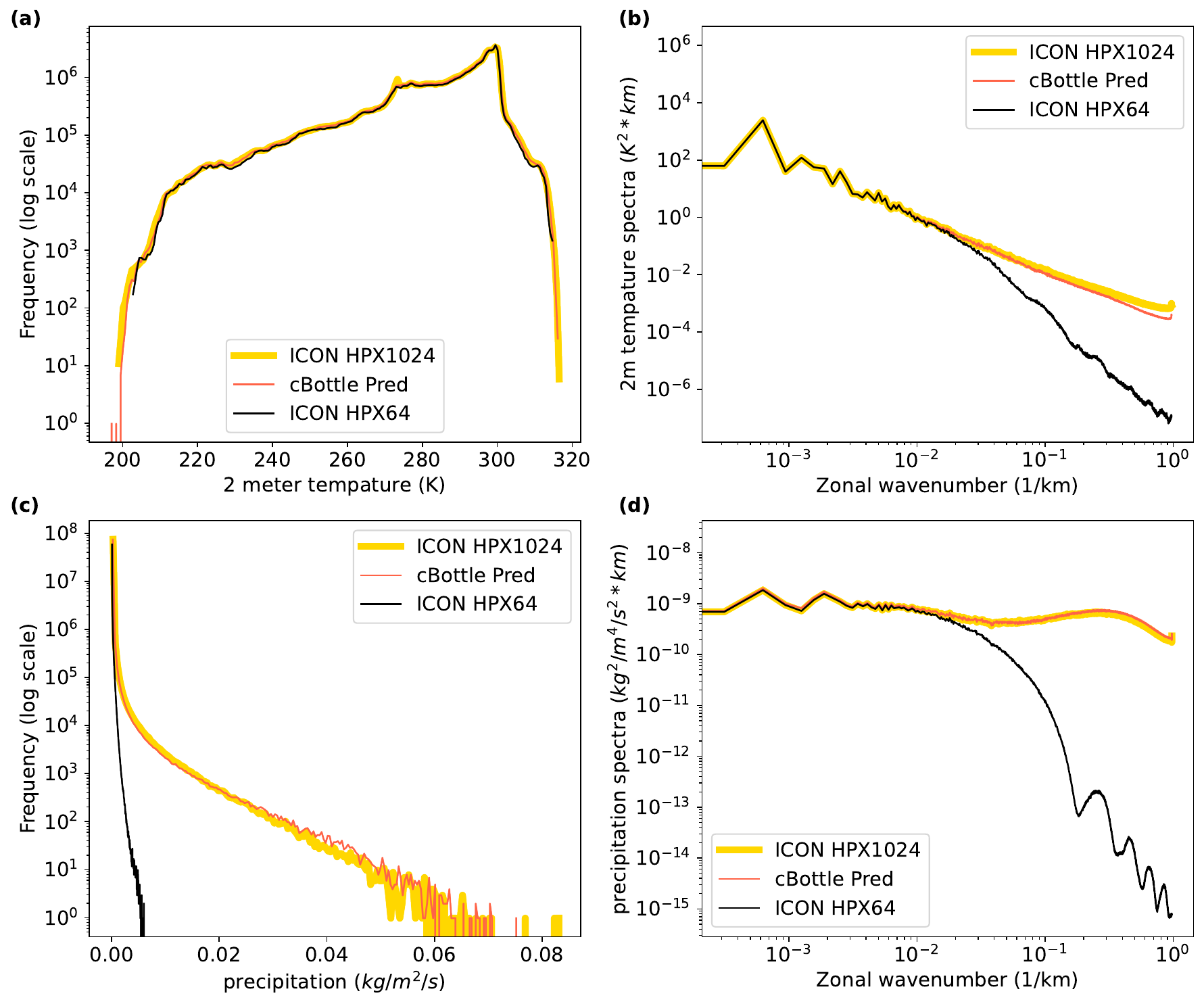}
    \caption{Spherical power spectra and distribution plots of ICON HPX64 super-resolution inferences across 32 validation samples. (a), (c): power spectra for 2-meter temperature and precipitation. (b), (d): log-scale distributions of 2-meter temperature and precipitation.}
    \label{fig:superres_spectra_pdf}
\end{figure}

A more quantitative assessment is shown in Figure \ref{fig:superres_spectra_pdf} from the view of distributions and spectra. The plots are generated using 32 randomly selected samples spanning from March 6, 2024, to July 23, 2025. The spherical power spectra are computed by averaging the squared magnitudes of the spherical harmonic coefficients over the azimuthal index $m$ and across the temporal dimension. Correct distribution of variance as a function of spatial scale is confirmed in the power spectra for temperature and precipitation.

Interestingly, similar quality results can be obtained when only training on small 4-week subset of the 5 year ICON dataset and patching artifacts can be mitigated by the choice of the patching overlap  (see Appendix \ref{sec:ablations}).

Figure \ref{fig:superres_spectra_pdf}(a,c) shows the probability distributions of surface air temperature and precipitation aggregating over all temporal and spatial locations. The super-resolution (red line) captures the shape of temperature and precipitation PDFs including the correct power law regulating the heavy tail of low-likelihood extreme precipitation occurrence (Figure \ref{fig:superres_spectra_pdf}c).). 

\subsubsection{End-to-end synthesis: Coarse generation + Super-resolution}

\begin{figure}
    \centering
    \includegraphics[width=0.98\textwidth]{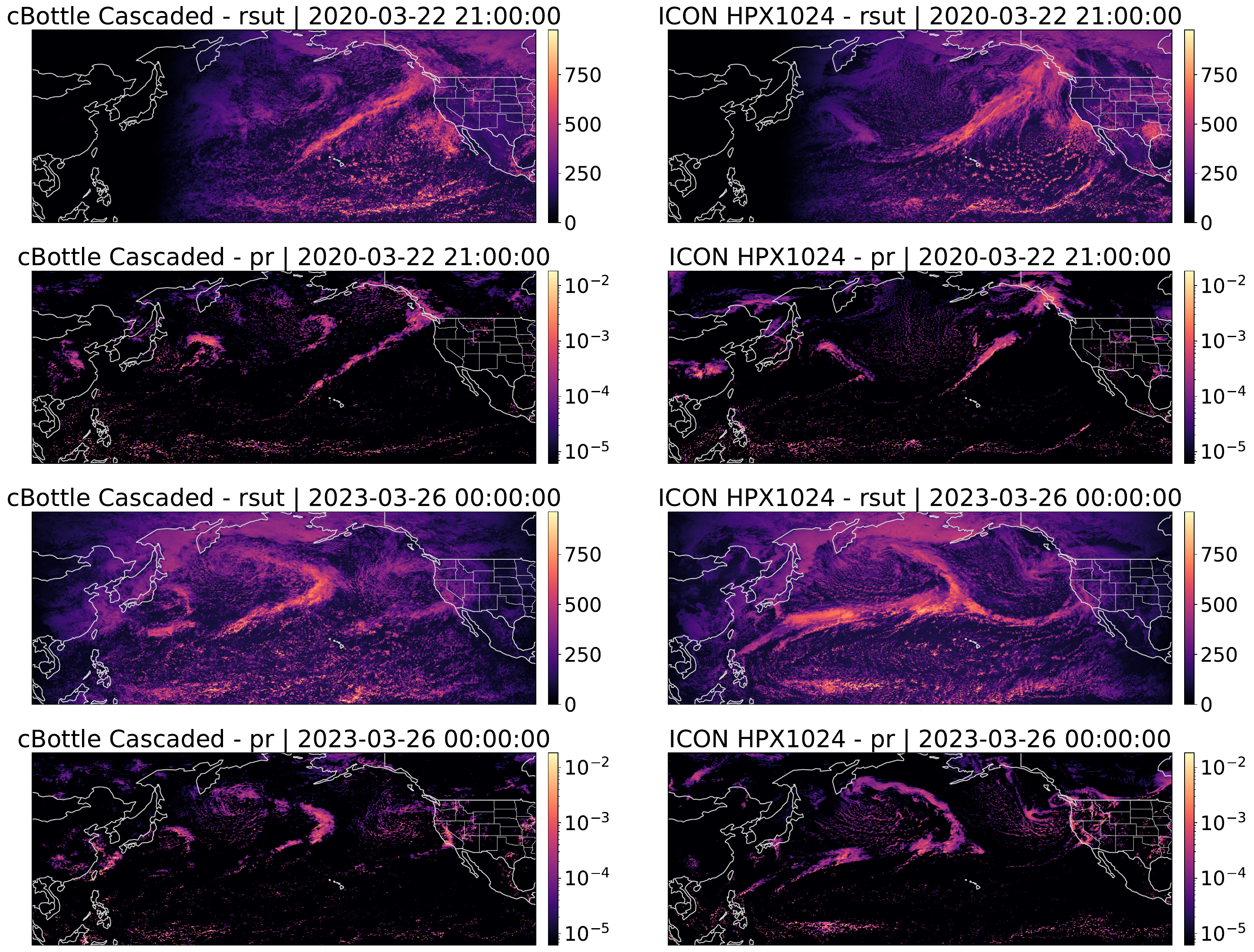}
    \caption{Demonstration of 2 HPX1024 samples from cascaded cBottle, showing outgoing shortwave radiation, and precipitation flux. Samples from cascaded and ICON are not matched.}
    \label{fig:cascade}
\end{figure}

We now turn to results from our capstone task of attempting to generate 12M-pixel multivariate ICON states purely from noise, time of year, time of day, and a monthly mean map of SST. The left column of Figure \ref{fig:cascade} provides a qualitative demonstration of the results (see also Figure \ref{fig:1}). Results are shown for an arbitrary subregion spanning the entire North Pacific ocean, across four arbitrary time periods having distinct weather states. It is important to note that the ground truth (corresponding right columns) should not be compared to the model output for any matched spatial positioning of individual weather events, since the task is unconditional generation. That is, the ground truth also has distinct weather states. Rather, what is noteworthy in the comparison is the common modulation of similar microscale weather detail by variably positioned macroscale weather events.

From this view, we find encouraging co-generation of outgoing shortwave radiation and precipitation channels. As in the ground truth, sub-sectors of midlatitude eddies exhibit km-scale organization patterns of non-precipitating shallow cloud (i.e. evident in the shortwave channel but not the precipitation channel). As in the ground truth, zonally elongated macroscale filaments of the intertropical convergence zone are embedded with fine-scale precipitation having its own characteristic length scale of organization. In contrast, midlatitude precipitation is comparatively stratiform -- lower intensity and more spatially coherent -- with the exception of fronts visible in both brightness and rainfall. While difficult to summarize statistically, such multi-scale, multi-variate spatial associations are readily visible to the trained human eye accustomed to inspecting weather data, and suggest successful synthesis. 

\begin{figure}
    \centering
    \includegraphics[width=0.9\textwidth]{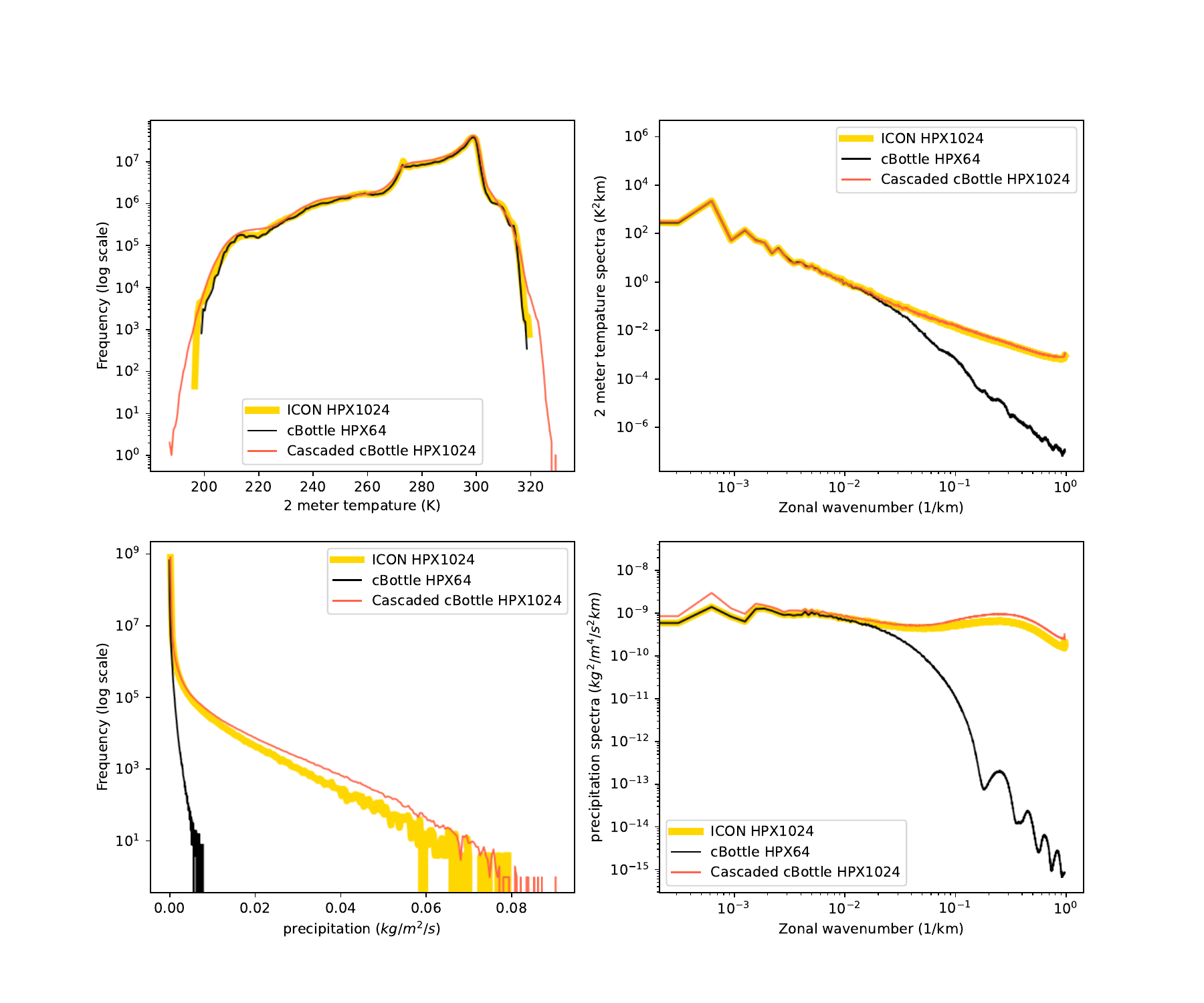}
    \caption{Distribution and spherical power spectra plots of cascaded \textit{cBottle} inferences across 64 validation samples. Column 1: log-scale distributions of 2-meter temperature and precipitation. Column 2: power spectra for 2-meter temperature and precipitation. }
    \label{fig:cascade_spectra_pdf}
\end{figure}

Figure \ref{fig:cascade_spectra_pdf} presents the probability distributions and spherical power spectra computed from 64 randomly selected samples generated by the cascaded cBottle model, spanning the period from January 3, 2020, to July 23, 2025. Spectra analysis again quantitatively confirm the finding. On the one hand, secondary imperfections are introduced that were not apparent when super-resolution is used in isolation of macro-scale generation (Figures \ref{fig:superres}) such as an increased presence of high-frequency components in the precipitation spectrum. On the other hand such imperfections are minor such that the results are overall of reasonable enough quality so as to suggest the viability of cascaded super-resolution as a means to generate multivariate ICON states. 

In general, the model-generated distributions exhibit good agreement with the target, though the cascaded approach shows a tendency to exaggerate extreme temperature values and slightly overpredict precipitation. The super-resolution (Figures \ref{fig:superres}) did not have these problems, which suggests that the bias likely comes from the coarse-resolution generations in this cascaded setup.

\subsection{Multimodal: ICON-Informed Downscaling of ERA5}
Although the super-resolution model is trained exclusively on ICON data, it demonstrates zero-shot generalization capability when applied to data from other sources. In this subsection, we assess its skills on ERA5 data regridded to HPX64 resolution. The corrupted radiation channels in ERA5 have been infilled using the method described in Section \ref{sec:channel_fill}.
Please note that this method differs slightly from the translate-then-super-resolve approach in Figure \ref{fig:multimodal}h, which translates ERA5 to ICON and then super-resolves instead.
The resulting super-resolved outputs are compared against both the 3-km High-Resolution Rapid Refresh (HRRR) dataset over CONUS and the native-resolution ICON model at HPX1024. All datasets used in this evaluation are regridded to a common 3-km Lambert conformal projection over the CONUS region to ensure fair spatial alignment.

\begin{figure}
    \centering
    \includegraphics[width=1\textwidth]{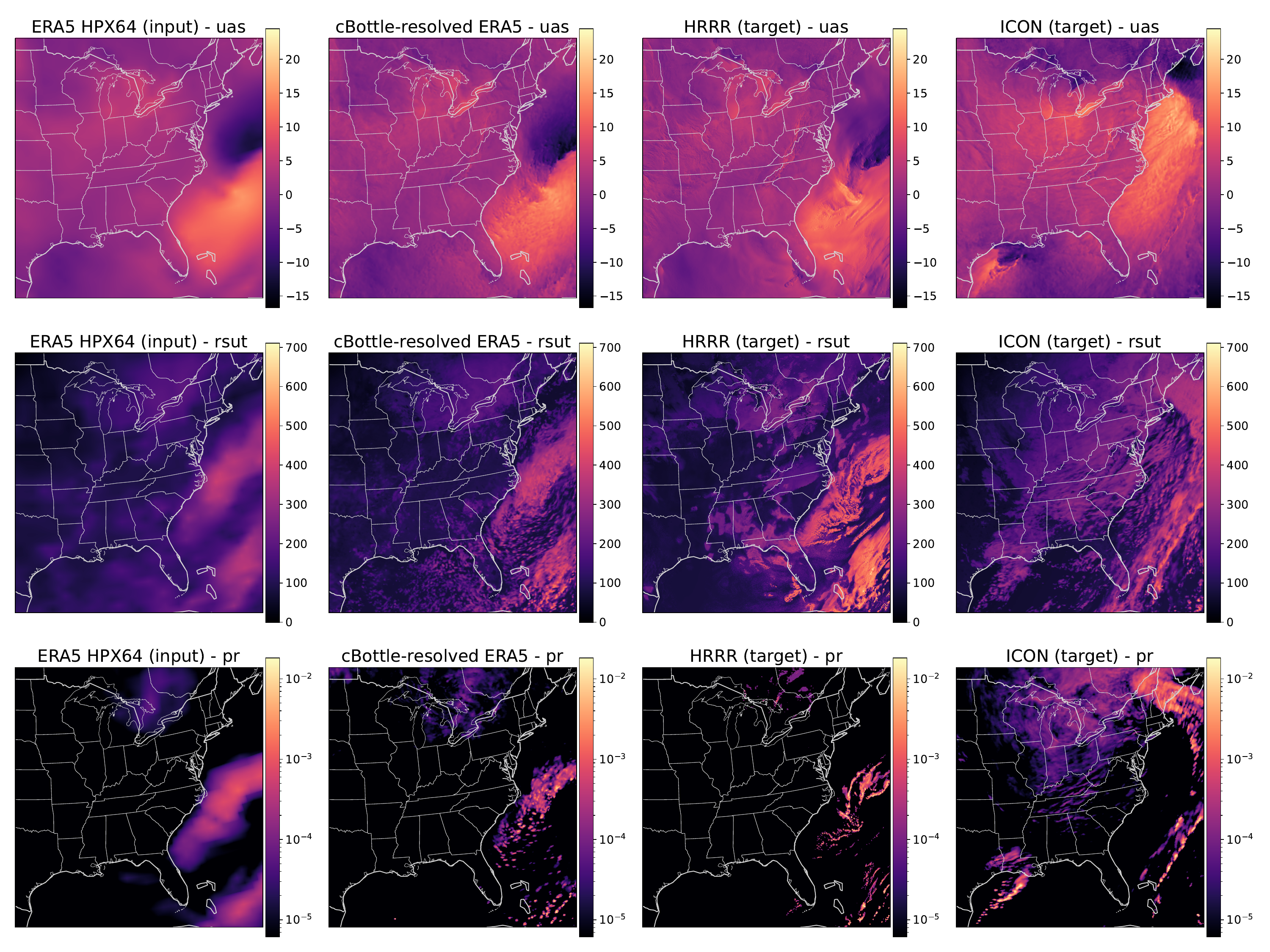}
    \caption{Illustration of the downscaled ERA5 output at 15:00 UTC on December 11, 2018, with radiative channels infilled using the \textit{cBottle} HPX64 generator. The first three columns correspond to the same timestamp, while the rightmost column (ICON) shows the free-run forecast for 15:00 UTC on December 11, 2024. From top to bottom, the rows represent 10-meter eastward wind velocity, outgoing shortwave radiation, and precipitation. }
    \label{fig:era5_superres}
\end{figure}

Figures \ref{fig:era5_superres} presents a qualitative comparison of three representative variables: 10-meter eastward wind velocity, outgoing shortwave radiation, and precipitation rate. The figure shows spatial subsets spanning the Midwest to the Eastern United States, with four columns corresponding to the original ERA5 HPX64 input, the super-resolved ERA5 output generated by our model, the HRRR reference, and the ICON HPX1024 reference. Note that the ICON simulation is a free-running model and is not temporally aligned with the ERA5 or HRRR data; however, it still serves as a useful reference for assessing small-scale structural similarity. While the large-scale patterns in the super-resolved ERA5 resemble those in the original ERA5 and HRRR data, the finer-scale structures more closely resemble those in the native-resolution ICON fields. This suggests that the model effectively emulates high-frequency features in ICON and is capable of generating high-resolution patterns and dynamics that are not explicitly present in the coarse-resolution data.

\begin{figure}
    \centering
    \includegraphics[width=0.8\textwidth]{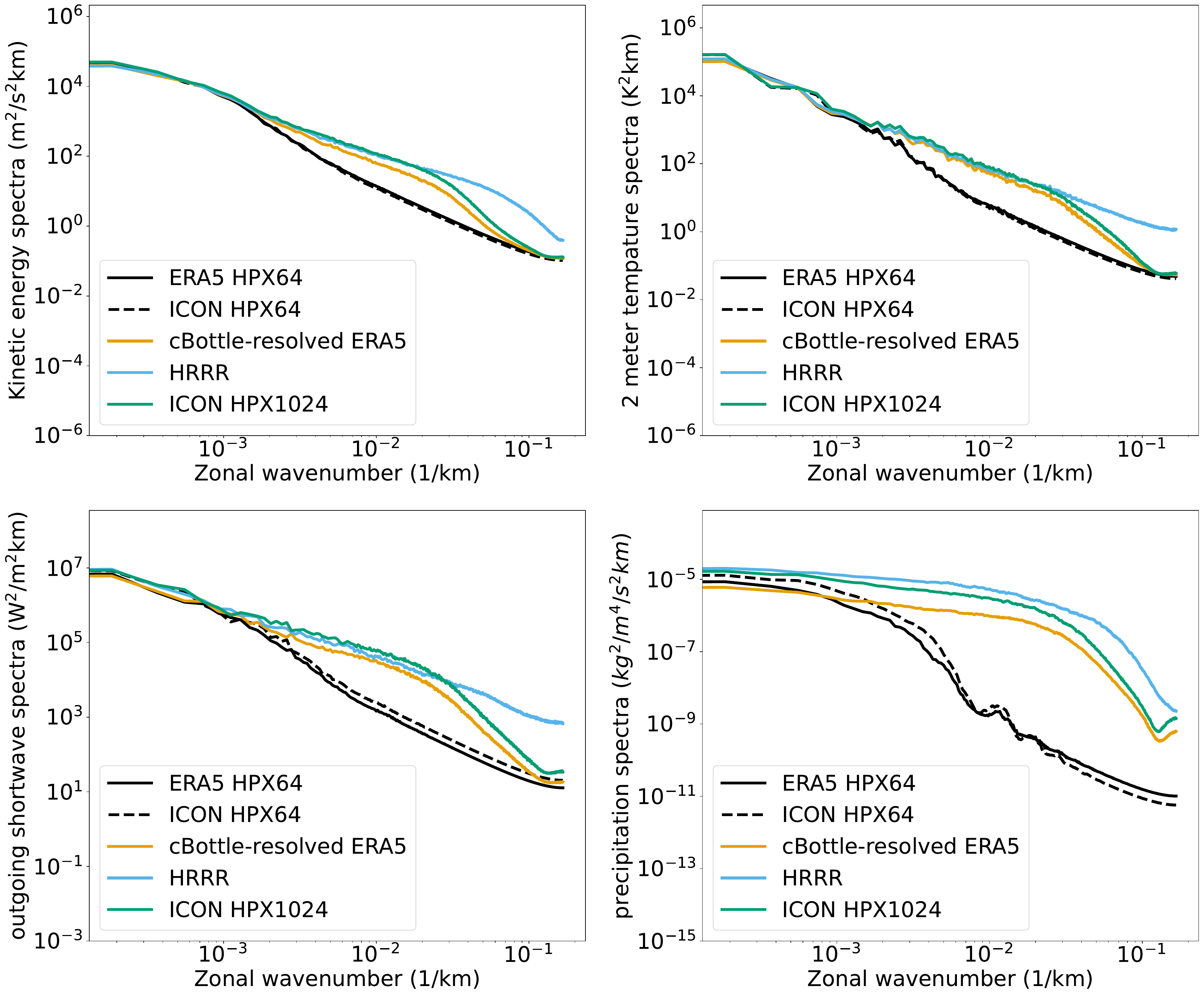}
    \caption{Emulation of native-grid ICON with ERA5: Power spectra of cBottle-resolved ERA5 and various baselines, computed over 10 samples. All datasets are regridded to a Lambert-Conformal projection for comparison. ERA5, cBottle-resolved ERA5, and HRRR share an identical set of evaluation timestamps. ICON HPX64 and ICON HPX1024 use matching times of day and days of the year, but from different years. }
    \label{fig:era5_superres_spectra}
\end{figure}

The zonal-mean spherical power spectra for four representative variables are shown in Figure \ref{fig:era5_superres_spectra}: 2-meter temperature, outgoing shortwave radiation, and precipitation. Comparisons are made across five datasets: the ERA5 HPX64 input, the super-resolved ERA5 output generated by our model (\textit{cBottle}), the HRRR reference, ICON at HPX64, and ICON at HPX1024 resolution.

For variables where ERA5 HPX64 and ICON HPX64 exhibit similar spectral characteristics—such as kinetic energy and temperature—the super-resolved outputs from \textit{cBottle} closely follow the spectral slope and amplitude of the ICON HPX1024 reference, indicating successful reconstruction of high-frequency content. This alignment demonstrates the model's capacity to synthesize physically plausible fine-scale variability. 

For channels where the ERA5 HPX64 input exhibits significant spectral differences from the ICON HPX64 input (precipitation), however, the model's ability to recover fine-scale variability is limited. While \textit{cBottle} partially bridges the gap, it underrepresents high-wavenumber energy relative to ICON HPX1024 and HRRR. This highlights the challenge of zero-shot generalization when there is a domain shift in spectral characteristics between the training and evaluation inputs.

To sum up, it is encouraging to observe that the super-resolution model can effectively super-resolve unseen data in a zero-shot setting, while also faithfully emulating the characteristics of its training data. Reducing spectral and distributional discrepancies caused by domain shifts—potentially through the inclusion of more diverse training datasets—may further improve the model's robustness and enhance its applicability in climate science.

\subsection{Further climate diagnostics \label{sec:si-climate-diagnostics}}

\begin{figure}
    \centering
    \includegraphics[width=0.5\linewidth]{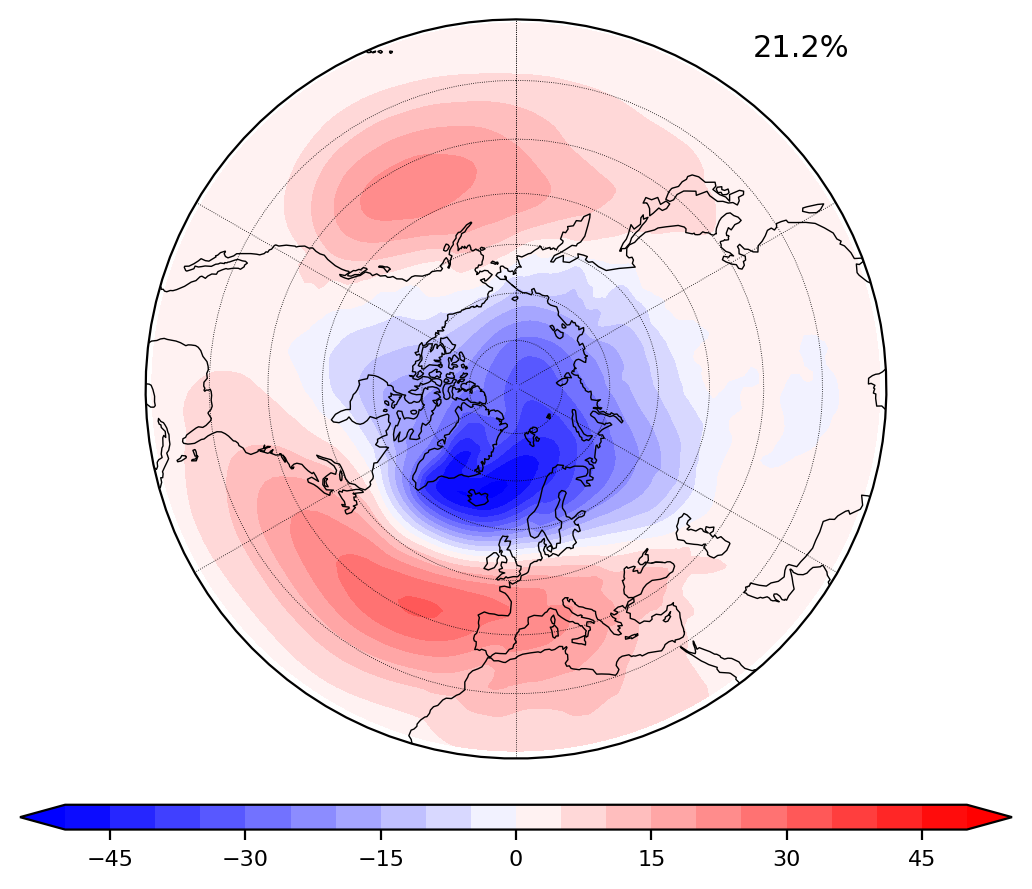}
    \caption{Loading pattern for the Northern Annular Mode computed from ERA5.  This is the first empirical orthogonal function of the ERA5 Z1000 monthly anomalies computed over 1970-2010, north of 20 degrees N, and for the winter months of November-April.}
    \label{fig:si:nam}
\end{figure}

\begin{figure}
    \centering
    \includegraphics[width=1\linewidth]{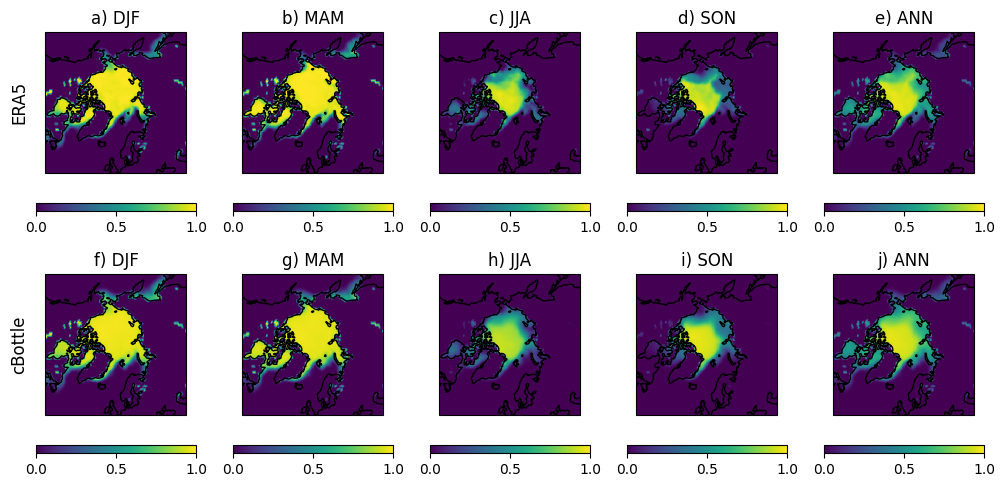}
    \caption{Cycle of sea ice concentration for ERA5 emulation and ground truth, computed over the test period}
    \label{fig:cycles}
\end{figure}

\begin{figure}
    \includegraphics[width=\linewidth]{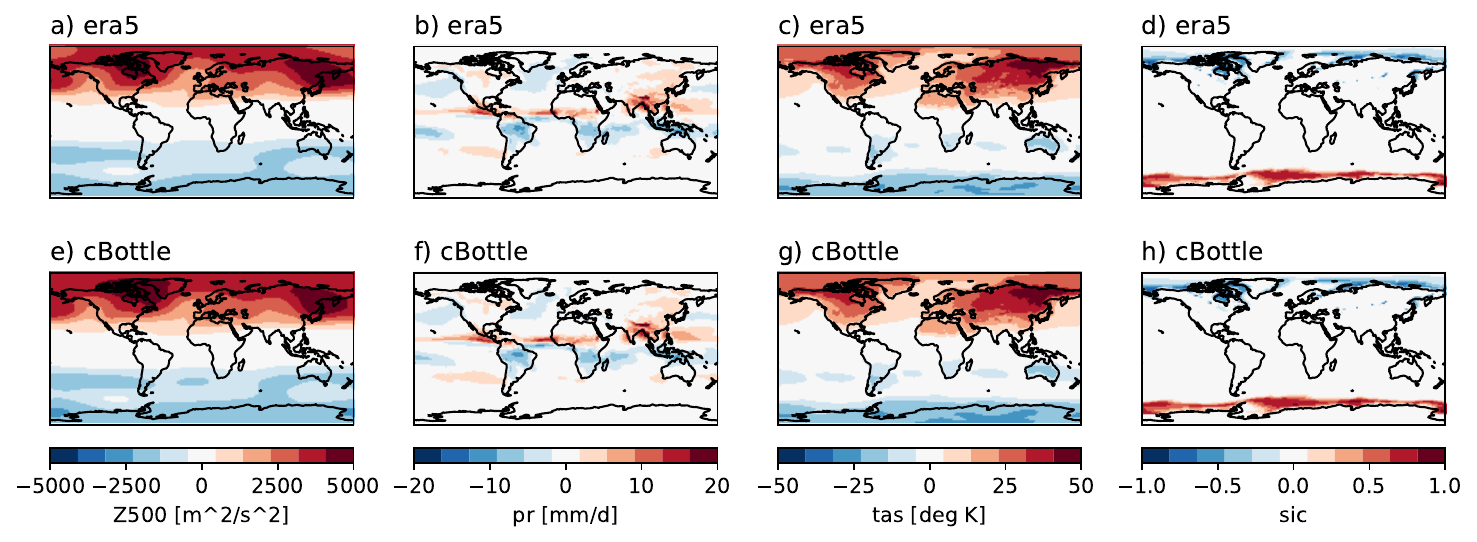}
    \caption{Seasonal cycle for (JJA- DJF) for ERA5 and \textit{cBottle}. Panels are shown for Z500, surface air temperature, precipitation, and sea ice concentration.\label{fig:seasons}}
\end{figure}

\subsubsection{Zonal Mean Precipitation}
\begin{figure}
    \centering
    \includegraphics[width=1.0\linewidth]{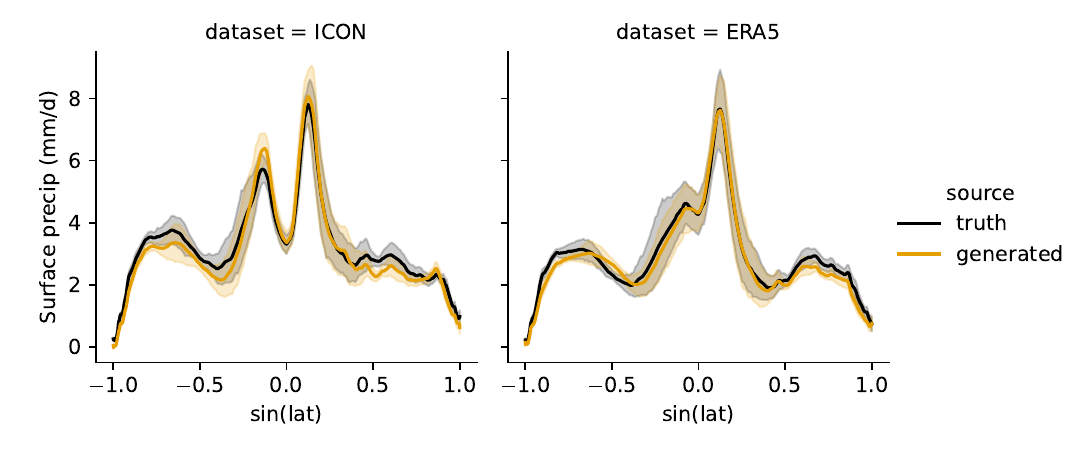}
    \caption{Zonal Average of precipitation ICON, ERA5, and their emulators. 95\% confidence intervals for the validation set means are estimated by bootstrapping over monthly averages. Only data from the validation periods for ICON and ERA5 are used.}
    \label{fig:pr-zonal}
    

    
\end{figure}

Figure \ref{fig:pr-zonal} examines mean precipitation in \texttt{\textit{cBottle} ERA5} and its relationship with latitude computed. Overall \texttt{\textit{cBottle} ERA5} captures the existence and correct positioning of local maxima associated with midlatitude storm tracks and intertropical convergence zones straddling the equator. Local imperfections include positive biases in the northern (+4\%) and southern (+11\%) tropical rain bands, and as much as 11\% underestimation of midlatitude rainfall. These differences can in part be explained partially by sampling error.

\subsubsection{Climate change}
\begin{figure}[h]
\centering
\begin{center}
\begin{tabular}{ccc}
\begin{minipage}[t]{0.3\textwidth}
\textbf{a}\\
\includegraphics[width=\linewidth]{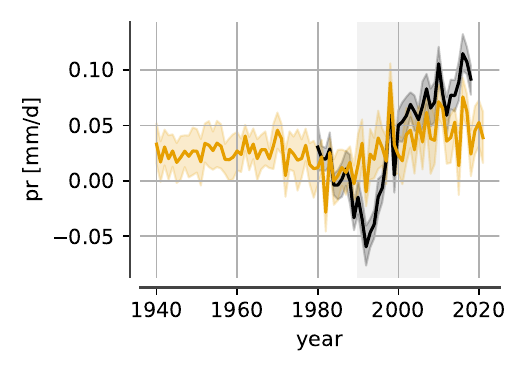}
\end{minipage} &
\begin{minipage}[t]{0.3\textwidth}
\textbf{b}\\
\includegraphics[width=\linewidth]{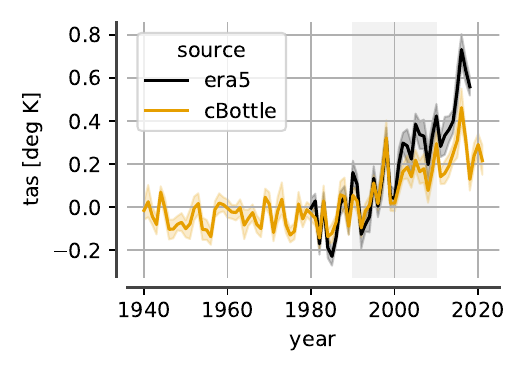}
\end{minipage} &
\begin{minipage}[t]{0.3\textwidth}
\textbf{c}\\
\includegraphics[width=\linewidth]{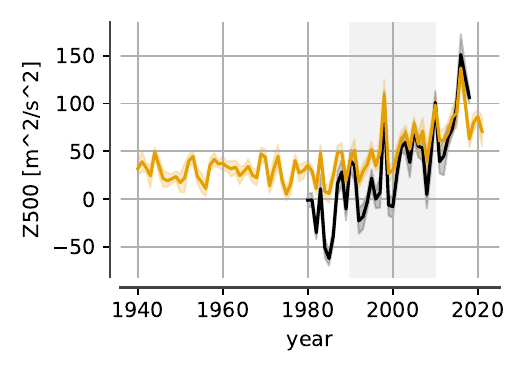}
\end{minipage}
\end{tabular}
\end{center}
    \caption{Climate change. Time series of global mean anomalies from ERA5's monthly climatology computed over 1980-2010. (a) surface air temperature, (b) precipitation, (c) 500-mb geopotential. 95\% confidence-intervals for the annual mean anomaly are computed by bootstrapping over the monthly averages. }
    \label{fig:si:global-mean}
\end{figure}

\subsubsection{Power-spectral fidelity of the coarse-resolution generator}

\begin{figure}
    \centering
    \includegraphics[width=\linewidth]{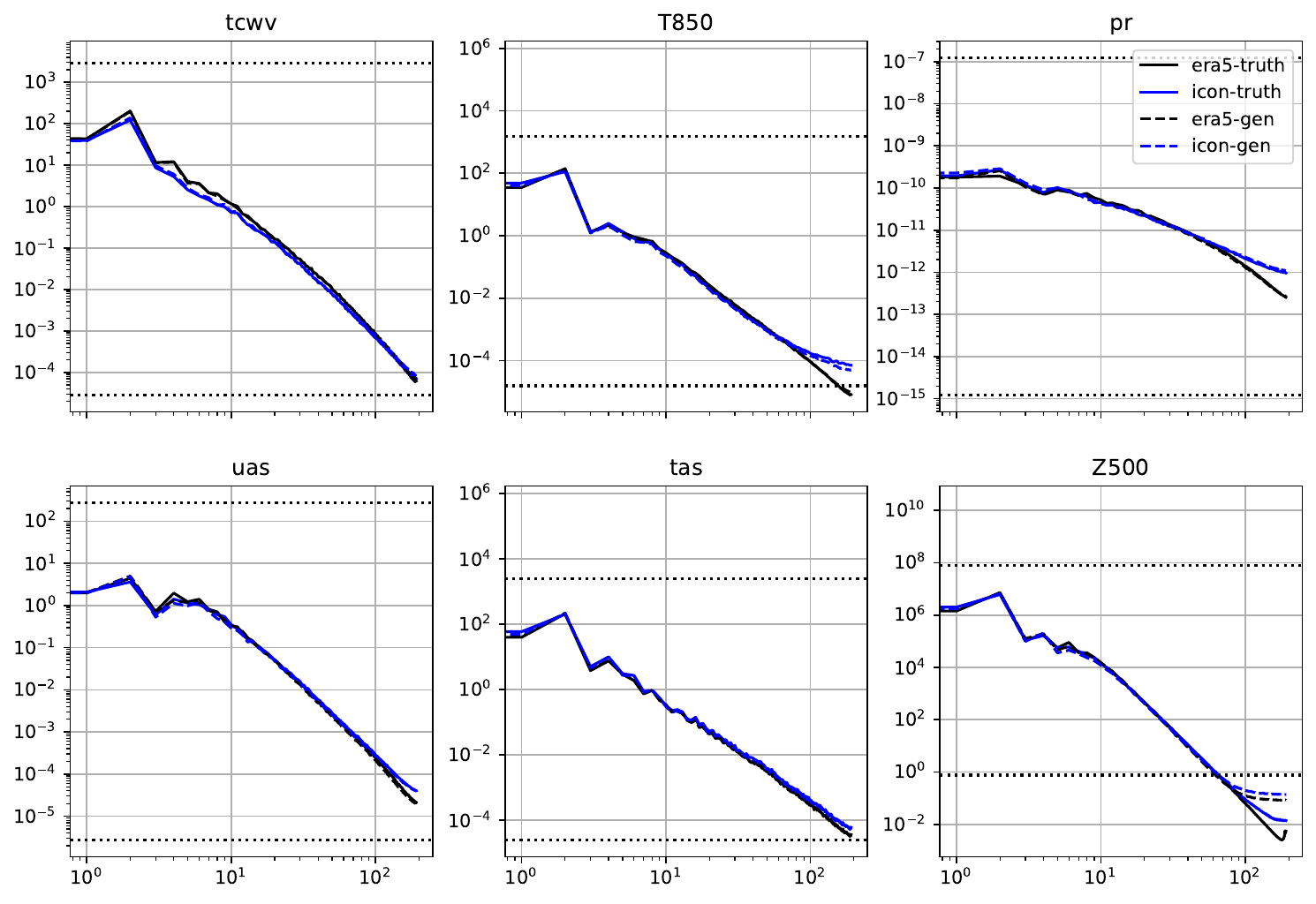}
    \caption{(left) Spherical power spectra of versus the truth. Averages over 100 random samples from the test period are shown. The dotted lines indicate the range of noise levels used for the sampling $\sigma_{\{min,max\}}^2 \frac{4\pi}{n_{pix}}$. The factor of $\frac{4\pi}{n_{pix}}$ scales the pixel-wise variance to the same units as the power spectra, which is variance per surface area of the unit sphere.}
    \label{fig:spectra}
\end{figure}

\begin{figure}
    \centering
    \includegraphics[width=0.5\linewidth]{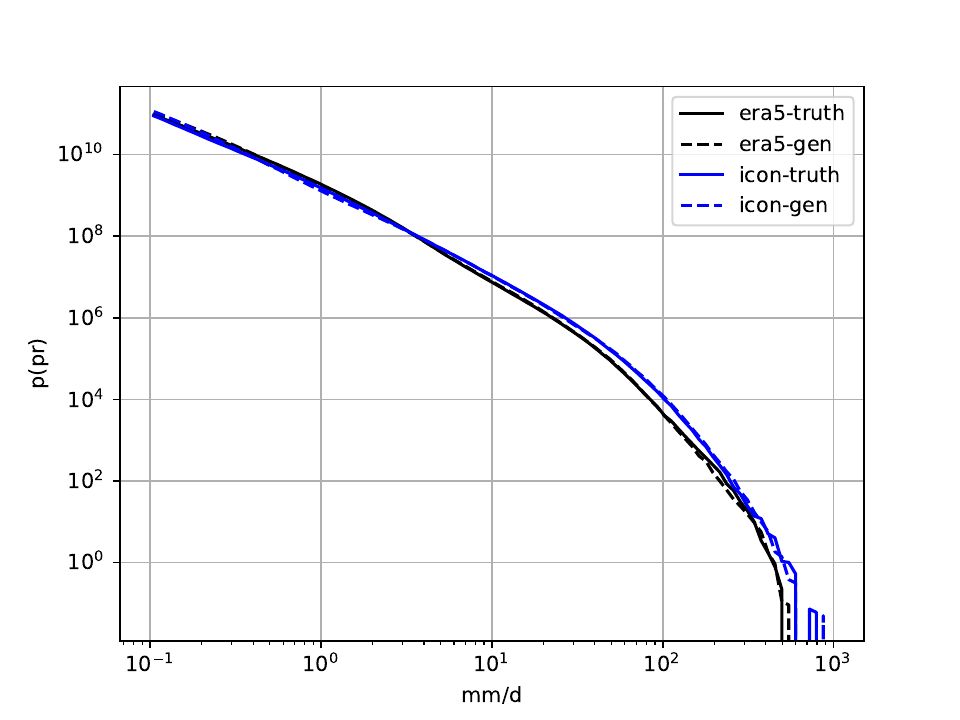}
    \caption{PDF of instantaneous precipitation. Computed for same samples as in \ref{fig:spectra}.}
    \label{fig:precip-pdf.}
\end{figure}

Figure \ref{fig:spectra} shows that the spectra of the generated coarse samples are consistent with their ICON and ERA5 ground truth. For some fields, the ICON data shows significantly more power than ERA5 at large wave numbers, but we have not confirmed if this comes from the aliasing of smaller scale modes due to the block averaging used to coarsen the native HPX1024 outputs for ICON or an actual difference.

\subsubsection{Variance maps}

\begin{figure}
    \centering
    \includegraphics[width=1\linewidth]{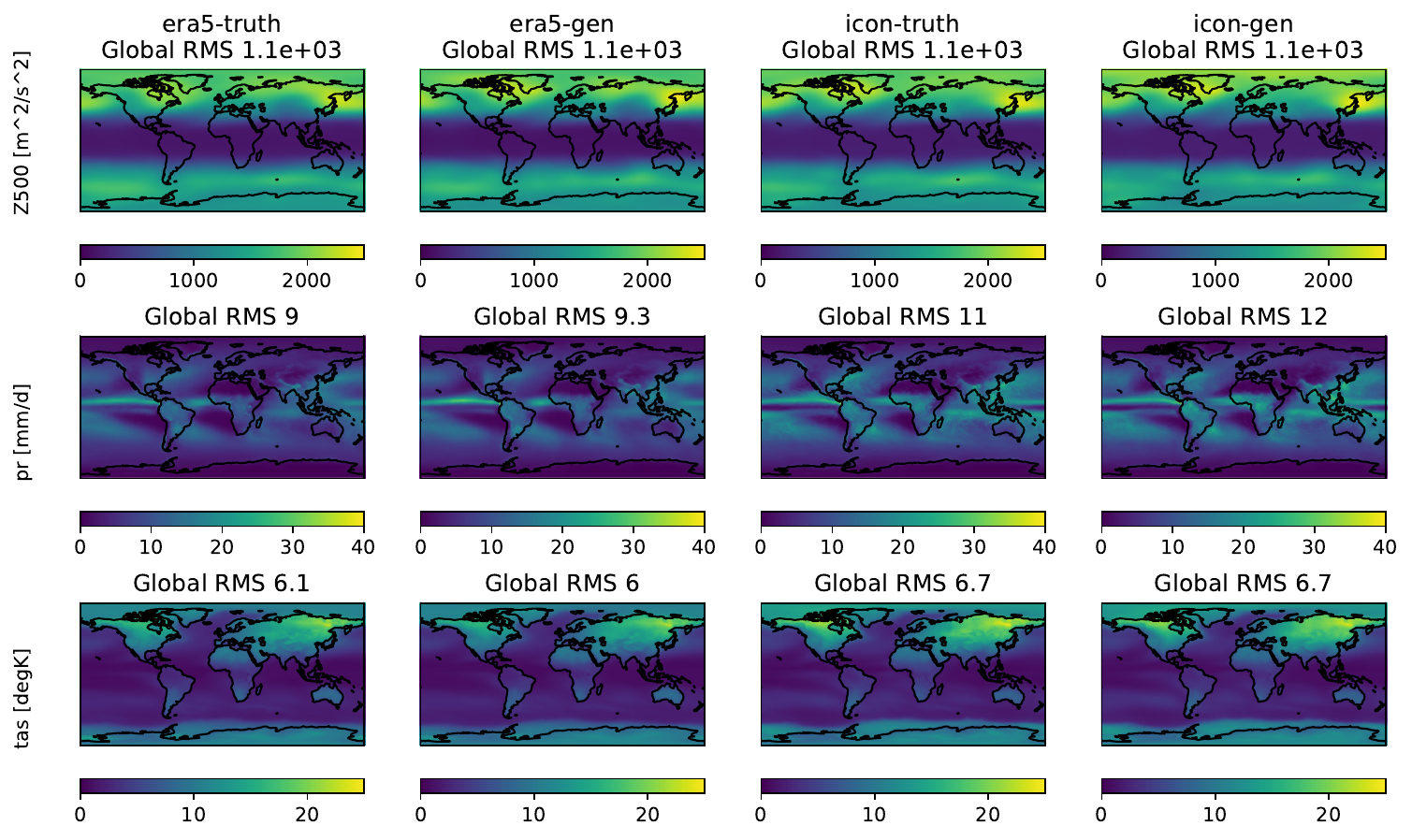}
    \caption{Total standard deviation maps with respect to the annual mean computed over the train and test periods for surface air temperature, and precipitation. Panels are shown for the ground truth and \textit{cBottle} for ICON and ERA5. The globally-averaged RMS is shown in the label for each subpanel.}
    \label{fig:variance-maps}
\end{figure}

Figure \ref{fig:variance-maps} maps summarize the pointwise variance of samples after removing the time mean (Figure \ref{fig:variance-maps}). This summarizes dispersion produced by a variety of sources including diurnal, synoptic variability, seasonal, and inter-seasonal. Variance of 500 hPa geopotential height (top) is maximized over the Southern Ocean, North Atlantic Sea and northeast Pacific due to midlatitude weather systems, but avoids the tropics due to the weak temperature gradient constraint there \citep{sobel2001weak}. Midlatitude rainfall variance occurs over the storm tracks including the Gulf Stream and Kuroshio currents. Rainfall variance is especially high within the intertropical convergence zones. In the special case of surface temperature, diurnal variability leads to a distinct land-sea contrast -- a result of solar heating causing strong day-night temperature cycles over land, where surface specific heat capacity is relatively low. 

The main point of Figure \ref{fig:variance-maps} is that \texttt{cBottle-ERA5} and \texttt{cBottle-ICON} faithfully generate all of these point-wise variance features summarizing all sources variability -- and with sufficient precision to detect differences between them, such as the ``double ITCZ'' -- which manifests in \texttt{cBottle-ICON} and \texttt{cBottle-ICON-Target} as a zonal band of excessive dispersion in tropical rainfall in the eastern Pacific just south of the equator.

\subsection{Ablations}
\label{sec:ablations}
\begin{figure}
    \centering
    \includegraphics[width=0.98\textwidth]{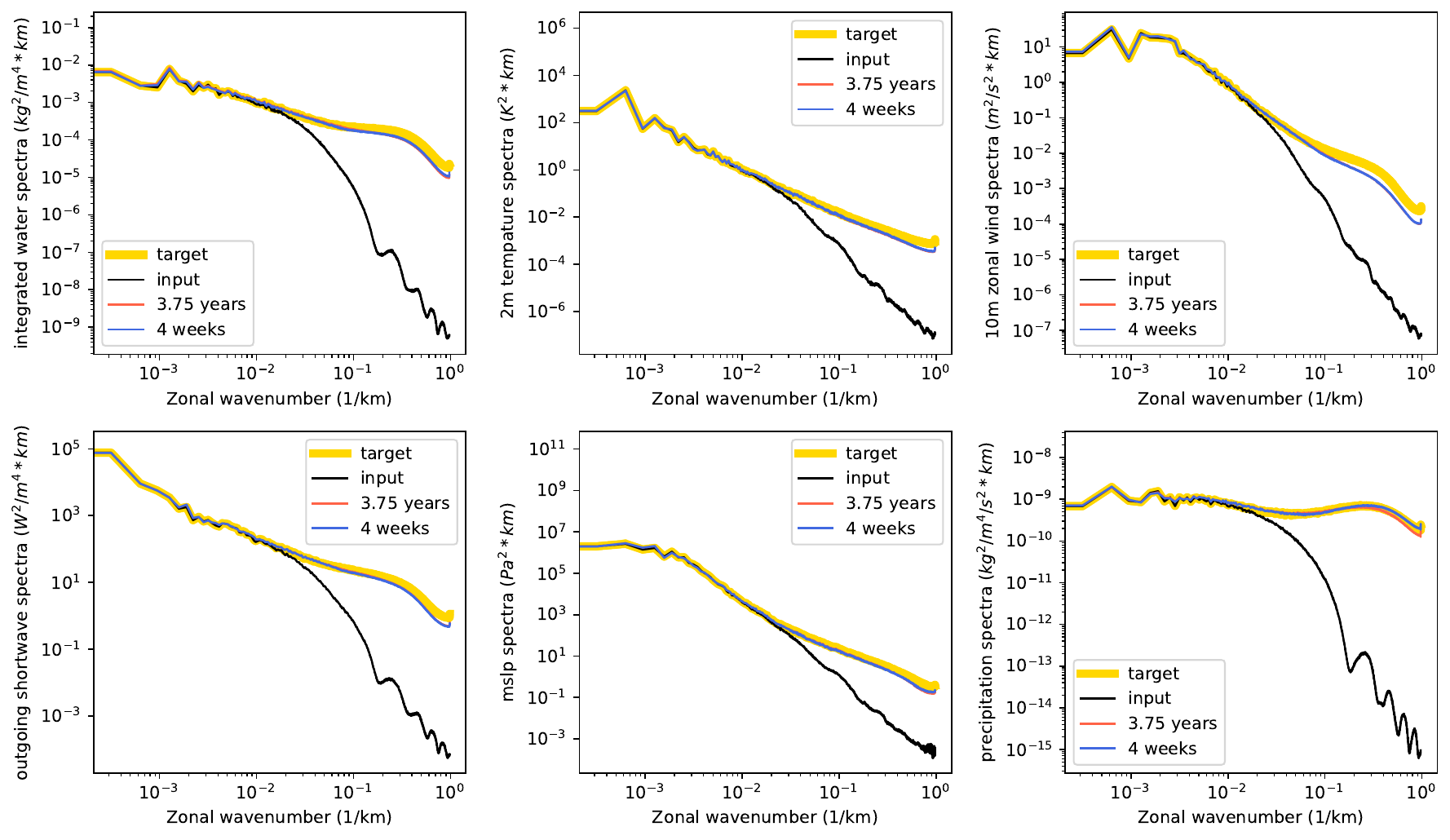}
    \caption{Training data size ablation study: Power spectra from global super-resolution models trained on 3.75 years versus 4 weeks of timesteps. }
    \label{fig:superres_data_ablation}
\end{figure}

\paragraph{Training data size} Figure \ref{fig:superres_data_ablation} illustrates the spherical power spectra of inference results from two models trained on datasets of differing sizes: one with 3.75 years of distinct samples (72,360 in total), and another with 4 weeks of distinct samples (1,344 in total), with each week randomly selected from a different season to ensure representative temporal coverage. Both models were trained using 95,000 total samples via repetition to match training iterations.

Across almost all variables, the \texttt{4 weeks} model closely reproduces the spectral characteristics of the \texttt{3.75 years} model. This alignment is particularly notable given the drastic reduction in training data diversity. The only substantial deviation appears in the precipitation spectra, where the \texttt{4 weeks} model exhibits slightly lower zonal energy at high wavenumbers, suggesting a modest underrepresentation of fine-scale variability. These results demonstrate the model’s strong data efficiency: even when trained on a small, seasonally representative subset, it retains robust downscaling performance and preserves key spectral structures across a range of physical variables.

\paragraph{Noise schedule}
\label{sec:results:noise}

\begin{figure}
    \centering
    \includegraphics[width=\linewidth]{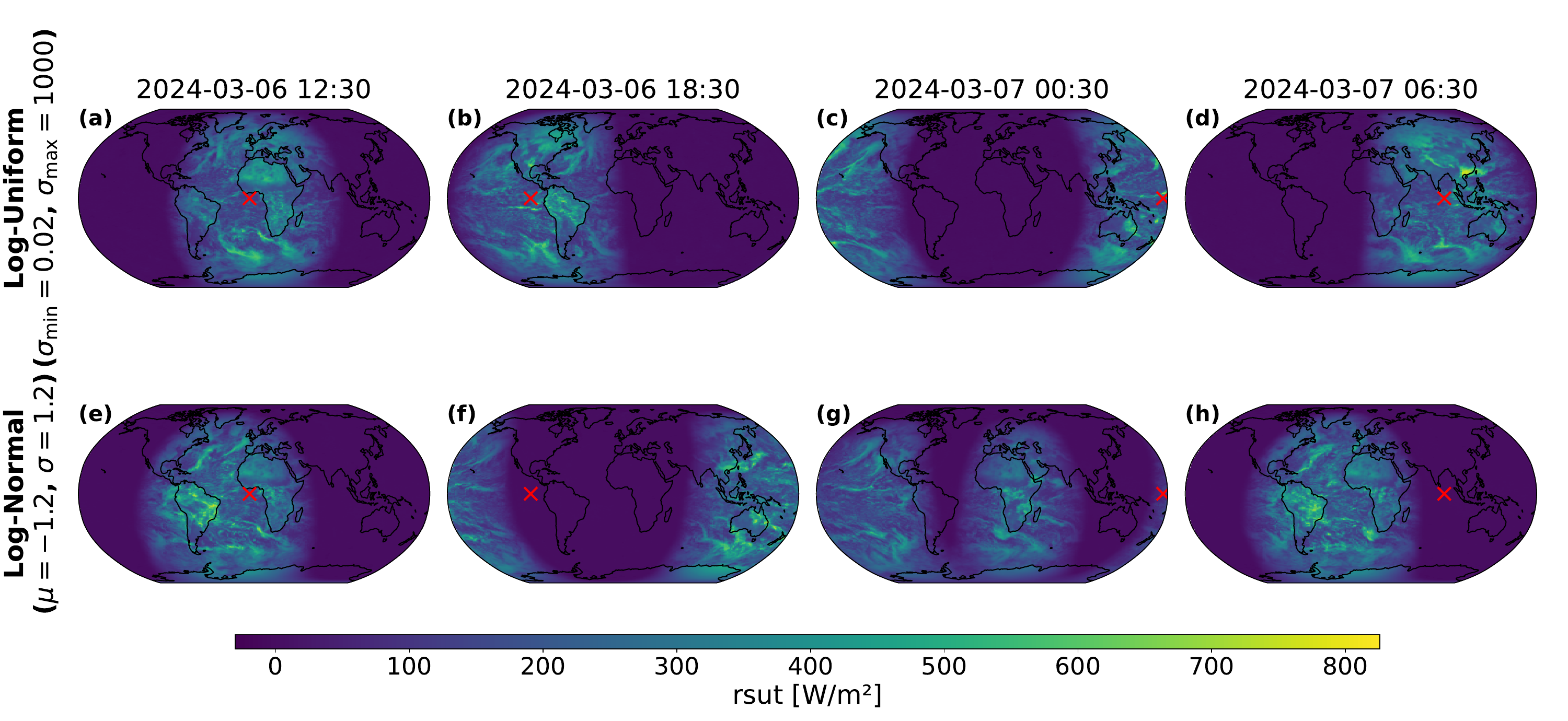}
    \caption{Four consecutive frames of reflective visible light (rsut) from a sequence generated by video diffusion models trained with either a log-uniform (top) or log-normal (bottom) noise schedule. Generations are conditioned on the calendar embedding and monthly average SST. Red × marks indicate the subsolar point at each timestep.}
    \label{fig:diurnal_conditioning_rsut}
\end{figure}

Figure \ref{fig:diurnal_conditioning_rsut} demonstrates the impact of noise schedule choice on diffusion models for atmospheric data. The bottom row shows results using a log-normal schedule ($P_{mean}=-1.2$, $P_{std}=1.2$) typically used for image generation \citep{Karras2022-edm}. This schedule introduces insufficient noise to mask the dominant diurnal cycle of outgoing shortwave radiation during training. Consequently, the model learns to detect and amplify residual signal patterns from noised inputs rather than using the time-of-day conditioning. However, during inference where starting from pure noise provides no such signal to extract, the model produces physically inconsistent results. Panel (g) shows this most clearly, with a ``double sun" artifact with radiation appearing in multiple incorrect locations.

In contrast, the log-uniform schedule (top row) with higher $\sigma_{max}$ provides sufficient noise to completely obscure the diurnal pattern during training, forcing the model to learn the relationship between time-of-day conditioning and the spatial distribution of solar radiation. This yields physically consistent results qualitatively with correctly positioned subsolar points. This visual comparison confirms the earlier theoretical argument (see Subsection\ref{sec:si:tuning-noise}) that atmospheric variables, some with planetary scale patterns, require substantially higher noise ceilings than is standard for diffusion to avoid \emph{signal leak bias}.

\paragraph{Multidiffusion overlap} The overlap size in multidiffusion sampling is a critical factor influencing both computational efficiency and output quality. Specifically, it determines the number of denoiser evaluations required: fewer evaluations reduce computational cost but can compromise visual consistency due to patch independence. A zero-overlap configuration, while minimizing runtime, often introduces visible artifacts at patch boundaries. Therefore, selecting an appropriate overlap size is essential to balancing reconstruction fidelity and efficiency.

\begin{figure}
    \centering
    \includegraphics[width=0.98\textwidth]{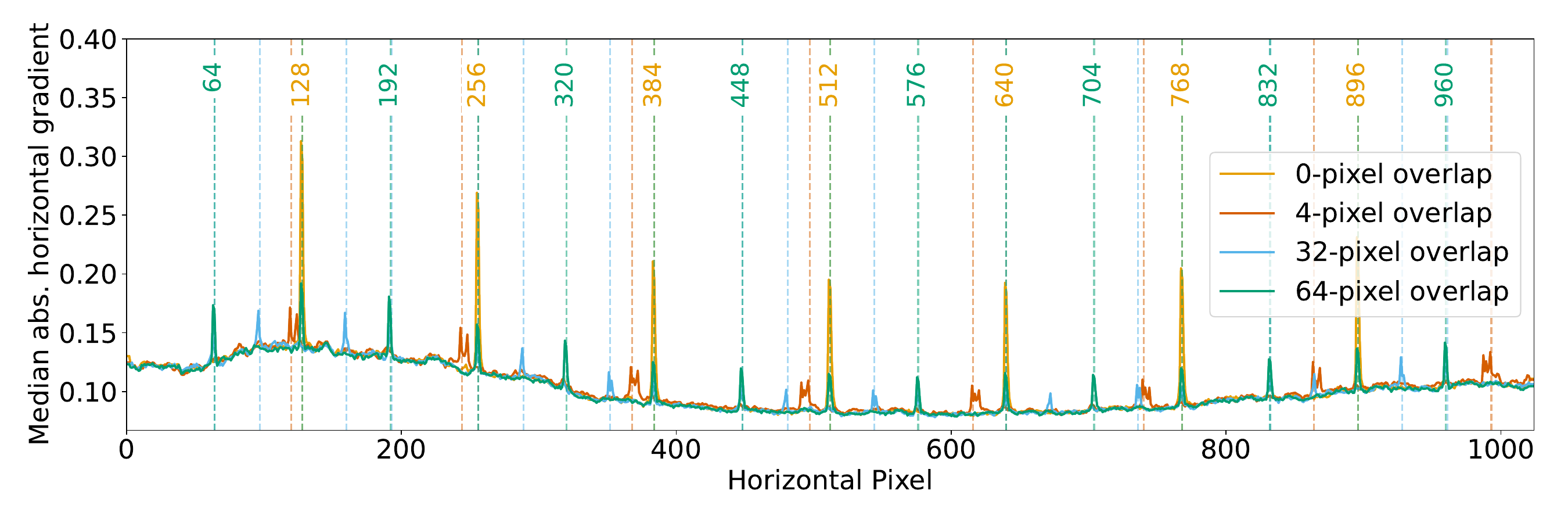}
    \caption{Severity of patch artifacts across different overlapping pixel sizes, visualized using the mean absolute Sobel-X gradient of surface temperature field as a function of horizontal pixel position. }
    \label{fig:overlap_gradient}
\end{figure}

To investigate this trade-off, we evaluate various overlap sizes using the median absolute horizontal Sobel gradient of surface temperature, as shown in Figure~\ref{fig:overlap_gradient}. This metric serves as a proxy for boundary artifacts, where higher gradient magnitudes indicate sharper discontinuities. To generate the plot, we aggregate all faces from 16 randomly sampled outputs. A Sobel-X filter is applied to the surface temperature field, and the median of the resulting gradient magnitudes is computed across the face, sample, and vertical dimensions.

As expected, the 0-pixel overlap configuration exhibits pronounced, periodic spikes at 128-pixel intervals, indicating strong discontinuities at patch edges. The 64-pixel overlap shows secondary spikes at 64-pixel intervals. The 4-pixel overlap yields slightly stronger artifacts than the 32-pixel case, suggesting that very narrow overlaps may be insufficient to effectively smooth patch boundaries.

\paragraph{Ensemble of experts and over-fitting}

\begin{figure}
    \centering
    \begin{tabular}{lll}
    a. ERA5 &b. ICON (overfit) & c. ICON (regularized) \\
       \includegraphics[width=0.3\linewidth]{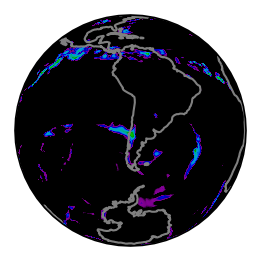} &
    \includegraphics[width=0.3\linewidth]{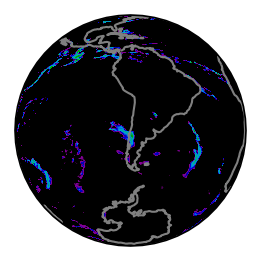} &
    \includegraphics[width=0.3\linewidth]{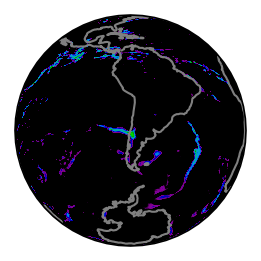} \\
    c.  &d. & e. \\
       \includegraphics[width=0.3\linewidth]{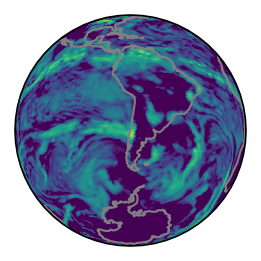} &
    \includegraphics[width=0.3\linewidth]{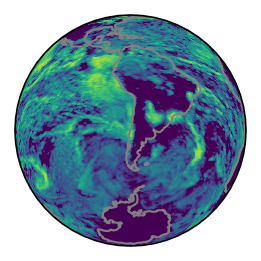} &
    \includegraphics[width=0.3\linewidth]{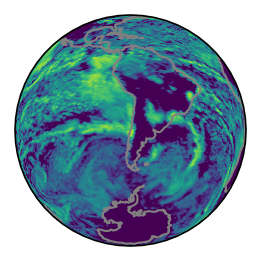} 
    \end{tabular}
    
    \caption{Impact of ensemble of experts regularization on the data-set translation example shown in Fig. \ref{fig:multimodal}. (a-c) dBR. (a) The original ERA5 dBR field. This sample translated to ICON by (b) the over-fit denoiser trained to 9.8M images and (c) the regularized ensemble of experts approach copied from Fig \ref{fig:multimodal}d. (d-e) is the same but for column integrated liquid water}
    \label{fig:si:overfit-translation}
\end{figure}

\begin{figure}
        \centering
        \includegraphics[width=1\linewidth]{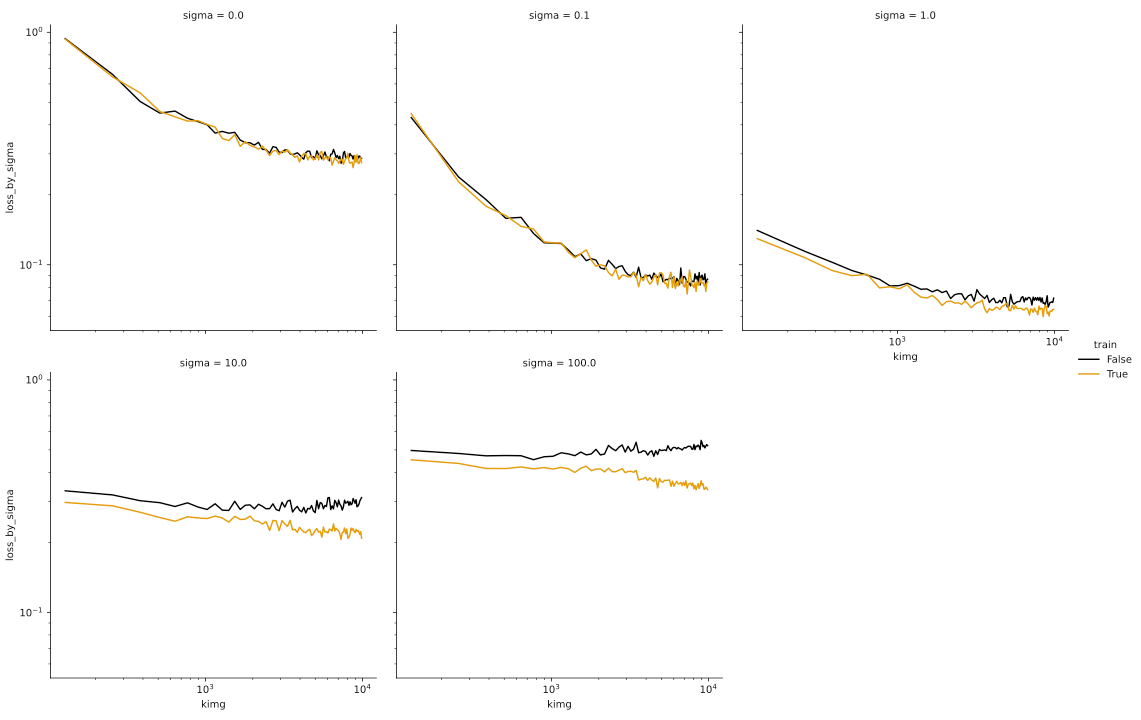}
        \caption{Diffusion loss versus $\sigma$ for different training iterations for the training and a subset of the validation set. From upper left, 
        $\sigma\in(0, 0.1]$, (0.1, 1], (10, 100], (100, 1000]. The x-axis is the training progress in 1000s of samples.}
        \label{fig:si:overfit}
\end{figure}

As discussed in Section \ref{sec:overfit} the coarse-resolution \texttt{cBottle-3d} is prone to over-fitting the largest modes of the data and propose a regularized ensemble of experts approach as a fix. This was initially pointed out to us by astute readers of an early version of this manuscript, who noticed the over-fitting in the Northern Annular Mode (see \ref{fig:ao-simple}b).  Intriguingly, this also improves the sample translation abilities demonstrated in Fig \ref{fig:multimodal}. The translated sample by the regularized approach Fig \ref{fig:si:overfit-translation}c much more closely resembles the original sample ( Fig. \ref{fig:si:overfit-translation}a). On the other hand, the over-fit model \ref{fig:si:overfit} hallucinates an entirely new storm system in the south pacific and elsewhere.

\end{document}